\documentclass[longauth]{aa} 

\usepackage{xspace}
\usepackage{natbib}
\bibpunct{(}{)}{;}{a}{}{,} 
\usepackage[colorlinks=true,citecolor=blue]{hyperref}

\usepackage{siunitx}

\DeclareSIUnit\parsec{pc}
\DeclareSIUnit\Year{y}

\usepackage{graphicx}
\usepackage{lscape}
\usepackage{threeparttable}
\usepackage{multirow}
\usepackage{footnote}
\usepackage{makecell}

\usepackage{txfonts}
\usepackage[dvipsnames]{xcolor}
\usepackage{caption}
\usepackage{subcaption}

\newcommand{\abratio}[2]{[\mathrm{#1}/\mathrm{#2}]\xspace}

\newcommand{\corr}[1]{#1\xspace}
\newcommand{\corevun}[1]{#1\xspace}
\newcommand{\corevdeux}[1]{#1\xspace}

\defcitealias{viscasillas22}{VV22}

\begin{document}

\title{The \emph{Gaia}-ESO survey: placing constraints on the origin of \emph{r}-process elements}
\titlerunning{The \emph{r}-process in the thin disc}
\authorrunning{Van der Swaelmen et al.}
\author{
  M. Van der Swaelmen\inst{\ref{oaa}}
  \and
  C. Viscasillas Vázquez\inst{\ref{vilnius}}
  \and
  G. Cescutti\inst{\ref{units},\ref{oats},\ref{infnts}}
  \and
  L. Magrini\inst{\ref{oaa}}
  \and
  S. Cristallo\inst{\ref{oaab},\ref{infnpg}}
  \and
  D. Vescovi\inst{\ref{guf}}
  \and
  S. Randich\inst{\ref{oaa}}
  \and
  G. Tautvai\v{s}ien\.{e}\inst{\ref{vilnius}}
  \and
  V. Bagdonas\inst{\ref{vilnius}}
  \and
  T. Bensby\inst{\ref{lund}}
  \and
  M. Bergemann\inst{\ref{mpiahe},\ref{nbia}}
  \and
  A. Bragaglia\inst{\ref{oabo}}
  \and
  A. Drazdauskas\inst{\ref{vilnius}}
  \and
  F. Jim\'enez-Esteban\inst{\ref{damadrid}}
  \and
  G. Guiglion\inst{\ref{postdam}}
  \and
  A. Korn\inst{\ref{dpasweden}}
  \and
  T. Masseron\inst{\ref{iac},\ref{datenerife}}
  \and
  R. Minkevičiūtė\inst{\ref{vilnius}}
  \and
  R. Smiljanic\inst{\ref{copernic}}
  \and
  L. Spina\inst{\ref{oapd}}
  \and
  E. Stonkutė\inst{\ref{vilnius}}
  \and
  S. Zaggia\inst{\ref{oapd}}
}

\institute{
  INAF - Osservatorio Astrofisico di Arcetri, Largo E. Fermi 5, 50125, Firenze, Italy \email{mathieu.van+der+swaelmen@inaf.it, laura.magrini@inaf.it}\label{oaa} 
  \and
  Institute of Theoretical Physics and Astronomy, Vilnius University, Sauletekio av. 3, 10257 Vilnius, Lithuania\label{vilnius} 
  \and
  Dipartimento di Fisica, Sezione di Astronomia, Università di Trieste, Via G. B. Tiepolo 11, 34143 Trieste, Italy\label{units}
  \and
  INAF, Osservatorio Astronomico di Trieste, Via Tiepolo 11, I-34143 Trieste, Italy\label{oats}
  \and
  INFN, Sezione di Trieste, Via A. Valerio 2, I-34127 Trieste, Italy\label{infnts}
  \and
  INAF - Osservatorio Astrofisico d'Abruzzo, via Maggini snc, 64100, Teramo, Italy\label{oaab}
  \and
  INFN - Sezione di Perugia, via A. Pascoli, 06123, Perugia, Italy\label{infnpg}
  \and
  Goethe University Frankfurt, Max-von-Laue-Strasse 1, Frankfurt am Main D-60438, Germany\label{guf}
  \and
  Lund Observatory, Department of Astronomy and Theoretical Physics, Box 43, SE-22100 Lund, Sweden\label{lund}
  \and
  Max Planck Institute for Astronomy, Königstuhl 17, 69117, Heidelberg, Germany\label{mpiahe}
  \and
  Niels Bohr International Academy, Niels Bohr Institute, University of Copenhagen Blegdamsvej 17, DK-2100 Copenhagen, Denmark\label{nbia}
  \and
  INAF - Osservatorio di Astrofisica e Scienza dello Spazio di Bologna, via Gobetti 93/3, 40129, Bologna, Italy\label{oabo} 
  \and
  Departamento de Astrof\'{\i}sica, Centro de Astrobiolog\'{\i}a (CSIC-INTA), ESAC Campus, Camino Bajo del Castillo s/n, E-28692 Villanueva de la Ca\~nada, Madrid, Spain\label{damadrid}
  \and
  Leibniz-Institut für Astrophysik Potsdam, An der Sternwarte 16, 14482 Potsdam, Germany\label{postdam}
  \and
  Department of Physics and Astronomy, Division of Astronomy and Space Physics, Angstrom laboratory, Uppsala University, Box 516, 75120 Uppsala, Sweden\label{dpasweden}
  \and
  Instituto de Astrofísica de Canarias, E-38205 La Laguna, Tenerife, Spain\label{iac}
  \and
  Departamento de Astrofísica, Universidad de La Laguna , E-38206 La Laguna, Tenerife, Spain\label{datenerife}
  \and
  Nicolaus Copernicus Astronomical Center, Polish Academy of Sciences, ul. Bartycka 18, 00-716, Warsaw, Poland\label{copernic}
  \and
  INAF - Padova Observatory, Vicolo dell'Osservatorio 5, 35122 Padova, Italy\label{oapd}
}

\date{Received 12/04/2022; accepted 08/07/2022}

\abstract
{A renewed interest about the origin of \emph{r}-process elements has been stimulated by the multi-messenger observation of the gravitational event \object{GW170817}, with the detection of both gravitational waves and electromagnetic waves corresponding to the merger of two neutron stars. Such phenomenon has been proposed as one of the main sources of the \emph{r}-process. However, the origin \corevun{of} the \emph{r}-process elements at different metallicities is still under debate.}
{We aim at investigating the origin of the \emph{r}-process elements in the Galactic thin disc population.}
{From the sixth internal data release of the \emph{Gaia}-ESO we have collected a large sample of Milky Way thin- and thick-disc stars for which abundances of Eu, O, and Mg are available. The sample consists of members of \num{62} open clusters, located at a Galactocentric radius from $\sim \SI{5}{\kilo\parsec}$ to $\sim \SI{20}{\kilo\parsec}$ in the disc, in the metallicity range $-0.5, 0.4$ and covering an age interval from \num{0.1} to \SI{7}{\giga\Year}, and about \num{1300} Milky Way disc field stars in the metallicity range $[-1.5, 0.5]$. We compare the observations with the results of a chemical evolution model, in which we varied the nucleosynthesis sources for the three considered elements.}
{Our main result is that Eu in the thin disc is predominantly produced by sources with short lifetimes, such as magneto-rotationally driven SNe. There is no strong evidence for additional sources at delayed times.}
{Our findings do not imply that there cannot be a contribution from  mergers of neutron stars in other environments, as in the halo or in dwarf spheroidal galaxies, but such a contribution is not needed to explain Eu abundances at thin disc metallicities.}

\keywords{Galaxy: abundances -- Galaxy: disc -- Galaxy: open clusters and associations: general -- stars: abundances}

\maketitle

\section{Introduction}
Most of the numerous chemical species making our Universe have been produced in the stellar interiors through nuclear processes occurring until the very last stages of a star's life. Chemical elements are classified in broad families depending on the nuclear process(es) and production site(s) responsible for their production. For instance, oxygen, magnesium, silicon, calcium are called $\alpha$-elements\footnote{titanium is often include in the list since its abundance behaves like an $\alpha$-element, though its atomic number is not a multiple of 4} since they are obtained by successive captures of $\alpha$ nuclei. However, a scrutiny shows that all of the aforementioned elements cannot be strictly treated as a whole since O and Mg, on the one hand, Si \corevun{and Ca (and Ti)}, in the other hand, are produced in stars of different mass-class and in different stages of stellar evolution. This difference translates in different yields and therefore, in a different pattern of chemical enrichment.

Elements with more protons than the iron nucleus are \corevun{mainly} produced by neutron accretion onto pre-existing iron seeds. This accretion is defined as slow (\emph{s}-process) or rapid (\emph{r}-process), with respect to the $\beta$-decay timescale \citep{Burbidge_1957}. The rapid neutron-capture process, which is responsible for about half of the production of the elements heavier than iron \citep[see, e.g.][]{Kajino19,2021RvMP...93a5002C}, is not yet fully understood, and an interdisciplinary analysis is needed to reach an adequate comprehension of all the facets of the issue. Such an approach should take into account nuclear astrophysics, observational results from stellar spectroscopy, gravitational waves, short gamma-ray bursts (GRBs), and galaxy formation theories \citep[see, e.g.][]{cote19}. 

A renewed interest in the origin of \emph{r}-process elements has been stimulated by the multi-messenger observation (detection of both gravitational waves and electromagnetic waves) of the the gravitational event \object{GW170817}, corresponding to the merger of two neutron stars (NSM; \citealp{2017ApJ...848L..12A,2017Natur.551...75S,2017Natur.551...80K}). The spectroscopic follow-up of the fading glow of the kilonova \object{AT 2017gfo}, associated to this NSM, showed that the radiation is powered by the radioactive decay of lanthanides. The modelling of the observed broad absorption features in the late-time spectra was shown to be compatible with bands of heavy \emph{r}-process elements such as cesium and tellurium \citep{2017Natur.551...75S}. On the other hand, the multi-epoch analysis of the early spectra revealed the presence of Sr \citep{watson19}, indicating this element as a common by-product of such events \citep{perego2022}, despite its production is mostly due to the \emph{s}-process \corevun{at solar metallicity} \citep{Prantzos_2020}. These studies have revived the interest towards NSMs as a credible production sites of \emph{r}-process elements \citep{2017Natur.551...67P}. However, numerous parameters controlling the production of \emph{r}-process by NSMs are yet to be estimated: yields, time-delay, frequency, merging rate \citep[e.g., see][for a discussion on the coalescence time]{2016MNRAS.455...17V,2018ApJ...865...87O}.

If \object{GW170817} is likely the first observation of \emph{in situ} production of heavy elements by the \emph{r}-process, it does not yet answer the question of the origin of \emph{r}-elements. Several possible sites of productions and physical mechanisms have been considered in the latest decades \citep[see][for a complete review and references therein]{Kajino19} and are still under study. Here we briefly recall the most popular ones\footnote{The literature on the \emph{r}-process sites being very rich, we tried to quote in this introduction the early works for each investigated \emph{r}-production site.}: {\em i)} neutrino-driven winds above proto-neutron stars in core-collapse supernovae (CCSNe), which is likely the site of production of the \emph{weak} \emph{r}-process and produce neutron rich nuclei up to about A$\sim$125 \citep{1994ApJ...433..229W}; {\em ii)} magnetic neutrino-driven wind, which provides a possible mechanism for nucleosynthesis of rare heavy elements \citep{2018MNRAS.476.5502T}; {\em iii)} shock-induced ejection of neutron-rich material in CCSNe with $M < 10 M_{\odot}$ \citep{1984A&A...133..175H}; {\em iv)} compact-object binary mergers, which can involve both two neutron-stars (NSM) or a neutron star and a black hole binary system (NS-BH) \citep{ls74,ros05,2011ApJ...738L..32G,2012MNRAS.426.1940K}. In these systems, the ejected matter can be very neutron-rich and it can produce elements up to A$\sim$300; {\em v)} magneto-hydrodynamic jet (MHDJ) supernova model, in which magnetic turbulence launches neutron rich material into a jet, undergoing \emph{r}-process nucleosynthesis \citep{2006ApJ...642..410N}; {\em vi)} collapsar (failed supernovae) might produce \emph{r}-process, through neutron-rich matter coming from the accretion disc and ejected into a relativistic jet along the polar axis \citep{2006ApJ...644.1040F}; {\em vii)} \emph{r}-process from dark matter induced black hole collapse \citep{2016ApJ...826...57B}; {\em viii)} truncated \emph{tr}-process from fall-back supernovae, in which there is a first collapse forming a neutron star and a subsequent infall causing the formation of a black hole. The \emph{r}-process is interrupted when the neutron star collapses to a black hole \citep{2008JPhG...35b5203F}. Moreover, the \emph{i}-process \citep[e.g.,][]{Mishenina15}, characterized by neutron densities intermediate ($n \approx 10^{14}-10^{18}\,\SI{}{\per\cubic\centi\metre}$) between those of the \emph{s}- ($n \approx 10^{6}-10^{10}\,\SI{}{\per\cubic\centi\metre}$) and \emph{r}-process ($n > 10^{20}\,\SI{}{\per\cubic\centi\metre}$; \citealp[e.g.,][]{2016ApJ...831..171H}), may play a role in the formation of the elements heavier than iron in low-mass, low-metallicity asymptotic giant branch (AGB) stars. 

What emerges from this long list of possible production sites is that the theoretical framework is extremely varied and complex, and strong observational constraints are needed to choose the dominant production scenarios. On the one hand, one of the most commonly adopted approaches to pose observational constraints on the \emph{r}-process nucleosynthesis are spectroscopic observations of the metal-poor stars in the halo of our Galaxy. They can be indeed used to trace the \emph{r}-process nucleosynthesis \citep[see, e.g.][]{frebel18, Horowitz19}, since the production of most neutron-capture elements is dominated by the \emph{r}-process in the early stage of formation of the Galaxy. The enhanced scatter of halo low-metallicity stars in the $\abratio{Eu}{H}$ vs. $\abratio{Fe}{H}$ plane, compared to the one of $\abratio{\alpha}{H}$ vs. $\abratio{Fe}{H}$, is a hint that the production of Eu in the early epochs of Galactic evolution might have been more stochastic compared to the production of the $\alpha$ elements, which are mainly produced by CCSNe \citep[see, e.g.][]{cescutti15}. On the other hand, spectroscopic observations of stellar populations in the thin and thick discs give us information about the contribution of the \emph{r}-process in more recent times. However, starting at $\abratio{Fe}{H} > -1.5$, stars do not present only \emph{r}-process enrichment, since the production of neutron-capture elements by the \emph{s}-process starts to widely contribute to their abundance pattern \citep[see, e.g.][]{gallino1998}. For this reason, the choice of chemical elements with a tiny production by the \emph{s}-process and therefore, with a production still largely dominated by the \emph{r}-process at solar metallicity is to be preferred to probe the evolution of the \emph{r}-process in the Milky-Way discs. Europium is an ideal element in this respect since \SI{95}{\percent} of Eu is predicted to be indeed produced by the \emph{r}-process at the time of formation of the Solar system \citep{Prantzos_2020}, given our knowledge of the \emph{s}-process yields \citep[see, e.g.][]{cri11,cri15,Bisterzo14, karakas16} and of the possible role of the \emph{i}-process \citep[e.g.][]{deni19}.

In this work, we will use the data from the sixth data-release (iDR6) of the \emph{Gaia}-ESO survey \citep{Gilmore_2012, Randich_2013} to study the origin and the role of the \emph{r}-process in the Milky-Way discs, analysing abundances of both field and cluster stars. We consider the abundances of Pr, Nd, Mo and Eu. Following \citet{Prantzos_2020}, the abundances of these elements had a strong to moderate contribution from the \emph{r}-process when the Solar system has formed: \SI{95}{\percent} for Eu, \SI{27}{\percent} for Mo, \SI{47}{\percent} for Pr and \SI{39}{\percent} for Nd. \corevdeux{Other elements are known to have a strong contribution from the \emph{r}-process, like Sm or Dy, but those elements could not be measured in the \emph{Gaia}-ESO spectra. On the other hand, the production of elements like Ba or La is dominated by the \emph{s}-process (see, e.g. \citet{arlandini99} for their s-process percentages in the Sun, ranging from 81 to 92\% for Ba and from 62 to 83\% for La) and are, therefore, out of the scope of this work.} We add to our analysis the abundance of Mg and O, elements mostly produced by CCSNe, which are a useful comparison for identifying the timescale of the \emph{r}-process.

The paper is structured as follows: in Section~\ref{sec_data} we describe the \emph{Gaia}-ESO dataset, and the two samples of open cluster stars and of field stars adopted in the present work. In Section~\ref{sec_model}, we describe the Galactic Chemical Evolution (GCE) model and its assumptions. We present our results both as a function of age and of metallicity in Section~\ref{sec_results}. In Section~\ref{sec_summary} we discuss the implication of our results for the sites, mechanisms, and timescales of the \emph{r}-process, providing our conclusions and summarizing our results.

\section{Data and sample selection}
\label{sec_data}

\subsection{The \emph{Gaia}-ESO survey}

For this work, we used the sixth data release of the \emph{Gaia}-ESO survey \citep{Gilmore_2012, Randich_2013}, selecting the highest resolution spectra obtained with UVES (resolving power $R = \num{47000}$ and spectral range $480 - \SI{680}{\nano\metre}$). The data reduction and analysis was done within the \emph{Gaia}-ESO consortium, which is organized in several working groups (WG). The spectral analysis was performed with a multi-pipeline approach: different nodes analysed the same dataset, and their results are combined to produce a final set of parameters and abundances. The homogenization process made use of calibrators (benchmark stars, open and globular clusters), selected following the calibration strategy described in \citep{pancino17}. The analysis of the UVES data for FGK stars is described in \citet{smi14}, and can be summarized in the following steps: INAF-Arcetri took care of the data reduction and of the radial and rotational velocity determination \citep{sacco14}; reduced spectra are distributed by the working group 11 (WG~11) to the analysis nodes, which performed their spectral analysis, providing stellar parameters; the nodes' stellar parameters were homogenized by WG~15, and then redistributed to the nodes for the elemental abundances (line-by-line); WG~11 homogenized and combined the line-by-line abundances, providing the final set of elemental abundances, which are finally validated and homogenized by WG~15. The recommended parameters and abundances were distributed in the iDR6 catalogue, internally to the \emph{Gaia}-ESO consortium, and they are publicly available through the ESO portal. In this work, we use: the atmospheric stellar parameters, such as effective temperature, $T_{\mathrm{eff}}$, surface gravity, $\log g$, metallicity\footnote{In this paper, we use metallicity, iron abundance, $\abratio{M}{H}$ and $\abratio{Fe}{H}$ as synonyms.} $\abratio{Fe}{H}$, and the abundances of four \emph{r}-process and two $\alpha$-elements.

One of the most important aspects of \emph{Gaia}-ESO, compared to other spectroscopic surveys, is that it dedicated about \SI{36}{\percent} of its observing time to open star clusters. As it is well known, open clusters offer the unique advantage of allowing a more precise measurement of their ages and distances than isolated stars. Moreover, the observation of several members of the same cluster also provides reliable measurements of their chemical composition. We can therefore reasonably consider open clusters among the best tracers of the chemical evolution of our Galaxy. On the other hand, open star clusters, by their intrinsic characteristics, are limited in the age and metallicity ranges they span, being a thin disc population. In this context, it is a benefit to complement the use of clusters with that of field stars also studied by the \emph{Gaia}-ESO, which reach older ages and lower metallicities. and whose abundances are on the same abundance scale as the ones of open clusters.

\subsection{The open cluster sample}
In this work, we use the \num{62} open clusters with age $\geq$ \SI{100}{\mega\Year} available in the \emph{Gaia}-ESO iDR6. Not including the youngest clusters does not affect our approach based on chemical evolution, and it also eliminates problems related to the analysis of the youngest stars, whose abundances may be affected by several issues, like stellar activity \citep[see, e.g.][]{spina20, Baratella_2020, baratella21}. For our sample clusters, we used the homogeneous age determination obtained in \citet{CG20} using the second data release of \emph{Gaia}. The $\abratio{Fe}{H}$ are from \citet{Randich_2022}, except for the clusters not present in that work for which they were calculated in this work. 

\begin{figure}
  \resizebox{\hsize}{!}{\includegraphics{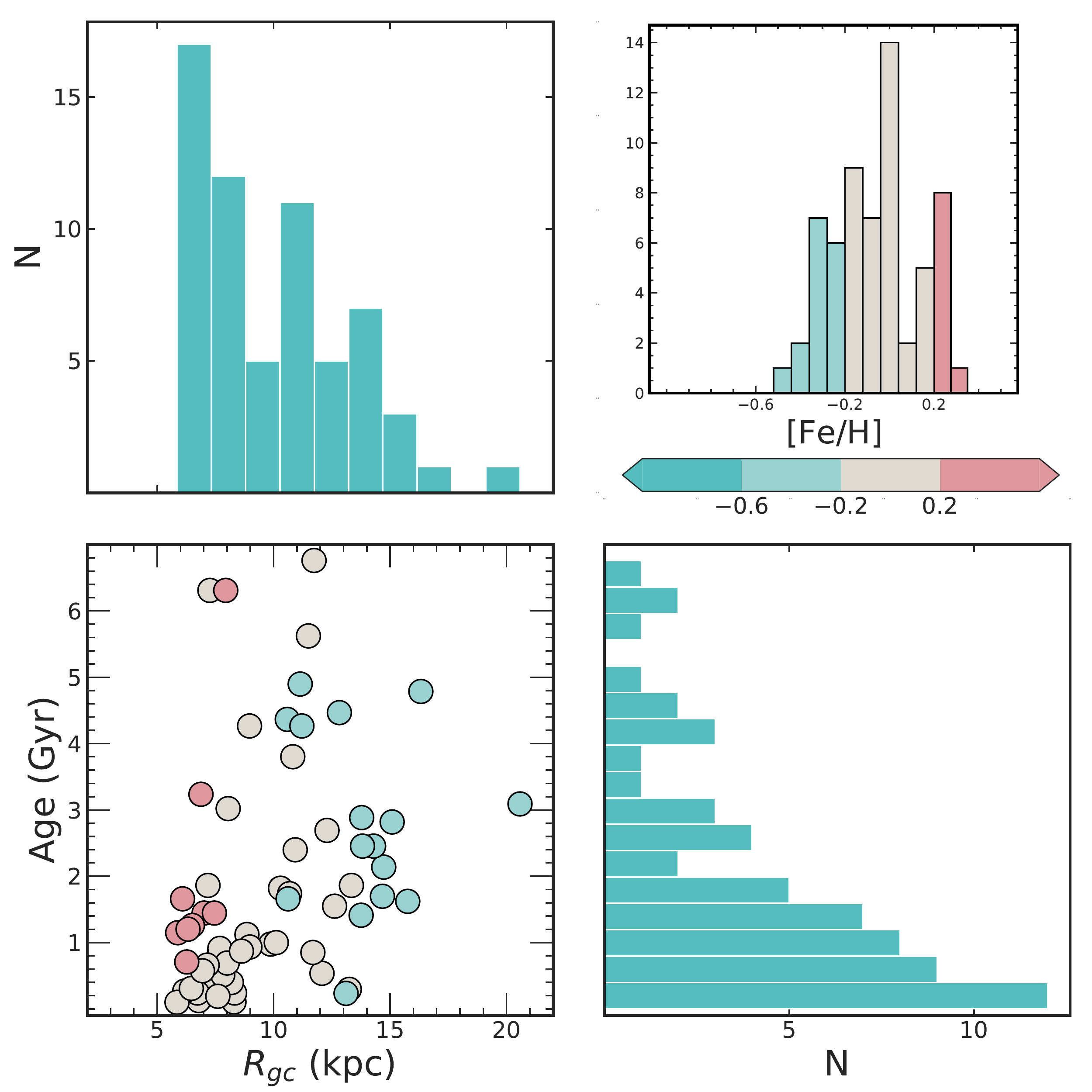}}
  \caption{\label{age_rgc_feh}Properties of our sample of \num{62} OCs. Upper left panel: histogram of $R_{\mathrm{GC}}$; lower left panel: distribution of clusters in the age vs. $R_{\mathrm{GC}}$ plane, colour-coded by $\abratio{Fe}{H}$; upper right panel: histogram of cluster metallicity; lower right panel: histogram of the ages.}
\end{figure}

The membership analysis was performed as in \citet[][hereafter VV22]{viscasillas22}. Figure~\ref{age_rgc_feh} shows the distributions of the properties of the sample of \num{62} OCs: the Galactocentric distance $R_{\mathrm{GC}}$, the age and the metallicity $\abratio{Fe}{H}$. The sample covers a wide range in $R_{\mathrm{GC}}$, from about \num{5} to \SI{20}{\kilo\parsec}, in age, from \num{0.1} to \SI{7}{\giga\Year}, and in metallicity $\abratio{Fe}{H}$, from \num{-0.45} to \num{0.35}. As explained in the next paragraphs, some clusters disappear from the analysis depending on the availability of the abundances for oxygen, magnesium and europium.

For any star, we remove the abundance of a given element if the uncertainty on the given abundance is $\geq 0.1$. We also removed the outliers from each cluster, adopting the same approach used in \citetalias{viscasillas22}, i.e. the interquartile range (IQR) method. This results in discarding \num{23} stars with Eu values out of range of the other stars in the same cluster: \num{10} of them extremely rich and \num{13} extremely poor compared to the other member stars of their respective OCs (see Fig.~\ref{fig:outlier_eu}). These stars, listed in table \ref{tab:outliers_eu} in the Appendix, will be analysed in a future work. In particular, we would like to mention one of them: the star with CNAME\footnote{the CNAMEs reported throughout this publication are the ID assigned by the \emph{Gaia}-ESO survey.} \object{06025078+1030280} in the open cluster NGC\,2141 (or \object{NGC\,2141\,4009}) was already mentioned in \citetalias{viscasillas22} for its extremely low abundance in all its \emph{s}-process elements, and now we recall it again for its low $A(\mathrm{Eu})$ value. 

After applying the selection cuts described above, the sample is reduced to \num{59} open clusters with Eu abundances, \num{62} OCs with Mg abundances and \num{38} OCs with O abundances. The reason why fewer clusters have data for oxygen is that the only measured atomic line -- the forbidden $[\ion{O\,I}]$ at \SI{6300}{\angstrom} -- is a weak line, potentially contaminated by telluric lines (depending on the radial velocity of the star). No telluric correction has been performed by the \emph{Gaia}-ESO data-reduction nodes; therefore, the forbidden O line shall be discarded when affected by the tellurics. In the case of a cluster, it means losing the whole set of member stars at a given epoch since all member stars have a similar radial velocity. We recall that the O line is also blended by a Ni line \citep{2003ApJ...584L.107J} whose contribution is accounted for by means of line profile fitting (see \citealp{2015A&A...573A..55T} for a description of the CNO determination method and see Fig.~6 in \citealp{2004A&A...415..155B} highlighting how the contribution of the Ni blend changes with the star's metallicity).

\begin{figure}
  \resizebox{\hsize}{!}{\includegraphics{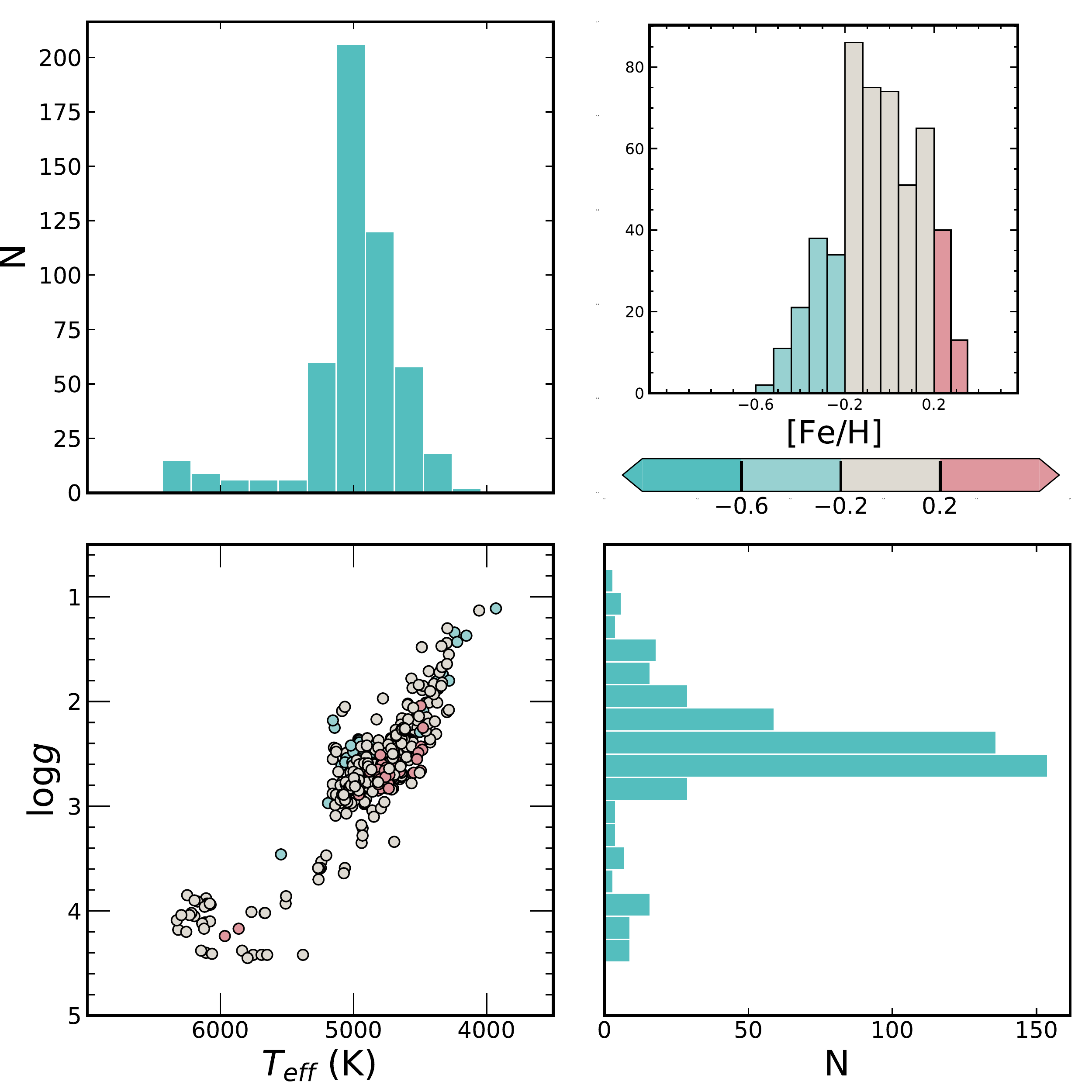}}
  \caption{\label{KD_clusters}Properties of the members of the cluster sample. Upper left panel: histogram of $T_{\mathrm{eff}}$; lower left panel: Kiel diagram colour-coded by $\abratio{Fe}{H}$; upper right panel: histograms of $\abratio{Fe}{H}$; lower right panel: histogram of $\log g$.}
\end{figure}

The Kiel diagram (KD) and the histograms of the distributions of stellar parameters ($\log g$, $T_{\mathrm{ eff}}$, $\abratio{Fe}{H}$) of member stars in the OC sample are shown in Fig.~\ref{KD_clusters}. The sample contains both dwarf and giant members, with a predominance of giants. The non-members are incorporated into our field-star sample.

In Tables~\ref{clusters_elh} and \ref{clusters_elfe} of the appendix, we provide the global metallicity of each cluster from \citet{Randich_2022}, together with $R_\mathrm{GC}$ and age \citep{CG20}, and the abundance ratios used along the paper with their $\sigma$. We provide both $\abratio{El}{H}$ and $\abratio{El}{Fe}$: the \corevun{computation of the latter using the former} is not straightforward since the \corevun{reported} overall metallicity $\abratio{Fe}{H}$ is generally calculated with a larger number of members.

\subsection{The field star sample}

The field star sample is made up of stars whose "GES{\_}TYPE" header keyword of the spectra in the \emph{Gaia}-ESO classification system corresponds to MW targets, which include  halo, thick disc and thin disc populations of the Milky Way. To that sample we also added benchmark stars ("SD") and the non-member stars of the OC sample (see above). We applied two quality cuts, the first one on the stellar parameters and on the signal-to-noise ratio (SNR): SNR $> 20$; $e[T_{\mathrm{eff}}] \le \SI{150}{\kelvin}$, $e[\log g] \le 0.25$, $e[\abratio{Fe}{H}] < 0.20$ and $e[\xi] \le \SI{0.20}{\kilo\meter\per\second}$; the second one on the abundances, considering only stars with $e[A(\mathrm{El})] \le 0.1$. We made a further selection, considering only the stars for which, at least, \ion{Eu\,II}\ and an $\alpha$-element \corevun{(\ion{Mg\,I} or \ion{O\,I})} could be measured. This reduces the sample to $\sim$\num{1300} stars.

\begin{figure}
  \resizebox{\hsize}{!}{\includegraphics{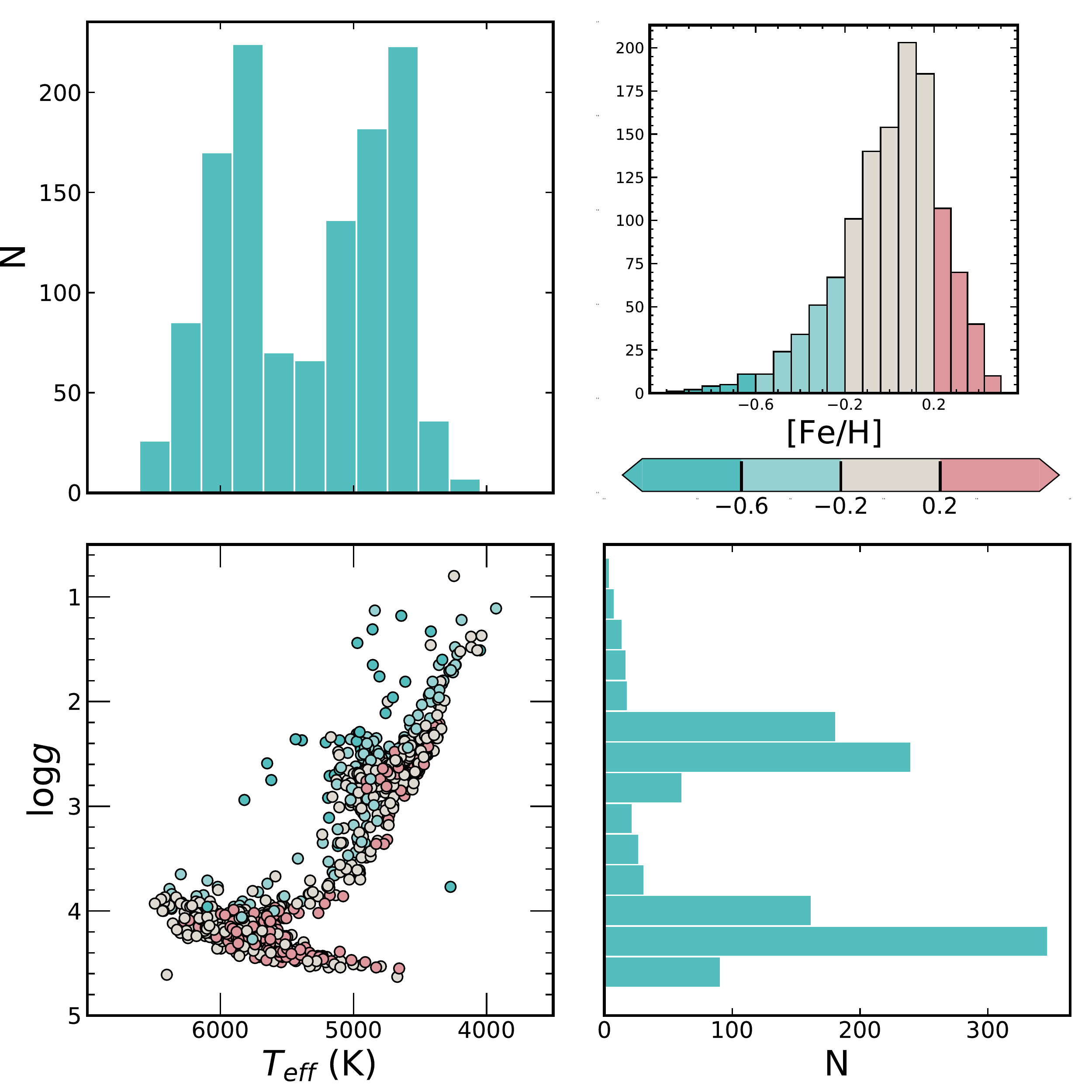}}
  \caption{\label{KD_field}Properties of the field-star sample. Definitions of the panels, symbols and colours as in Figure~\ref{KD_clusters}.}
\end{figure}

We did not apply to field stars any cut for possible outliers, which might be indeed stars of particular interest. \corevdeux{However, we checked the barium and carbon content of our selection: we find a solar mean $\abratio{Ba}{Fe}$ (standard deviation of $\sim 0.1$) and a slightly sub-solar $\abratio{C}{Fe}$ (standard deviation of $\sim 0.15$). For both elements, \SI{99}{\percent} of the sample has $\abratio{C, Ba}{Fe} \in [-0.2, 0.2]$, which is comparable to what is observed for the MW discs in other studies \citep[e.g., with GALAH data,][]{2021MNRAS.506..150B}. In addition, carbon-enhanced metal-poor stars (CEMP) with possibly enhanced \emph{s}- (e.g., Ba) or \emph{r}- (e.g., Eu) abundances are not expected in the metallicity range of this study \citep[e.g., see][]{2010A&A...509A..93M,2021A&A...649A..49G}. Barium stars (main-sequence and red giant stars that have accreted the \emph{s}-rich envelop of the former AGB companion now an extinct white dwarf; e.g., \citealp{2019A&A...626A.127J,roriz21}) can be found at our metallicities but there is no sign of it from individual abundances as shown above (though the thresholds are not settled, mild Ba stars are expected to have $\abratio{Ba}{Fe} \ge \sim 0.25$ and Ba stars often have $\abratio{Ba}{Fe}$ ranging from \num{1} to \num{2}).} The KD and the distribution of stellar parameters of our sample of field stars are shown in Fig.~\ref{KD_field}. We computed the ages of the field stars, which are predominantly main sequence stars at the turn off, using the \textsc{aussieq2} tool that is an extension of the qoyllur-quipu (\textsc{q2}) \textsc{Python} package \citep{ramirez14a}. It calculates stellar ages by isochrone fitting, starting from the stellar parameters, and adopting a grid of isochrones. In the calculation, the code also takes into account the uncertainties on the stellar parameters. 

\subsection{The definition of the solar scale}

In Table \ref{tab:solarabundance}, we show the abundances of \emph{r}-process dominated and mixed elements and of the two $\alpha$ elements, O and Mg in the Sun (for \emph{Gaia}-ESO iDR6 and from \citealt{Grevesse_2007}) and in the open cluster \object{M67} (mean values obtained for the whole sample of \object{M67} member stars, and for giant and dwarf stars, separately). The cluster \object{M67} has indeed a chemical composition very similar to the solar one \citep[see, e.g.][]{onehag11, liu16}, and it is often used to normalize the abundances to the solar scale in sample containing both giant and dwarf stars \citep[see, e.g.][VV22]{Magrini_2018}. The \emph{Gaia}-ESO measurements for the solar abundances agree with those of \citet{Grevesse_2007}. The most discrepant element is Mo, but nevertheless in agreement within $2 \sigma$. The iDR6 abundances in the Sun and in \object{M67} (mean value) are in agreement, within $1 \sigma$. Small differences can be appreciated between the average abundances for the \object{M67} giants and the \object{M67} dwarfs, particularly for Pr. Following \citetalias{viscasillas22}, we normalized the abundances of the dwarf and giant stars in our samples by the mean abundances of the \object{M67} dwarf and \object{M67} giant stars respectively (we refer to \citetalias{viscasillas22} for more details). For Mo, for which we have only abundances in the Sun and in the giants of \object{M67}, we used the former to normalise the abundances of dwarf stars, and the latter for the giant ones.

\begin{table*}
  \caption{\label{tab:solarabundance}iDR6 solar and \object{M67} abundances for \emph{r}-process elements and $\alpha$-elements.}
  \centering
  \begin{tabular}{l c c c c c}
    \hline\hline
    Species & Sun (iDR6) & Sun & \object{M67} (iDR6) & \object{M67} (iDR6) & \object{M67} (iDR6) \\ 
    & & \citep{Grevesse_2007} & & (giants) & (dwarfs)\\
    \hline
    $\ion{O}{i}$ & 8.66 $\pm 0.05$& 8.66 $\pm 0.05$ & 8.74 $\pm 0.08$  & 8.73 $\pm$ 0.06  & 8.8 $\pm$ 0.01\\
    $\ion{Mg}{i}$ & 7.51 $\pm 0.02$& 7.53 $\pm 0.09$ & 7.50 $\pm 0.05$  & 7.53 $\pm$ 0.04  & 7.49 $\pm$ 0.04\\
    \hline 
    $\ion{Mo}{i}$ & 2.01 $\pm 0.06$ & 1.92 $\pm 0.08$ & 1.92$\pm 0.03$ & 1.92 $\pm$ 0.03 & - \\    
    $\ion{Pr}{ii}$ & 0.57$\pm 0.02$ & 0.58 $\pm 0.10$ & 0.57$\pm 0.07$ & 0.54 $\pm$ 0.03 & 0.61 $\pm$ 0.04\\    
    $\ion{Nd}{ii}$ & 1.49$\pm0.02$ & 1.45 $\pm 0.05$ & 1.44$\pm0.06$ & 1.41 $\pm$ 0.04 & 1.45 $\pm$ 0.07\\  

    $\ion{Eu}{ii}$ &0.52$\pm0.02$  & 0.52 $\pm 0.06$ & 0.54$\pm 0.11$ & 0.56 $\pm$ 0.08 & 0.53 $\pm$ 0.11\\  
    \hline
  \end{tabular}
\end{table*}

\section{The Galactic chemical evolution model}
\label{sec_model}

The chemical evolution model adopted is based on the two-infall model \citep{chiappini1997}; there is a first and brief infall mimicking the formation of the thick-halo component followed by a hiatus in the star formation and by a more extended infall promoting the formation of the thin disc. Open clusters are formed during the second episode, therefore a different modelling of the first infall should not change our results \citep[see for example the  recent paper by][]{spitoni2019}. On the contrary, the inside-out formation of the Galactic disc plays a fundamental role, and for this, we follow the model B described in \citet{chiappini01}, shown to be the best model in the comparison with cepheids stars in \citet{cescutti07}. As highlighted in \citet{cescutti07}, the timescale of the enrichment dictates the steepness of the Galactocentric gradient for the chemical element. Flatter gradients are expected for elements produced on short timescales such as $\alpha$-elements, produced by massive stars and ejected in the interstellar medium (ISM) by CCSNe on timescales of few tens of million years \citep{woosley1995}. On the other hand, elements produced mostly on longer timescales as for example iron, which is mostly produced by SNe\,Ia \corr{\citep{Nomoto_1997}}, tend to present steeper gradients.

The original yields used for our simulations are based on the yields described in \citet{francois04} for oxygen, magnesium and iron. These elements are produced by massive stars and SNe\,Ia. At the solar metallicity, most of the enrichment of magnesium and oxygen comes from massive stars; on the contrary, Fe is mostly produced by SNe\,Ia. For the europium yields, we assume in this work two possible production modes.

In the first \corevun{model (model A)}, all the production takes place on a short timescale, so with no delay in the enrichment of the interstellar medium. In particular, we consider the same yields for Eu adopted in \citet{cescutti14}, where the main producers were the magneto-rotationally driven (MRD) SNe \citep[see][]{nishimura15}, \corevun{so a yield of $1 \cdot 10^{-6} M_{\odot}$ per MRD SNe assuming that only \SI{10}{\percent} of all the simulated massive stars explode as MRD SNe}. This production is compatible with the enrichment by neutron star mergers having short delay \citep{matteucci14,cescutti15}.

We run a second set of simulation (model B) with a second set of yields where we consider the substantially increased production (a factor of 5) of magnesium coming from SNe\,Ia. Since the model needs to respect the constraint dictated by the solar value, we have to decrease accordingly by a factor of 0.7 the yields for magnesium from CCSNe.
The main consequence of this change is to have a larger fraction of magnesium produced on longer timescale. This has an impact on the chemical evolution trend of this element in the $\abratio{Mg}{Fe}$ vs. $\abratio{Fe}{H}$; the typical enhancement at low metallicity is less pronounced and the subsequent slope is also less steep. This possibility was already discussed in \citet{magrini17}. 

We consider a model C where the yields for magnesium produced by SNe\,Ia have also a metal dependency which we impose empirically with this equation:

\begin{equation}
  Y_{Mg}^{SNeIa}=0.255\frac{z}{z_{\odot}}[M_{\odot}]
\end{equation}

With this metal dependency, the solar ring simulated by our GCE model is not expected to have significant variation; on the other hand, the outer rings tend to end their evolution with lower $\abratio{Mg}{Fe}$ compared to model B. In fact, due to the inside-out formation, the progenitors of SNe\,Ia present lower metallicity and this will inhibit the formation of Mg. 

\corevun{Finally, we run a fourth model (model D) considering the enrichment of europium from both neutron star mergers and the same short time-scale source as in the original set of yields. The original yields are evenly split between these two sources (\SI{50}{\percent} from NSMs and \SI{50}{\percent} from MRD SNe); the magnesium yields are the same as model C. We do not show results assuming a single production for europium from NSMs since \citet{cote19} and \citet{simonetti19} have already proved this scenario not compatible with the chemical evolution of europium in the Galactic disc. We present the results with a fixed delay of \SI{3}{\giga\Year} since we have already introduced a degree of elaborateness with this double Eu production. In this way, we want also to produce results comparable to the model described in \citet{skuladottir20} with a similar delay time (\SI{4}{\giga\Year}). The yields for each of these objects in our model is $1.5 \cdot 10^{-6} M_{\odot}$}.

The main assumptions of the four models for the yields of O, Mg and Eu are reported in Table~\ref{Table:summary_models}. 

\renewcommand{\arraystretch}{2.5}
\begin{table*}
  \caption{\label{Table:summary_models}Overview of the underlying assumptions for the production of O, Mg and Eu for model A, B, C and D. The words 'increased' and 'reduced' qualify the contribution of a given nucleosynthetic source and should be understood as relative to the assumptions in model A.}
  \centering
  \begin{tabular}{l c c c}
    \hline\hline
    Model & Source of oxygen & Source of magnesium & Source of europium\\
    \hline
    Model A & CCSNe & \makecell{CCSNe\\ (+ marginal contribution by SNe\,Ia)} & \corevun{MRD SNe}\\
    \hline
    Model B & CCSNe & \makecell{CCSNe (reduced)\\ + SNe\,Ia (increased)} & \corevun{MRD SNe}\\
    \hline
    Model C & CCSNe & \makecell{CCSNe (reduced)\\ + SNe\,Ia (increased\\ and metal-dependent yields)} & \corevun{MRD SNe}\\
    \hline
    Model D & CCSNe & \makecell{CCSNe\\ (+ marginal contribution by SNe\,Ia)} & \makecell{\corevun{MRD SNe} (\SI{50}{\percent})\\ + NSMs (\SI{50}{\percent})}\\
    \hline
  \end{tabular}
\end{table*}
\renewcommand{\arraystretch}{1}

\section{Results}
\label{sec_results}

To investigate the origin of Eu in the Galactic disc, we compare its evolution with that of two $\alpha$-elements which are expected to be mainly produced by CCSNe, on short timescales, i.e. Mg and O. The aim of our approach is to reveal possible differences in the production timescales of Eu with respect to the production timescales of these two $\alpha$-elements, and to possibly highlight the need of a delayed nucleosynthetic channel for Eu, as expected by neutron star mergers \citep{korobkin2012}. Although O and Mg are essentially produced by stars with masses in the same range, they are generated during different burning phases: oxygen is produced during the hydrostatic burning in the He-burning core and in the C shell and it is expelled during the pre-supernova phase, while magnesium is produced during the hydrostatic burning in the C shell and in the explosive burning of Ne \citep[see, e.g.][]{maeder05}. Therefore, we can expect differences in the evolution of these two elements. Moreover, for Mg, observational evidence has shown that the production from massive stars is not sufficient to explain its behaviour at high metallicity. Several attempts have been made to explain the evolution of Mg, and its difference from that of oxygen, such as the use of metallicity-dependent yields of massive stars, the production from hypernovae at solar and/or higher than solar metallicity, a larger contributions from SNe\,Ia, or significant Mg synthesis in low- and intermediate-mass stars, or a mixture of all these production sites \citep[see, e.g.][]{ chiappini05, romano10, magrini17}. As described in Section~\ref{sec_model}, to take into account the complexity of Mg production, we considered three different representations of the production of Mg: only CCSNe (model A), CCSNe and SNe\,Ia (model B), CCSNe and metal-dependent SNe\,Ia production (model C). As for the Eu production, we investigate two scenarios: only \corevun{MRD SNe} (model A, B, C) and an evenly-mixed production by \corevun{MRD SNe} and NSMs (model D).

\subsection{The evolution of Eu}
\label{Sec:Evolution_EuFe}

\begin{figure}
  \centering
  \includegraphics[width=8cm]{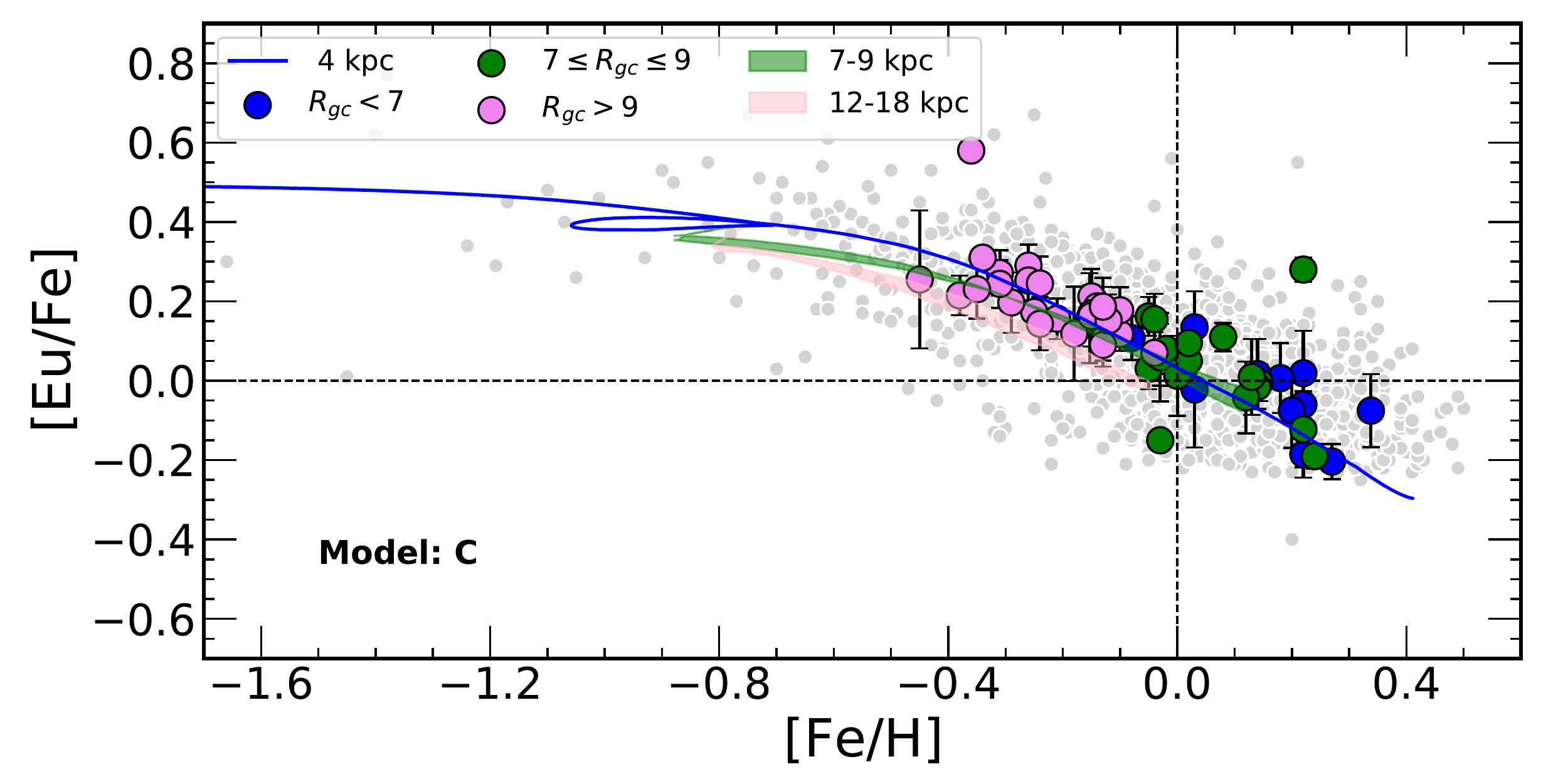}
  \includegraphics[width=8cm]{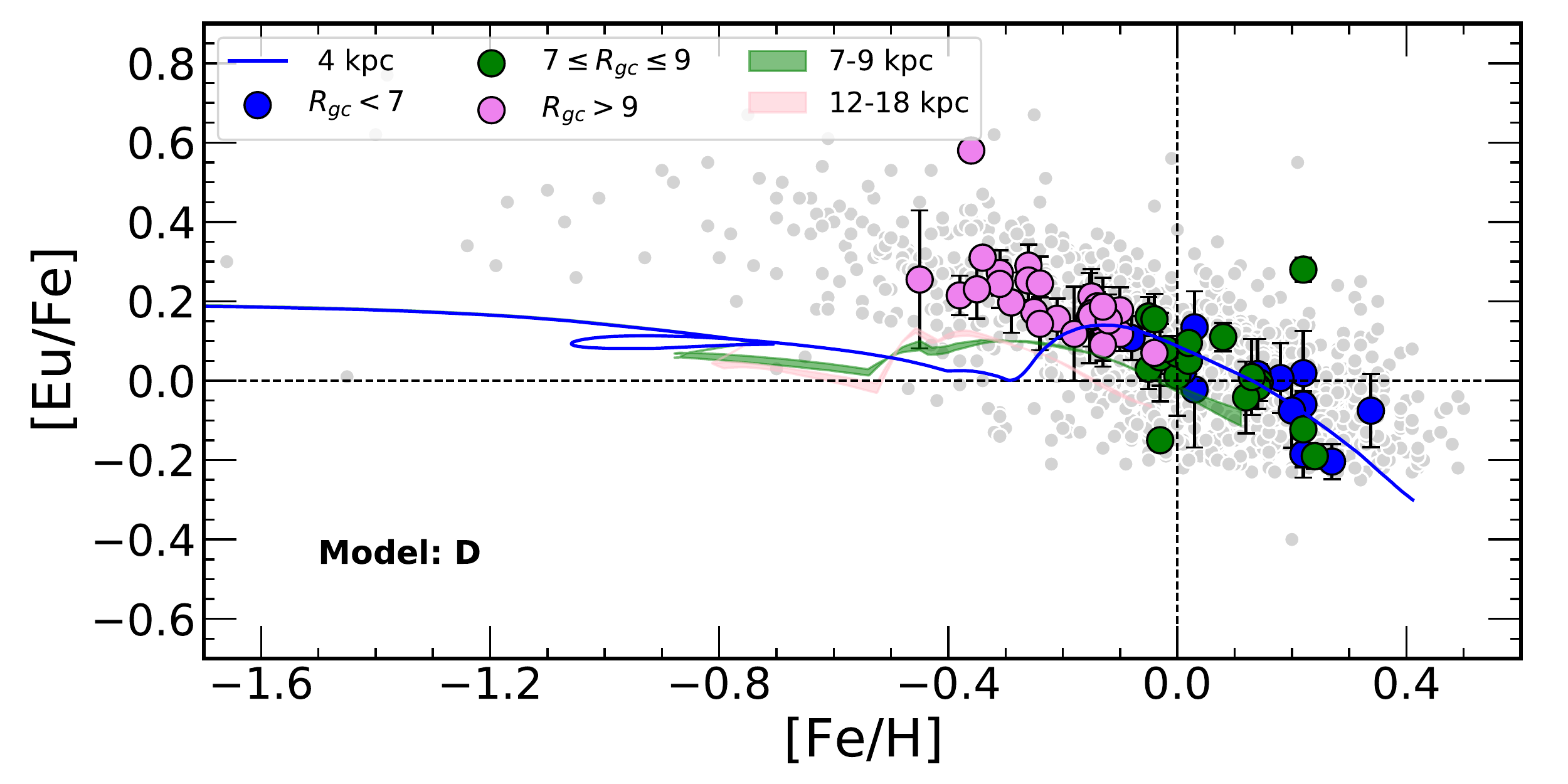}
  \caption{\label{fig:eufe_feh}$\abratio{Eu}{Fe}$ vs. $\abratio{Fe}{H}$ for the field-star sample and the cluster sample. The data are colour-coded by Galactocentric bin, and compared with three curves of a given model, corresponding to the same radial regions and coloured in the same way as the data: inner disc (blue), solar-ring (green) and outer disc (pink). We report only model C and D, since the prescriptions for Eu in model A and B are the same as in model C. Small grey dots stand for the field-star sample. Top: comparison with model C; bottom: comparison with model D. The error bars for the y-axis are displayed for the cluster-sample.}
\end{figure}

Figure~\ref{fig:eufe_feh} shows the behaviour of Eu in the $\abratio{Eu}{Fe}$ vs. $\abratio{Fe}{H}$ plane for the field-star sample (grey dots) and the open-cluster sample (coloured dots). For a metallicity lower than \num{-0.8}, only a dozen of field stars outline the well-known plateau at $\abratio{Eu}{Fe} \sim 0.4$, while at larger metallicity, we note a decrease of $\abratio{Eu}{Fe}$ with increasing $\abratio{Fe}{H}$, reaching $\abratio{Eu}{Fe} \sim -0.2$ at the super-solar metallicity $\abratio{Fe}{H} \sim 0.4$. Over the metallicity range $[-0.4, 0.4]$, the distribution for the OC sample overlaps that of the field-stars sample and exhibits the same decrease. 
\corevdeux{While the \emph{Gaia}-ESO Mg abundances allow us to disentangle the thin and the thick disc sequences in the $\abratio{Mg}{Fe} - \abratio{Fe}{H}$ plane, it is less obvious for Eu. However, like in \cite{delgadomena17}, and if we base our thin/thick disc separation on Mg, we note that (Mg-selected) thick disc stars tend to have higher $\abratio{Eu}{Fe}$ and lower $\abratio{Fe}{H}$ while (Mg-selected) thin disc stars tend to have solar $\abratio{Eu}{Fe}$ and solar $\abratio{Fe}{H}$. The fact that the frontier between the thin and thick disc sequences is blurred could be due to measurement random errors, keeping in mind that the Eu line is more difficult to measure than the Mg line.
In addition, at the typical metallicity of the Galactic discs, we do not expect to detect the remnants of the stochastic enrichment of Eu which are instead recognisable in the high $\abratio{Eu}{Fe}$ spread in the low-density and low-metallicity halo environment for $\abratio{Fe}{H} < -2.5$ \citep[e.g][]{cescutti15, nainman18, brauer21}.
}

On the other hand, the OC sample defines a thinner sequence since, in a given metallicity bin, one finds only open clusters with a similar chemical history.

We also overplot the predicted evolution by our models C and D of $\abratio{Eu}{Fe}$ with $\abratio{Fe}{H}$ for the three radial rings defined for the OC sample, i.e. for $R_{\mathrm{GC}} = \SI{4}{\kilo\parsec}$ (inner disc; \corevun{blue curve}), $\SI{7}{\kilo\parsec} \le R_{\mathrm{GC}} \le \SI{9}{\kilo\parsec}$ (solar ring; \corevun{green curve}) and $R_{\mathrm{GC}} \ge \SI{9}{\kilo\parsec}$ (outer disc; \corevun{pink curve}). While our solar-neighbourhood field-star sample shall be compared with the solar-ring curves, a finer analysis must be adopted for our OC sample since open clusters in this study probe Galactocentric radii from \SI{5}{\kilo\parsec} to \SI{20}{\kilo\parsec}. Therefore, in the followings, we will compare the inner-disc curve to the inner OCs, the solar-ring curve to the solar-neighbourhood OCs and the outer-disc curve to the outer OCs. Though we discuss four different models of chemical enrichment in this work, we recall here that the prescription for the Eu nucleosynthesis is identical in the three models A, B and C, i.e. a rapid production of Eu by magneto-rotationally driven SNe, and only differs in model D, i.e. a evenly-mixed production of Eu by short-timescale \corevun{MRD SNe} and delayed NSMs.

For the model C, the inner-disc and solar-ring curves overlap over the metallicity range $[-0.15, 0.15]$ and differ from each other at lower metallicity. The outer-disc curve gives lower $\abratio{Eu}{Fe}$ ratios than the inner-disc and solar-ring curves at any metallicity over the metallicity range $[-0.8, 0]$, except at a $\abratio{Fe}{H} \sim -0.8$ where both the solar-ring and the outer-disc predictions yield $\abratio{Eu}{Fe} \sim 0.4$. We note that the solar-ring curve is compatible with the mean trend of the field-star sample: it exhibits a flattening compatible with the plateau for $\abratio{Fe}{H} \le -0.8$ and the slope of the decrease matches the observed one for $\abratio{Fe}{H} \ge -0.8$. The overall shape of the predictions is also similar to the observed trends for the OC sample. While the solar-ring curve matches the observed ratios for the solar-neighbourhood OCs, the outer-disc and inner-disc curves are about \num{0.1} below the central trend but still agree at the $1\sigma$ level with the measured $\abratio{Eu}{Fe}$.

For the model D, the three curves exhibit a decrease of $\abratio{Eu}{Fe}$ with metallicity until $\abratio{Fe}{H} \sim -0.55$ for the outer-disc and the solar-ring and $\abratio{Fe}{H} \sim -0.3$ for the inner-disc, where a rapid increase of $\abratio{Eu}{Fe}$ occurs corresponding to the onset of the second source of Eu, namely NSMs, and then $\abratio{Eu}{Fe}$ decreases again until super-solar metallicities. This bump in $\abratio{Eu}{Fe}$ is not supported at all by the observations, indicating that if NSMs do contribute to the production of Eu in the thin disc then this contribution should be small enough to not compensate the decrease of $\abratio{Eu}{Fe}$ due to the release of Fe by SNe\,Ia. Moreover, the model D always under-predicts the Eu abundance for the outer-disc and the solar-ring; only the inner-disc curve matches the inner-disc OC data. 

\begin{figure}
  \centering
  \includegraphics[width=8cm]{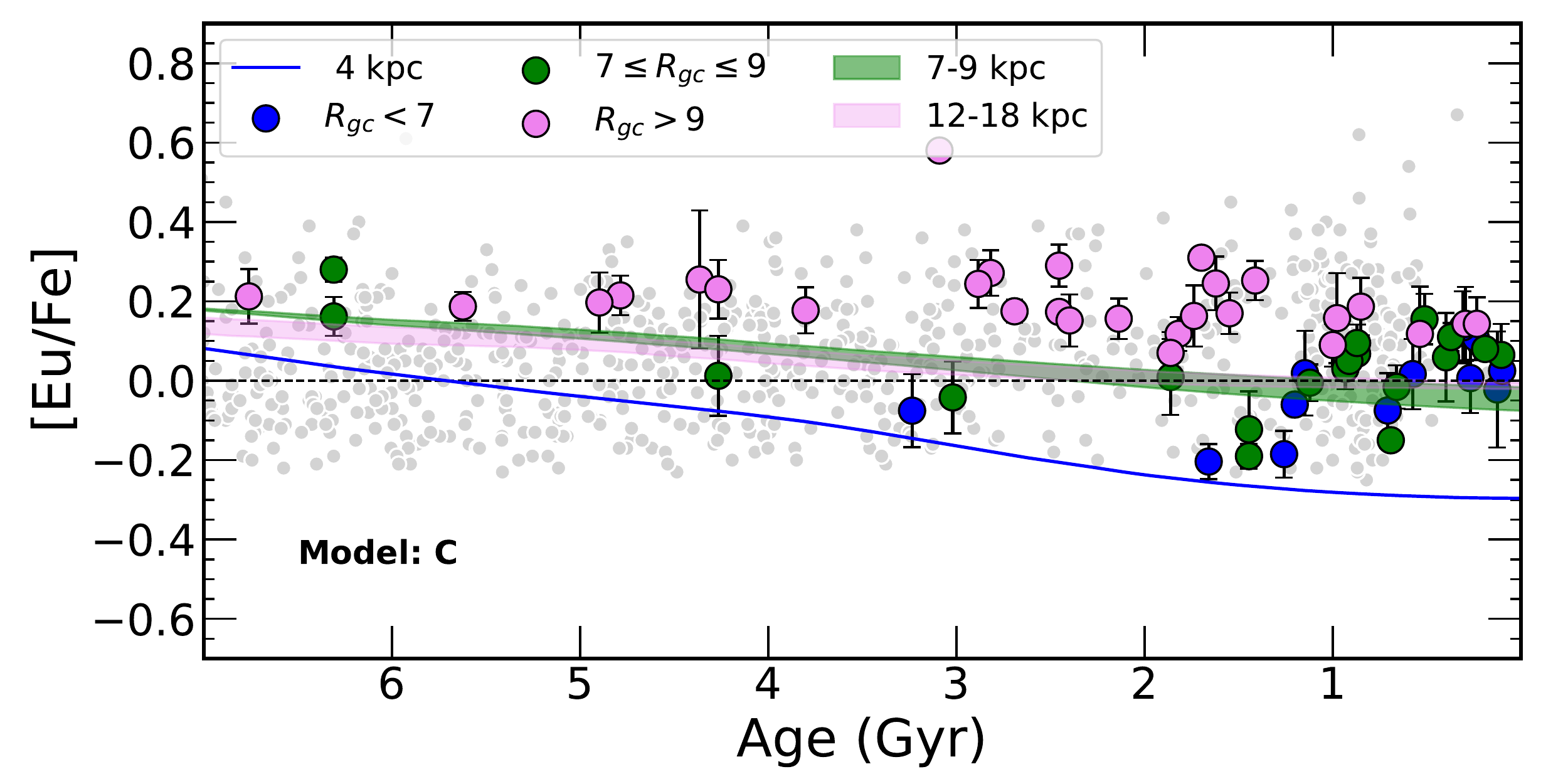}
  \includegraphics[width=8cm]{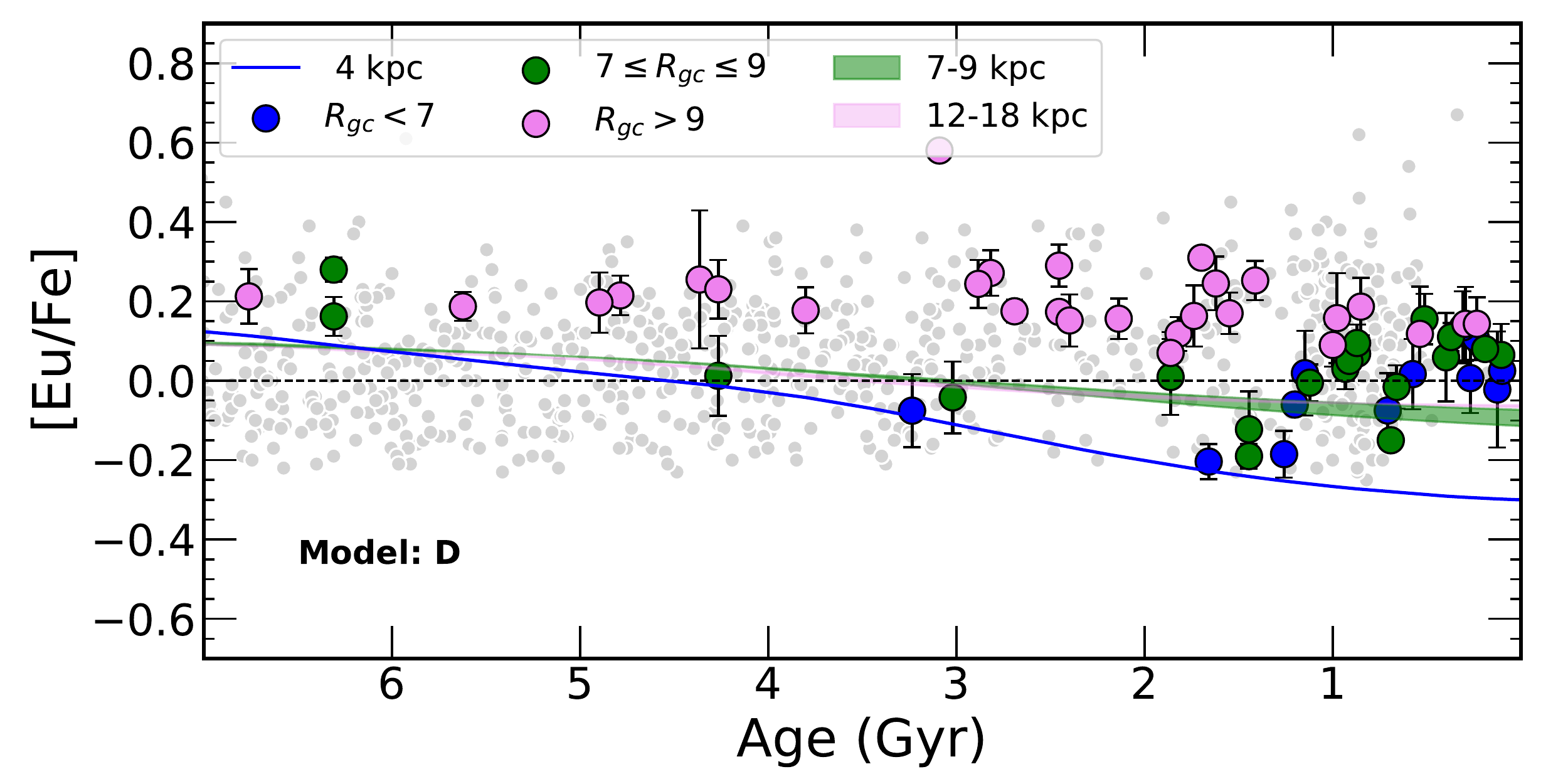}
  \caption{\label{fig:eufe_age}$\abratio{Eu}{Fe}$ vs. age for the field-star and open-cluster samples. Same symbols and colours as in Fig.~\ref{fig:eufe_feh}.}
\end{figure}

Figure~\ref{fig:eufe_age} displays the field-star and OC samples, and the models C and D in the $\abratio{Eu}{Fe}$ vs. age plane. We find OCs with enhanced $\abratio{Eu}{Fe}$ ($\ge 0.2$) of any age between \num{1} and \SI{7}{\giga\Year} and they tend to be located in the outer part of the Galaxy. OCs with solar or sub-solar $\abratio{Eu}{Fe}$ tend to be younger (less than \SI{4}{\giga\Year} old) and located in the solar-neighbourhood/inner Galaxy. This is inline with the model of inside-out formation of the Galactic disc \citep{2014A&A...565A..89B}. We remark that the agreement between the model C and observations in this parameter plane is not as good as in the $\abratio{Eu}{Fe}$ vs. $\abratio{Fe}{H}$ plane. The global trend is correct: $\abratio{Eu}{Fe}$ increases with increasing age for each Galactocentric region, the solar-ring/outer Galaxy exhibit larger $\abratio{Eu}{Fe}$ compared to the inner Galaxy at any age bin. However, the inner-disc and outer-disc curves underestimate the $\abratio{Eu}{Fe}$ ratio compared to the inner-disc and outer-disc OC sub-samples respectively. The solar-ring curve yields the most satisfactory match with the observed data. Though this discrepancy was already noticed in Fig.~\ref{fig:eufe_feh}, it is more visible in the $\abratio{Eu}{Fe}$ vs. age plane. Keeping in mind that the typical uncertainty on the age for the field-star sample is about \SI{1.7}{\giga\Year} (mean of age uncertainty from isochrone fitting), we find a flat $\abratio{Eu}{Fe}$ vs. age distribution for the field-star sample, indicating a mixing of stellar population with different chemical history. On the other hand, the disagreement between the model D and the data is worst: the offset between the solar-ring and outer-disc curves and the observed OC data is larger than with model C at any age.

\subsection{The evolution of Mg}
\label{Sec:Evolution_MgFe}

\begin{figure}
  \centering
  \includegraphics[width=8cm]{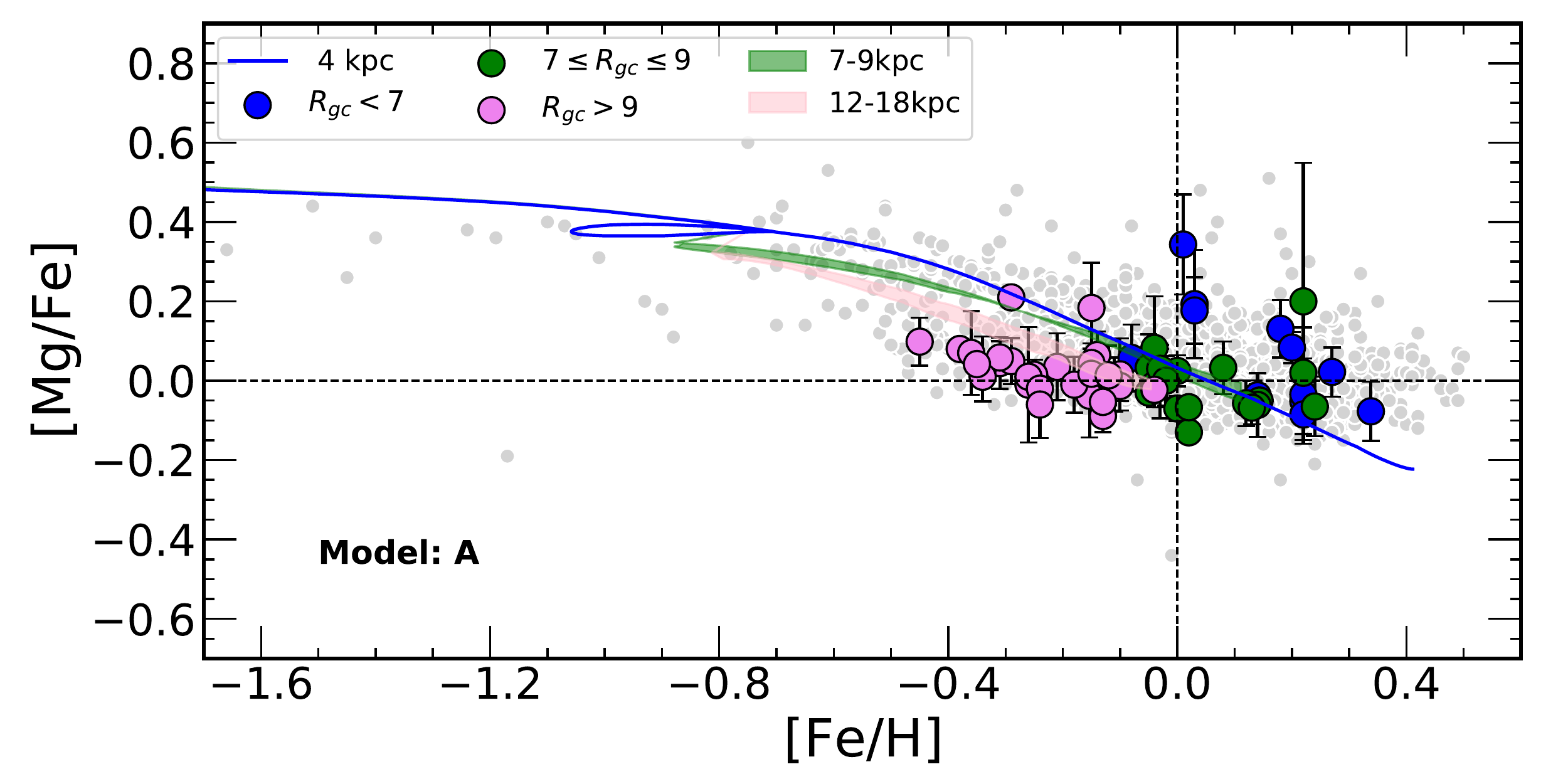}
  \includegraphics[width=8cm]{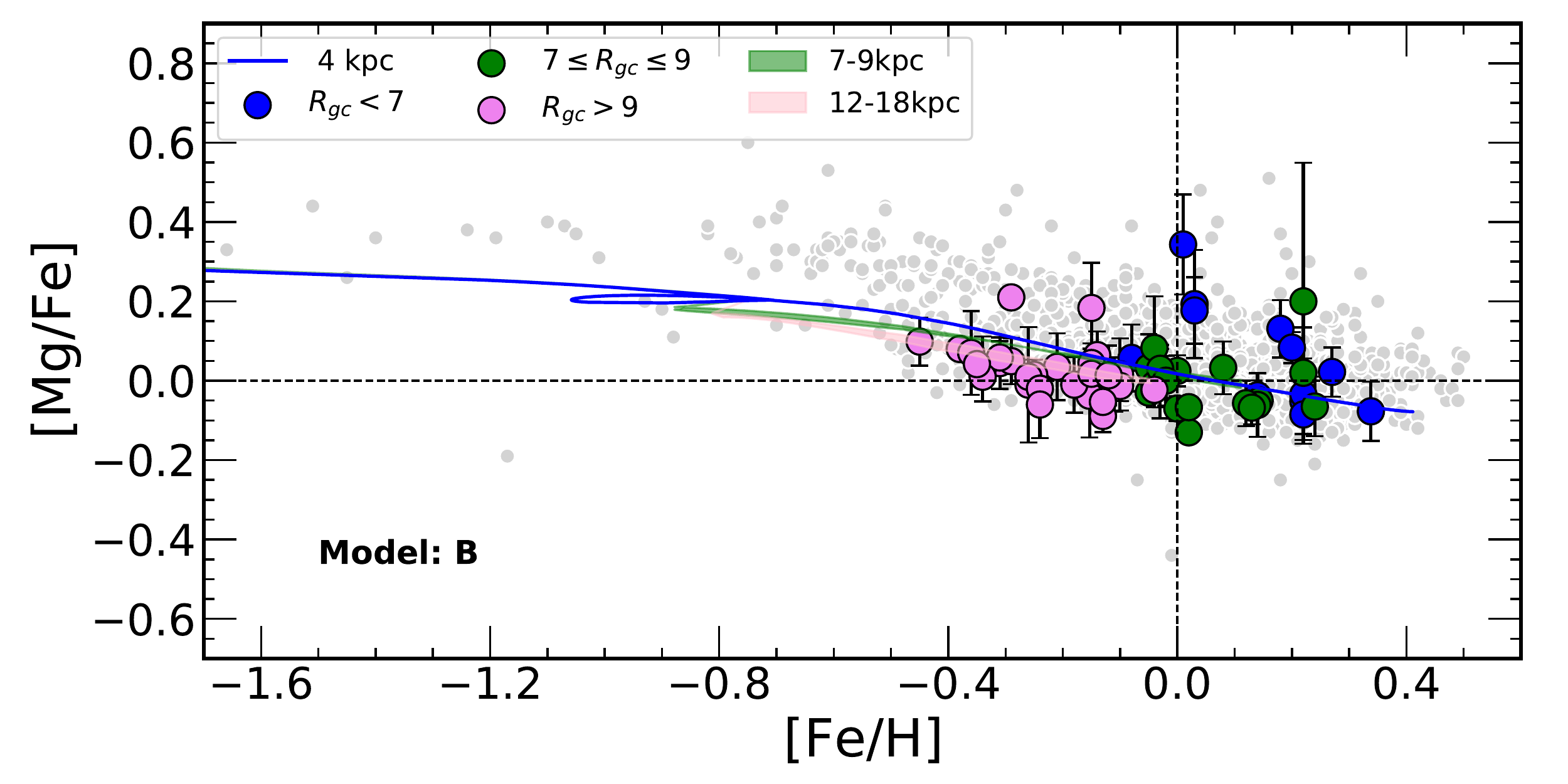}
  \includegraphics[width=8cm]{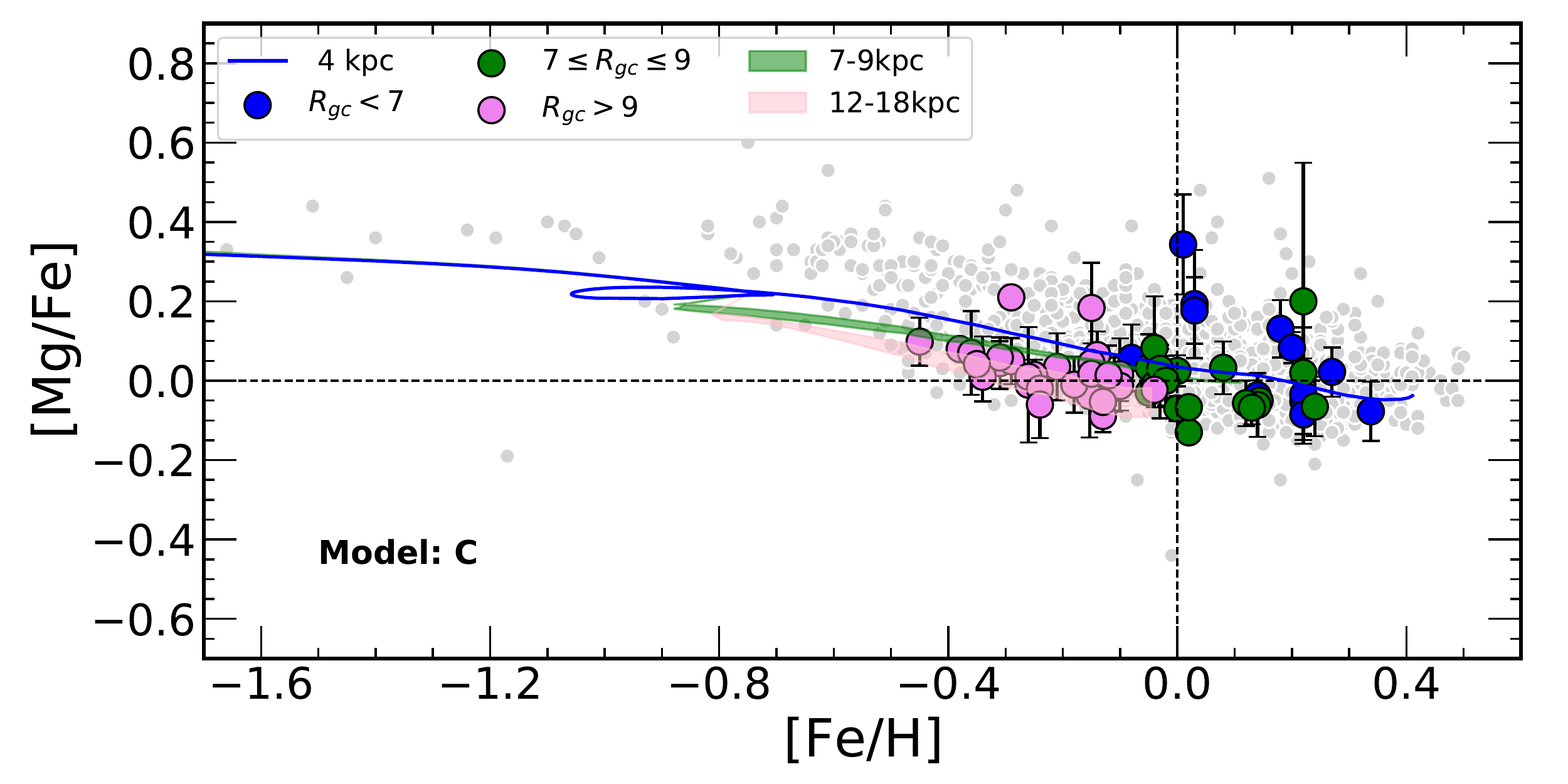}
  \caption{\label{fig:mgfe_feh}$\abratio{Mg}{Fe}$ vs. $\abratio{Fe}{H}$ for the field-star sample and the cluster sample. Same symbols and colours as in Fig.~\ref{fig:eufe_feh}. Three models are considered: model A (upper panel), model B (middle panel), model C (bottom panel).}
\end{figure}

\begin{figure}
  \centering
  \includegraphics[width=8cm]{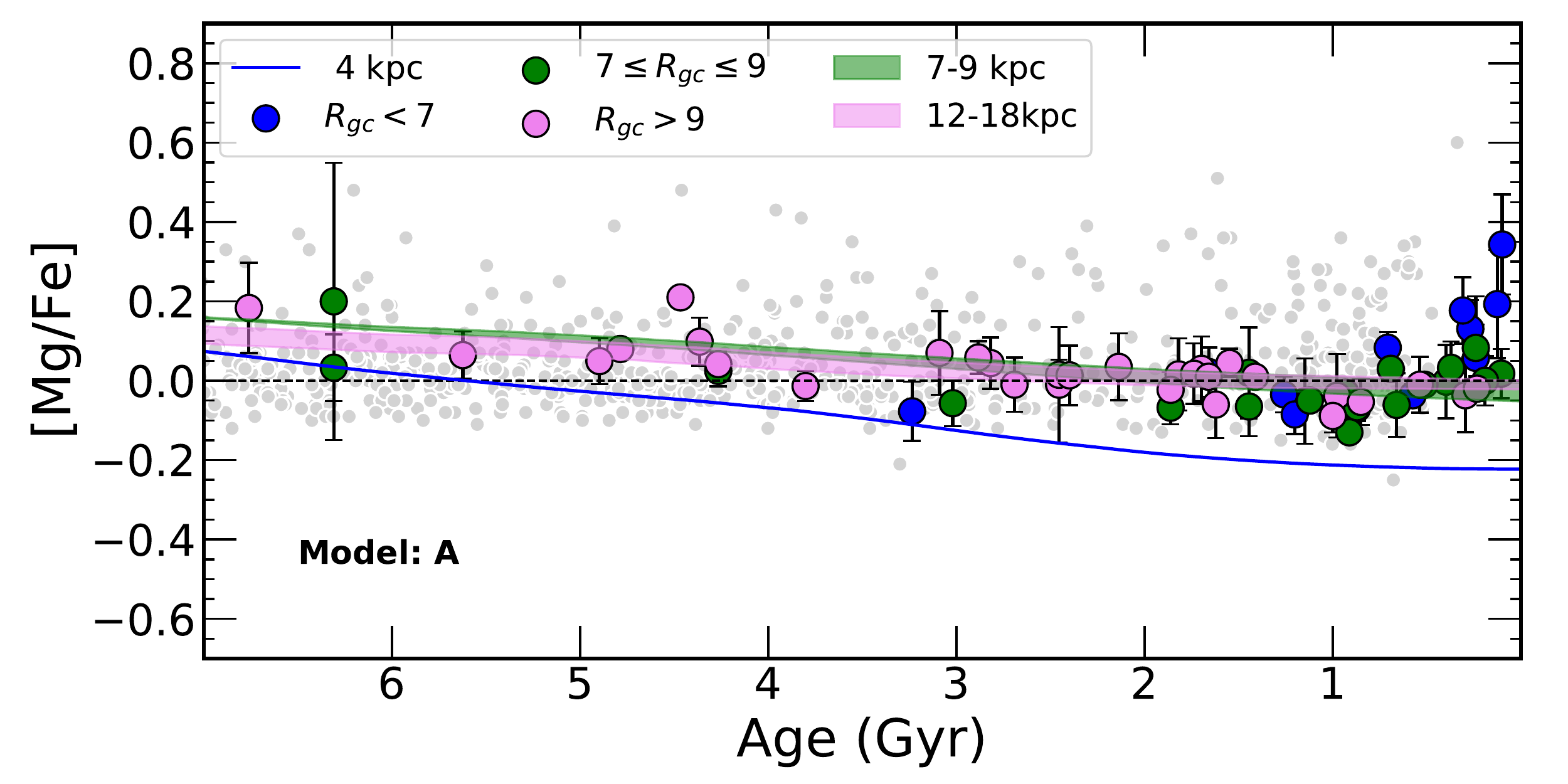}
  \includegraphics[width=8cm]{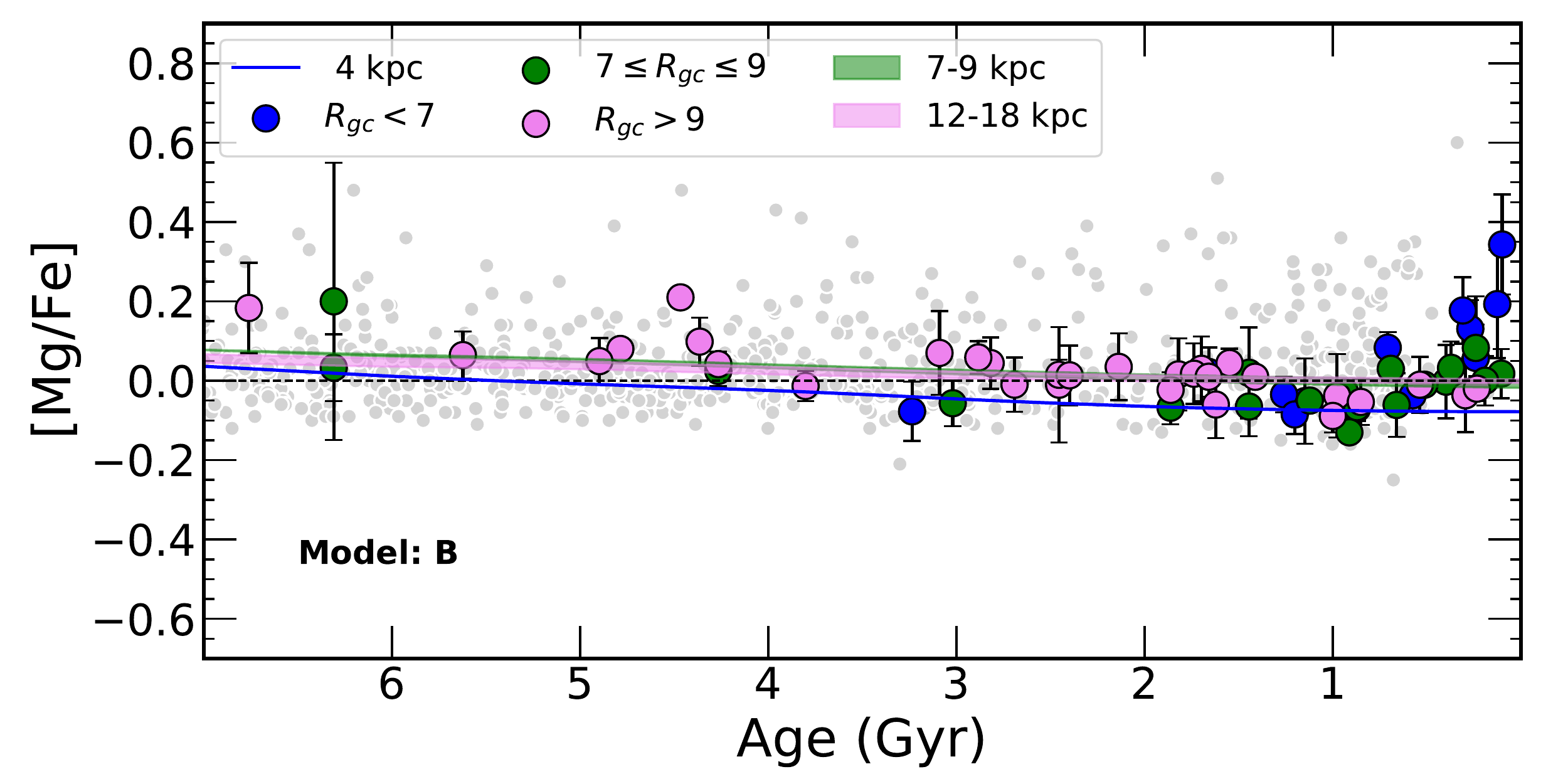}
  \includegraphics[width=8cm]{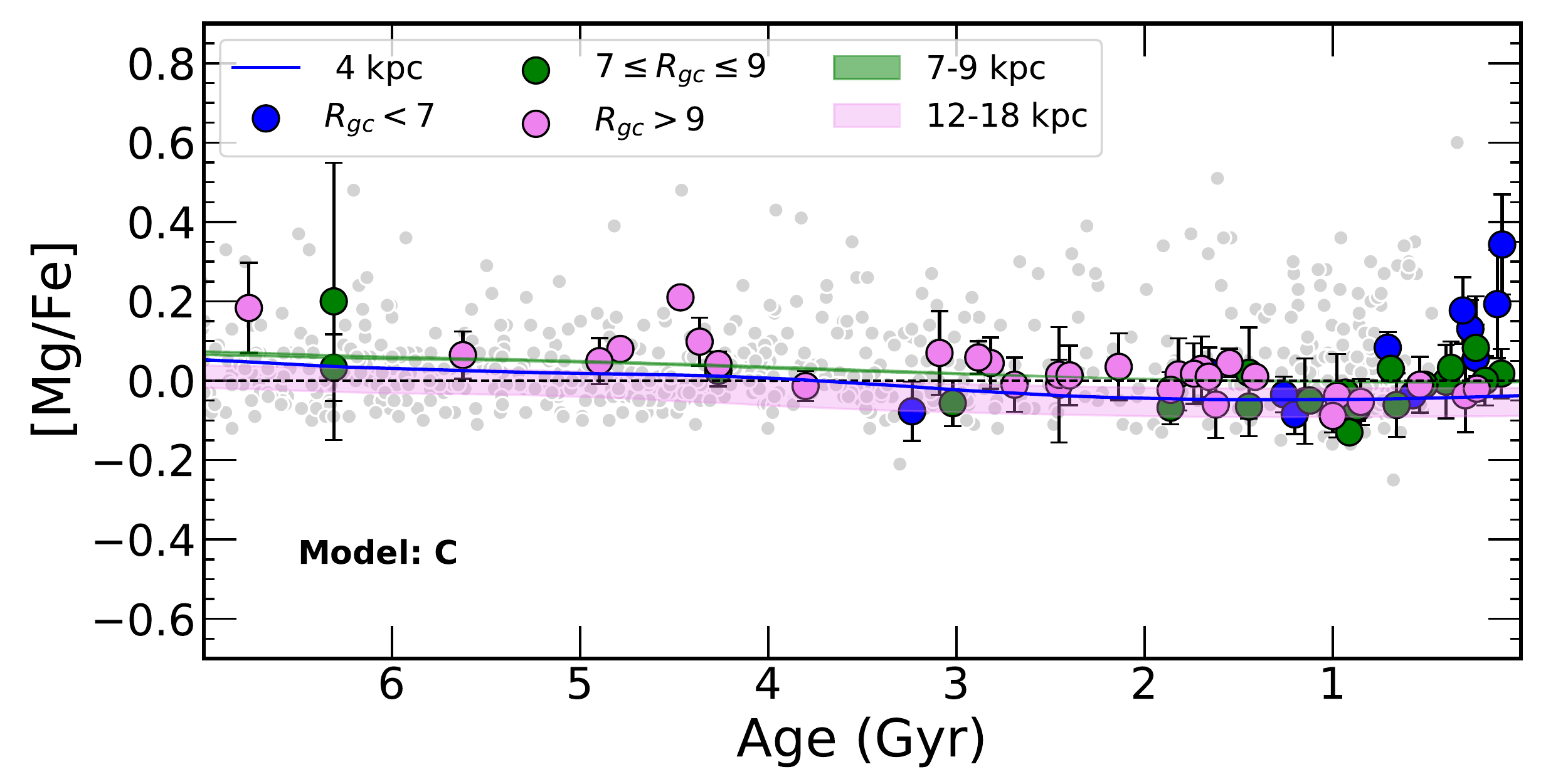}
  \caption{\label{fig:mgfe_age}$\abratio{Mg}{Fe}$ vs. age for the field-star sample and the cluster sample. Same symbols and colours as in Fig.~\ref{fig:eufe_feh}. Three models are considered: model A (upper panel), model B (middle panel), model C (bottom panel).}
\end{figure}

Figures~\ref{fig:mgfe_feh} and \ref{fig:mgfe_age} show the observed $\abratio{Mg}{Fe}$ as a function of $\abratio{Fe}{H}$ and of stellar age, respectively, along with the three models tested for the Mg production, namely the models A, B and C described in Section~\ref{sec_model}. We recall that the main change between the three models is how much SNe\,Ia contribute to the Mg production.

In Fig.~\ref{fig:mgfe_feh}, the field-star Mg-to-Fe ratio displays the well-known pattern for an $\alpha$-element in the Milky Way: for thick disc stars, a plateau at $\abratio{Mg}{Fe} \sim 0.4$ up to a metallicity of $-0.8$; for both thin and thick disc stars, a decrease of $\abratio{Mg}{Fe}$ with increasing metallicity, with a possible flattening around $\abratio{Mg}{Fe} \sim 0$ for super-solar metallicities. The open-cluster Mg-to-Fe ratios also exhibit a decreasing trend with increasing metallicity, overlapping the thin disc sequence. The best agreement between the observations and our models is reached for model C, in which the production of Mg is due to both CCSNe and SNe\,Ia. In this model, the yields of SNe\,Ia are metallicity-dependent in order to reproduce the behaviour of younger, metal-rich clusters. This choice affects not only the super-solar region, where, as already noted in \citet{magrini17}, the decline of $\abratio{Mg}{Fe}$ is not observed, but also the sub-solar region, with a lower enhancement at low metallicities. The value of $\abratio{Mg}{Fe}$ for thin-disc MW field stars in the super-solar metallicity regime is a debated topic, both by observers and theoreticians. It is well known from spectroscopists that abundance determination is not an easy task and that, despite their careful work, it is difficult to identify and correct any bias introduced during the spectral analysis \citep[e.g., see the discussion in ][]{2017A&A...601A..38J}. \citet{2020A&A...639A.140S} investigated the role of the continuum placement in the derived Mg abundances. After a thorough testing of the pseudo-normalisation procedure, they claim that $\abratio{Mg}{Fe}$ continues to decrease for $\abratio{Fe}{H} \ge 0$ instead of flattening. However, their conclusion is weakened by the fact that among their selected Mg lines, only the four saturated lines exhibit the decrease while the five weak lines show a flattening (their Fig.16). On the other hand, Galactic chemical evolution models are not robust enough to point at the most likely solution: for instance, \citet{2019MNRAS.487.5363M} was able to reproduce the flattening of the Mg-to-Fe ratios observed in the APOGEE dataset \citep[e.g.,][]{2020AJ....160..120J} by increasing the contribution of SNe\,Ia to the Mg production (similar to what is done in this work), but \citet{2020MNRAS.494.5534M} still wondered whether the flattening is an artefact or not.

Figure~\ref{fig:mgfe_age} is showing the $\abratio{Mg}{Fe}$ vs. age plane. \corevun{In this plane}, we do not separate the sequences for the inner-disc, solar-neighbourhood and outer-disc open clusters: for most OCs, $\abratio{Mg}{Fe}$ appear compatible with a single linear function of age. The only exception is observed for a handful of inner-disc, young, $\alpha$-enhanced open clusters (see \corevun{next paragraph}). The best agreement between data and models is obtained with the model C. In model A, the curve for the inner-disc differs greatly from the data of inner-disc open clusters: at an age of $\approx \SI{1}{\giga\Year}$, the inner-disc curve of model A predicts $\abratio{Mg}{Fe} \approx -0.2$, compared to the observed ratio of $\approx 0$; for the youngest open clusters, the disagreement is even larger. We note that when the contribution of SNe\,Ia to Mg is increased (model B and C), the theoretical curves for the three Galactic regions considered here come closer to each other, which is compatible with our OC data.

As noted in earlier works \citep{magrini14, magrini17, casmiquela18}, there is a population of inner clusters that are $\alpha$-enhanced, which is also clearly visible in our data. \citet{chiappini15} was among the first papers to report the existence of a young $\abratio{\alpha}{Fe}$-enhanced population in the CoRoT \citep{miglio13} and APOGEE \citep{majewski17} samples. They discovered several young stars with unexpectedly high $\abratio{\alpha}{Fe}$ abundances, located in the inner disc. A similar population is also present in other works \citep[e.g.,][]{haywood13, Bensby_14, bergemann14, martig15}. For field stars, several works investigated the role of mass transfer and binarity to explain their chemical pattern \citep[e.g., see][]{jofre16, hekker19,sun20, zhang21}. But, while for field stars, there is still the possible ambiguity in the determination of their ages and masses, even when it is done with asteroseismology, such uncertainty disappears when it concerns the determination of the ages of stars in clusters. For the $\alpha$-enhanced clusters, we need a different explanation, as chemical evolution and migration. A possible interpretation is that the $\alpha$-enhanced clusters might have been born in a region near the corotation of the bar where the gas can be kept inert for a long time and in which the enrichment is due only to CCSNe \citep{chiappini15}. Further migration might have moved them to their current radius.

\subsection{The evolution of O}

\begin{figure}
  \centering
  \includegraphics[width=8cm]{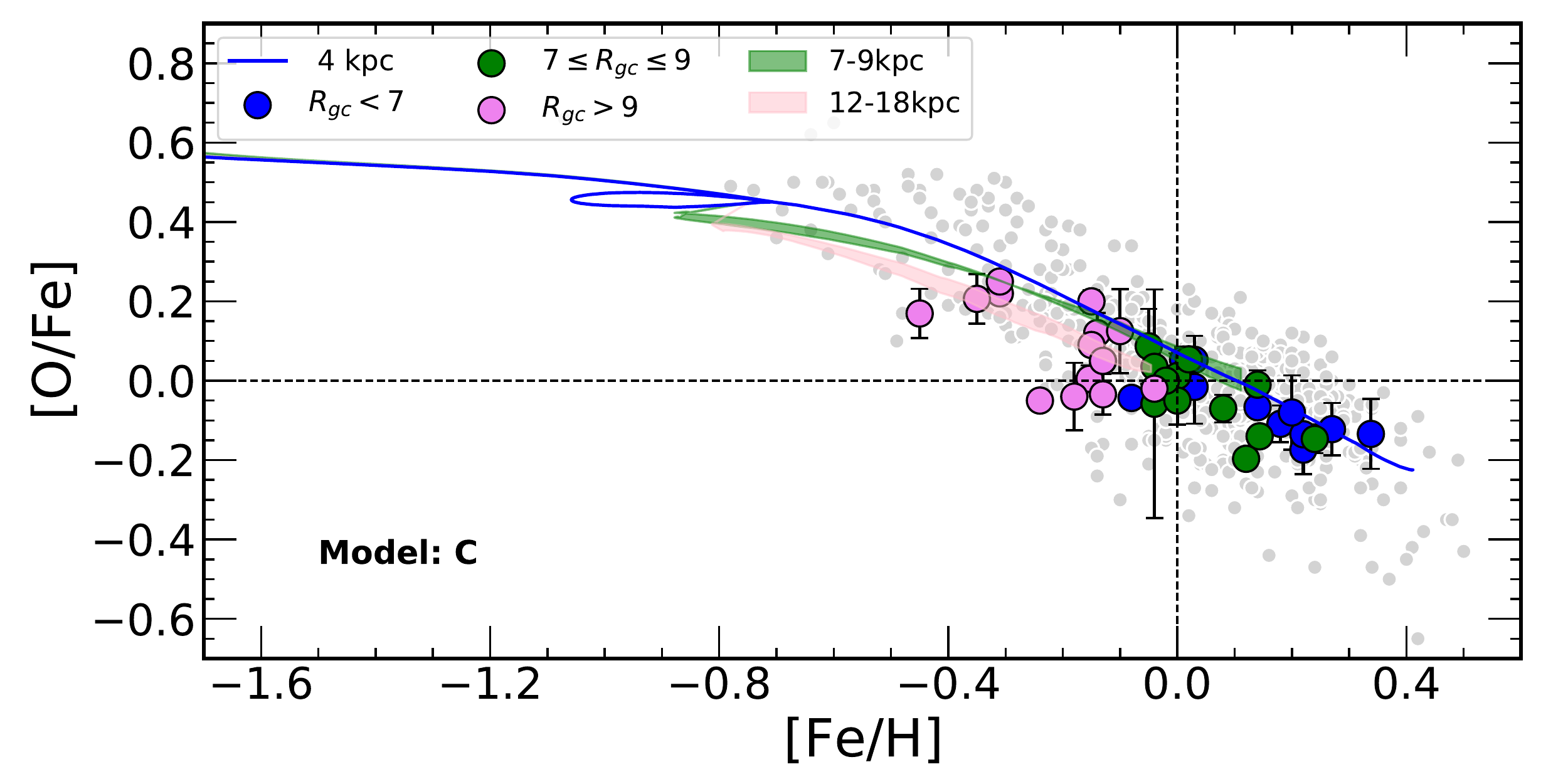}
  \caption{\label{fig:ofe_feh}$\abratio{O}{Fe}$ vs. $\abratio{Fe}{H}$ for the field-star sample and the cluster sample. Same symbols and colours as in Fig.~\ref{fig:eufe_feh}. Only model C is shown.}
\end{figure}

\begin{figure}
  \centering
  \includegraphics[width=8cm]{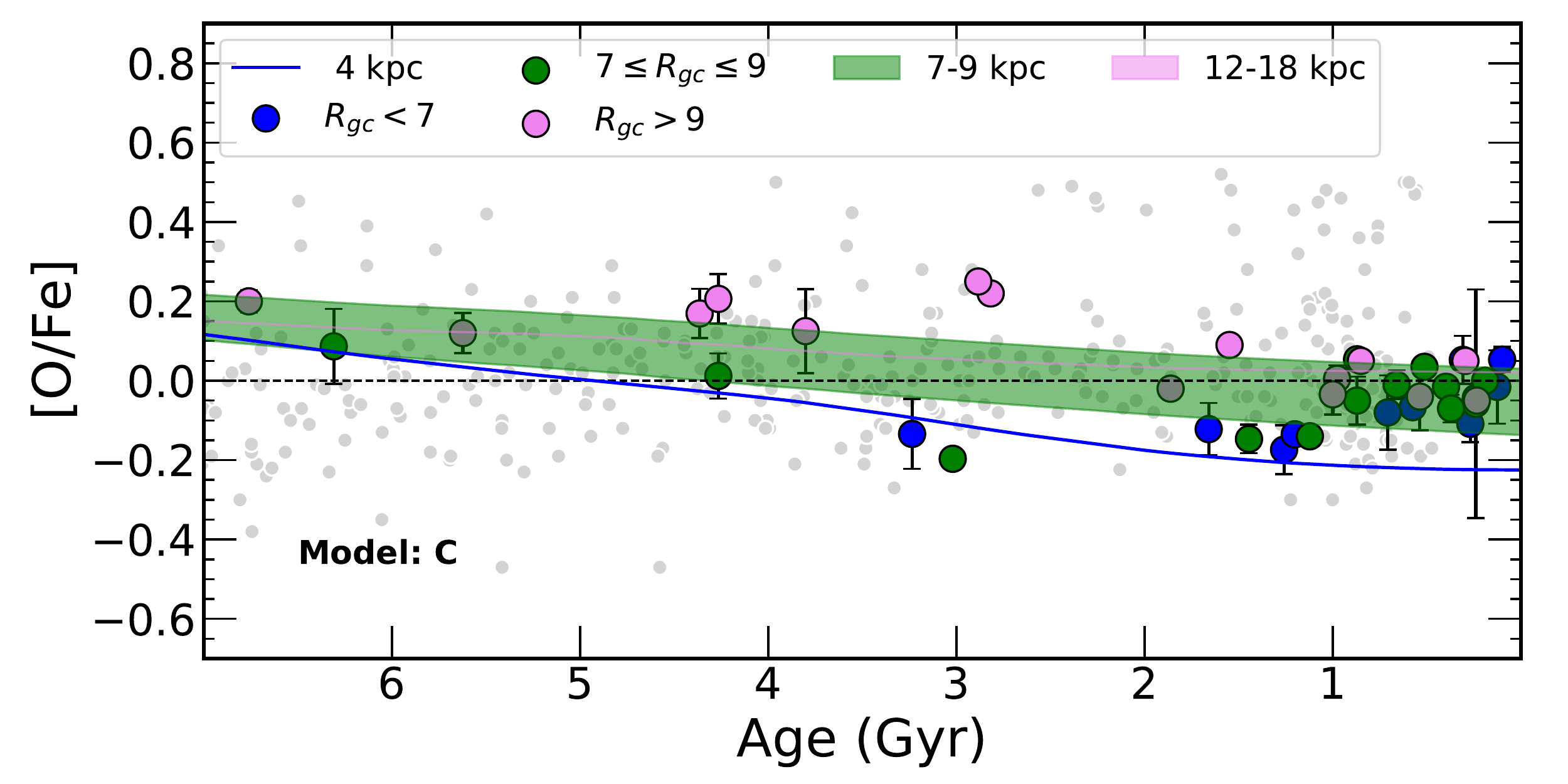}
  \caption{\label{fig:ofe_age}$\abratio{O}{Fe}$ vs. age for the field-star sample and the cluster sample. Same symbols and colours as in Fig.~\ref{fig:eufe_feh}. Only model C is shown.}
\end{figure}

We show the evolution of $\abratio{O}{Fe}$ as a function of $\abratio{Fe}{H}$ and age in Fig.~\ref{fig:ofe_feh} and \ref{fig:ofe_age}, respectively. For oxygen, we considered only its production by CCSNe, with short time scales; therefore, we display only the set of curves for model C. We remind the reader that the forbidden $[\ion{O\,I}]$ line may be affected by telluric lines preventing a reliable abundance measurement under specific conditions, hence the reduced number of data points for this chemical species \citep[e.g.,][and in particular their Fig.~2 displaying such an \ion{O\,I} line affected by the telluric blend]{1992A&A...261..255N}. Field stars exhibit a decrease from $\abratio{O}{Fe} \approx 0.4$ at $\abratio{Fe}{H} \le -0.7$ (upper limit because of the paucity of metal-poor stars with O determination) to $\abratio{O}{Fe} \approx -0.3$ at $\abratio{Fe}{H} \approx 0.4$. The open-cluster sample exhibit also a decrease of the O-to-Fe ratio with metallicity; the outer-disc OCs tend to be more O-enhanced than the inner-disc OCs. The three curves for the model C corresponding to the inner-disc, solar-ring and outer-disc are compatible with our OC data given the observational error bars. In the $\abratio{O}{Fe}$ vs. age plane, data and models are also in good agreement: $\abratio{O}{Fe}$ decreases with decreasing age; at a given age, the outer-disc OCs tend to be more O-enhanced than the inner-disc OCs; young clusters (less than \SI{2}{\giga\Year} old), no matter their Galactocentric radius, have a solar or sub-solar $\abratio{O}{Fe}$.

We note that the young, inner-disc, Mg-enhanced open clusters with solar or super-solar metallicity are not O-enhanced. Among the six OCs with $\abratio{Mg}{Fe} \ge 0.05$, four have a metallicity very close to solar, i.e. a metallicity where the determination of Mg should not be concerned by the issues briefly discussed in the previous section. We cannot explain this difference by analysis systematic effects and we think this difference between Mg and O is genuine for this population of open clusters. Thus, this observational fact may be another evidence of the different nucleosynthetic paths needed to produce oxygen on the one hand and magnesium on the other hand and it reminds us that the so-called $\alpha$-elements are not interchangeable when doing Galactic archaeology.

\subsection{The evolution of $\abratio{Eu}{Mg}$ and of $\abratio{Eu}{O}$}

The study and comparison of Eu with O and Mg is crucial to understand if this \emph{r}-process element and those two $\alpha$-elements share the same production sites or are released to the ISM over the same timescales. Such comparisons are in particular useful to probe the chemical enrichment of the early Galaxy. The interest of the $\abratio{Eu}{Mg}$ ratio has also increased in recent years, due to its potential to unveil the extragalactic origin of some MW stars with unusual values \citep[e.g.][]{McWilliam2013,Lemasle2014,xing19,skuladottir19,Matsuno2021}. An increasing number of studies about $\abratio{Eu}{Mg}$, based on larger and larger samples of stars, are being published \citep[see, e.g.][]{mashonkina2001, mashonkina2003, delgadomena17}. More recently, \citet{guiglion18} addressed the subject for the AMBRE project using a large sample of about \num{1400} FGK Milky Way disc stars, reporting a decreasing $\abratio{\emph{r}}{\alpha}$ trend with increasing metallicity and concluding that supernovae involved in the production of Eu and Mg should have different properties. \citet{tautvasiene21} also found that the $\abratio{Eu}{Mg}$ ratio decreases with metallicity for both thin and thick-disc stars, the gradient being steeper for the thick disc.

\begin{figure}
  \centering
  \includegraphics[width=8cm]{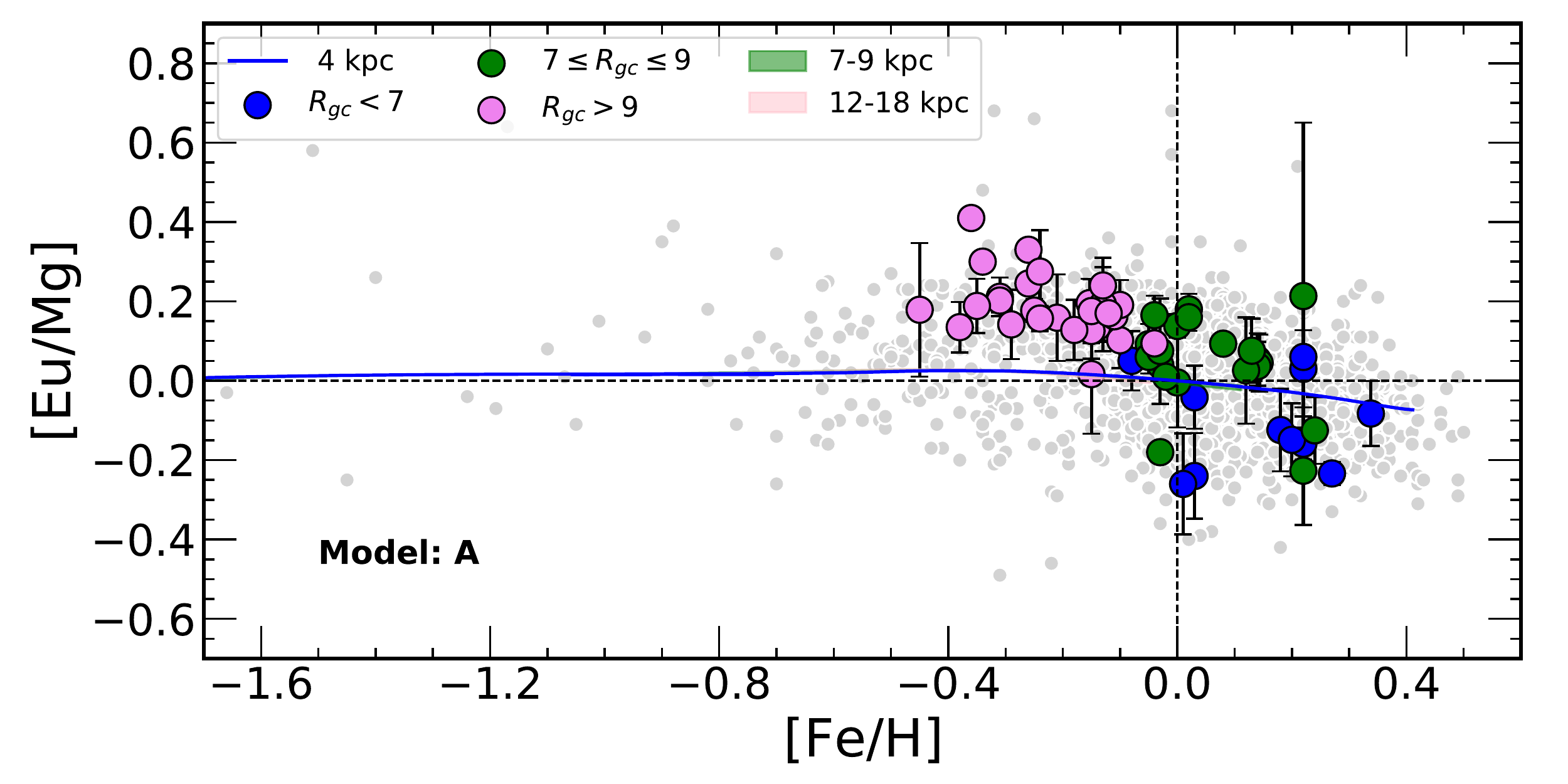}
  \includegraphics[width=8cm]{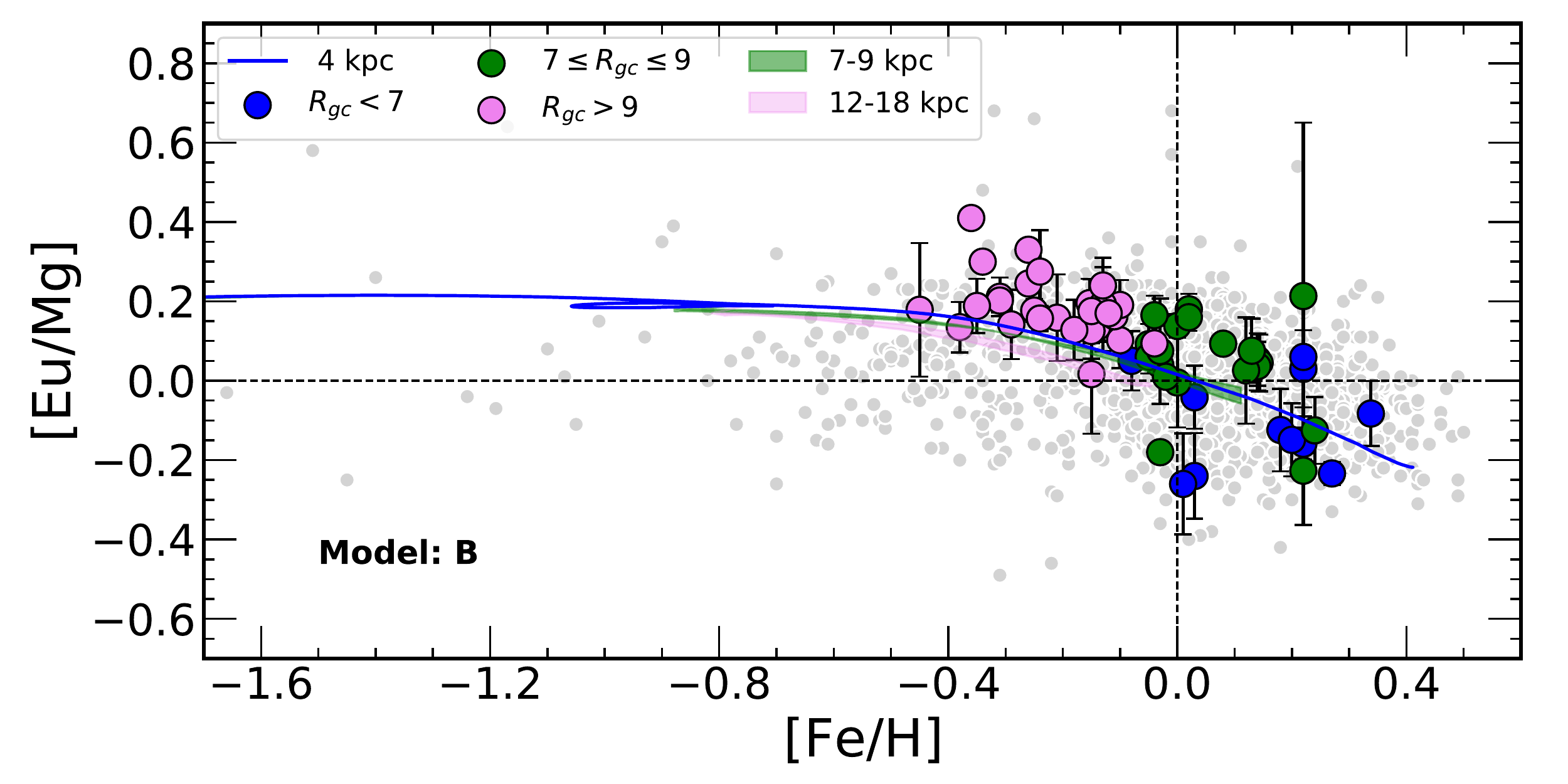}
  \includegraphics[width=8cm]{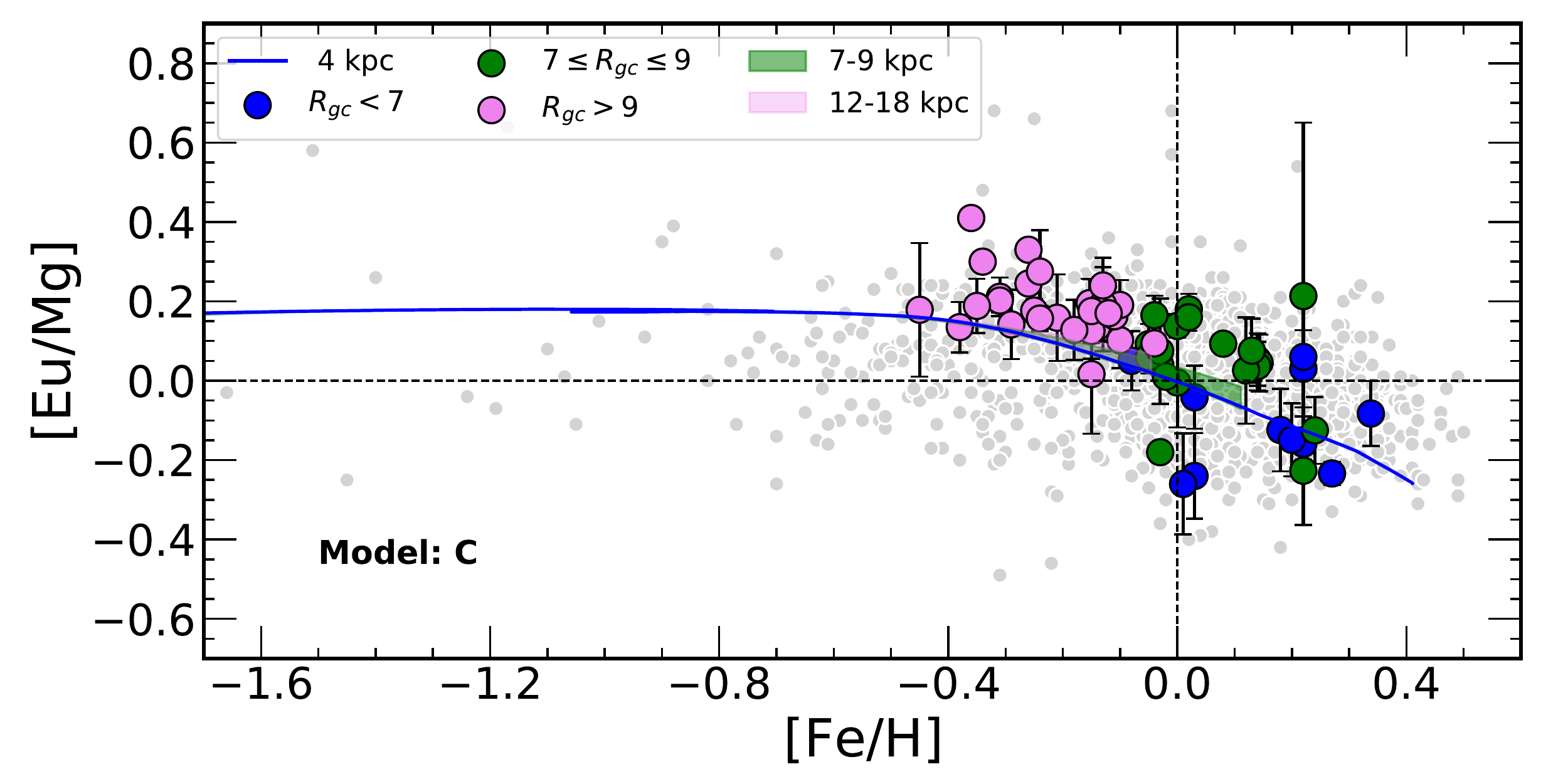}
  \includegraphics[width=8cm]{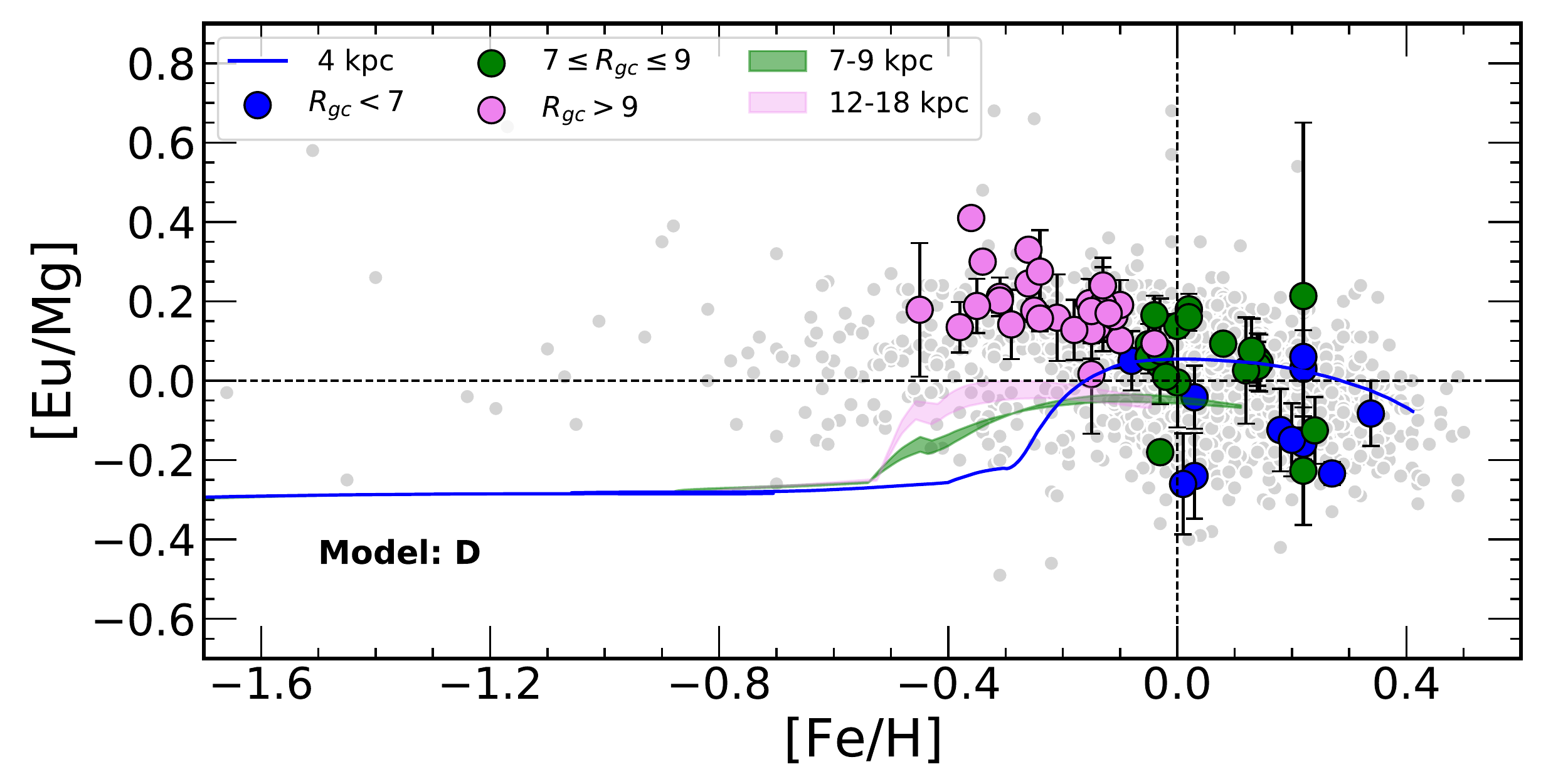}
  \caption{\label{Fig:EuMg_vs_FeH}$\abratio{Eu}{Mg}$ vs. $\abratio{Fe}{H}$ for the field-star sample and the cluster sample. Same symbols and colours as in Fig.~\ref{fig:eufe_feh}. Four models are considered. From top to bottom panel: model A, model B, model C and model D.}
\end{figure}

\begin{figure}
  \centering
  \includegraphics[width=8cm]{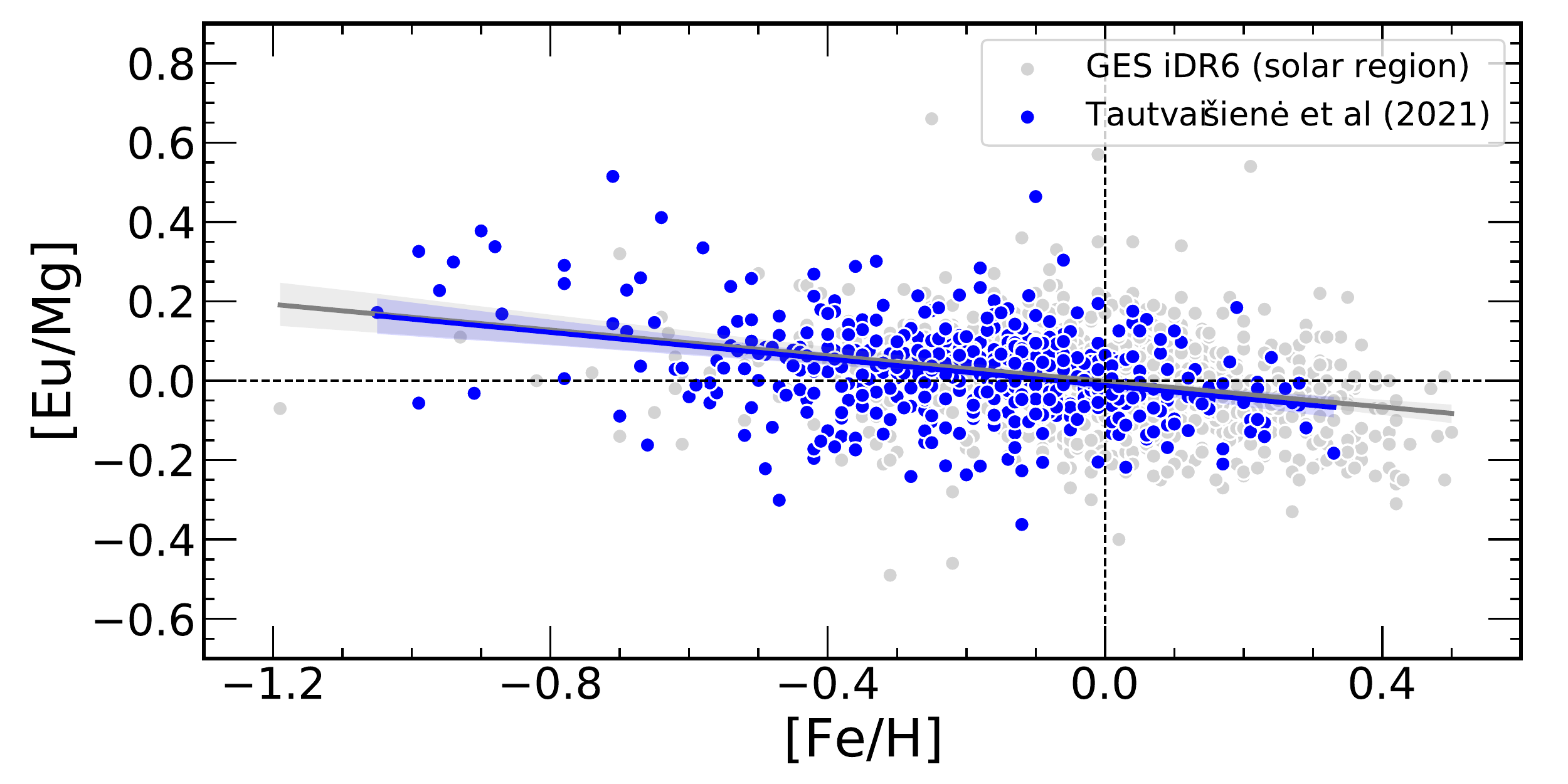}
  \caption{\label{fig:eumg_feh_ges_vs_tautvaisiene21}$\abratio{Eu}{Mg}$ vs. $\abratio{Fe}{H}$ for our sample of field stars (grey dots) in the solar region, compared to those of \citet{tautvasiene21} in the solar neighborhood (blue dots). The solid lines represent the linear regression lines, which seem to coincide for both samples, and the shaded regions the confidence interval.}
\end{figure}

Figure~\ref{Fig:EuMg_vs_FeH} shows the evolution of $\abratio{Eu}{Mg}$ as a function of $\abratio{Fe}{H}$ for the field disc stars, the thin-disc OCs and the four models A, B, C and D. Our field-star sample displays a large scatter; however, $\abratio{Eu}{Mg}$ tends to be around \num{0.2} at $\abratio{Fe}{H} \sim -0.4$ and tends to be around \num{-0.1} for $\abratio{Fe}{H} \sim 0.3$. The linear regression yields a slope of $-0.163$, a $y$-intercept of $0.015$ and a Pearson correlation coefficient (PCC) of $-0.30$. If we restrict our field stars to the solar region (\num{7} to \SI{9}{\kilo\parsec}), our sample is reduced to \num{741} field stars with a slope of $-0.162$ ($y$-intercept $= -0.002$). These regression parameters are almost the same as those obtained using the sample of \num{506} stars from \citet{tautvasiene21}: a slope of $-0.167$ and a $y$-intercept of $-0.012$ with a PCC of $=-0.33$ (see Fig.~\ref{fig:eumg_feh_ges_vs_tautvaisiene21}). On the other hand, our OC sample exhibits a steeper decreasing trend with a slope of $-0.535$ and a $y$-intercept of $0.055$ with a PCC of $= -0.69$. The slope of the linear regression for field stars and OCs are not directly comparable because the OC sample encompasses a much larger region of the disc.

The model A, with a pure CCSNe production of Mg results in a nearly constant $\abratio{Eu}{Mg}$ as a function of metallicity for the three Galactocentric regions. The model D, with a mixed production of Eu by \corevun{MRD SNe} and NSMs and a pure CCSNe production of Mg, under-predicts the Eu-to-Mg ratios at almost any metallicity bin. Only the model B and C, with a pure \corevun{MRD SNe} production of Eu and mixed production of Mg by CCSNe and SNe\,Ia, yield a satisfactory match to the OC data. The model C gives slightly better results: it minimises the under-prediction of the Eu-to-Mg ratio for the outer-disc OCs, it predicts slightly lower Eu-to-Mg ratios at $\abratio{Fe}{H} \sim 0.3$ than model B. Given that the model C was also the best-matching model in the plane $\abratio{Eu}{Fe}$ vs. $\abratio{Fe}{H}$ and $\abratio{Mg}{Fe}$ vs. $\abratio{Fe}{H}$ (see Sec.~\ref{Sec:Evolution_EuFe} and \ref{Sec:Evolution_MgFe}), we conclude from Fig.~\ref{Fig:EuMg_vs_FeH} that 
\begin{itemize}
\item the production of Eu in the thin disc can be explained solely by a production by \corevun{MRD SNe};
\item the production of Mg should involve at least two sources, namely CCSNe and SNe\,Ia with metal-dependent yields.
\end{itemize}

\citet{trevisan14} studied the $\abratio{Eu}{O}$ ratio vs. metallicity, age, and Galactocentric distance in a sample of \num{70} old and metal-rich dwarf thin/thick-disc stars selected from the NLTT catalogue. Combined with the literature data, they found a steady increase of Eu-to-O with metallicity over the metallicity range $[-1, 0.5]$. On the other hand, \citet{haynes19} provided galactic simulations of \emph{r}-process elemental abundances, comparing them with observations from the HERMES-GALAH survey. These observations show a flat $\abratio{Eu}{O}$ trend as a function of $\abratio{Fe}{H}$, suggesting that europium is produced primarily at the same rate as oxygen is.

\begin{figure}
  \centering
  \includegraphics[width=8cm]{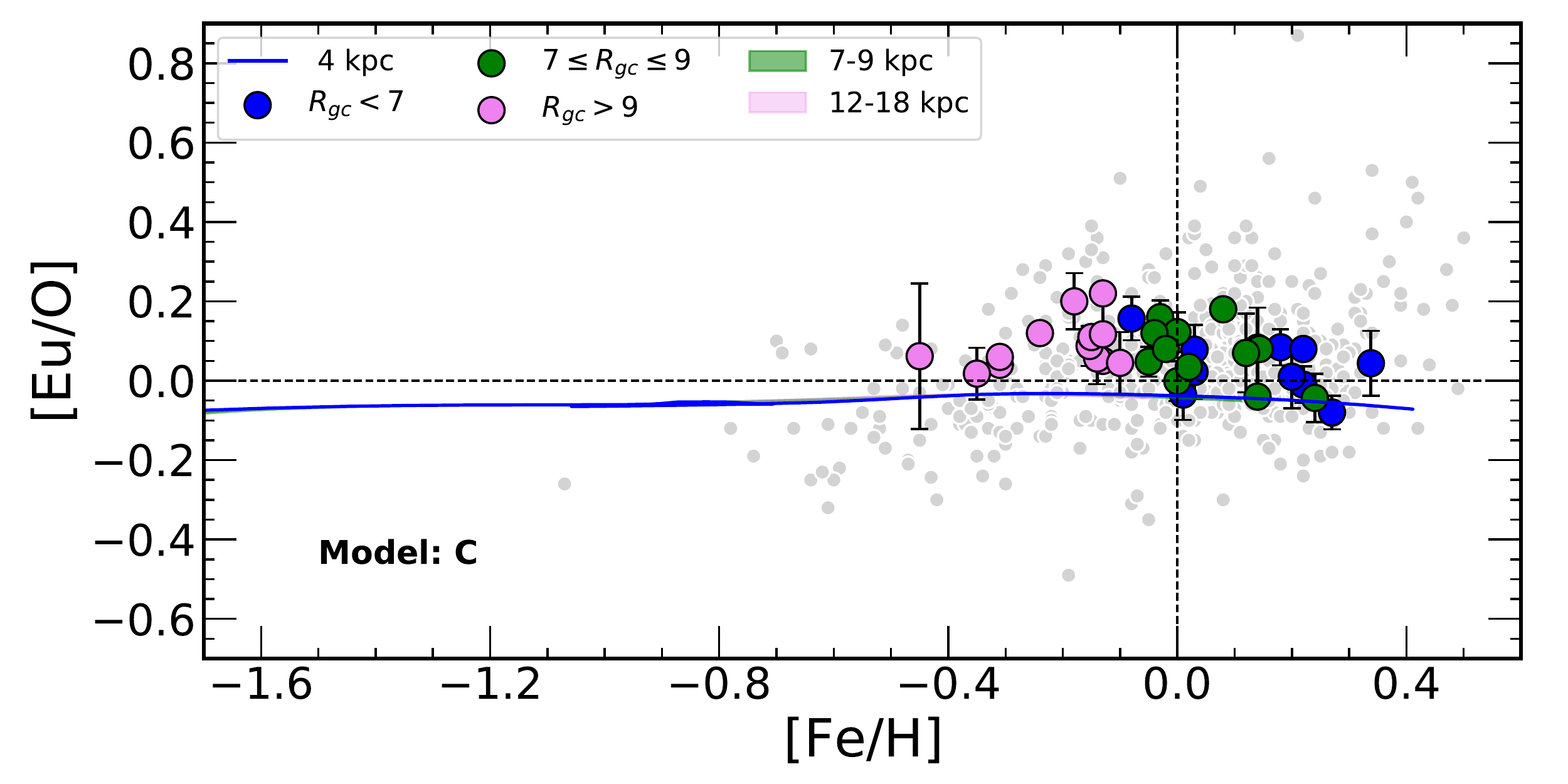}
  \includegraphics[width=8cm]{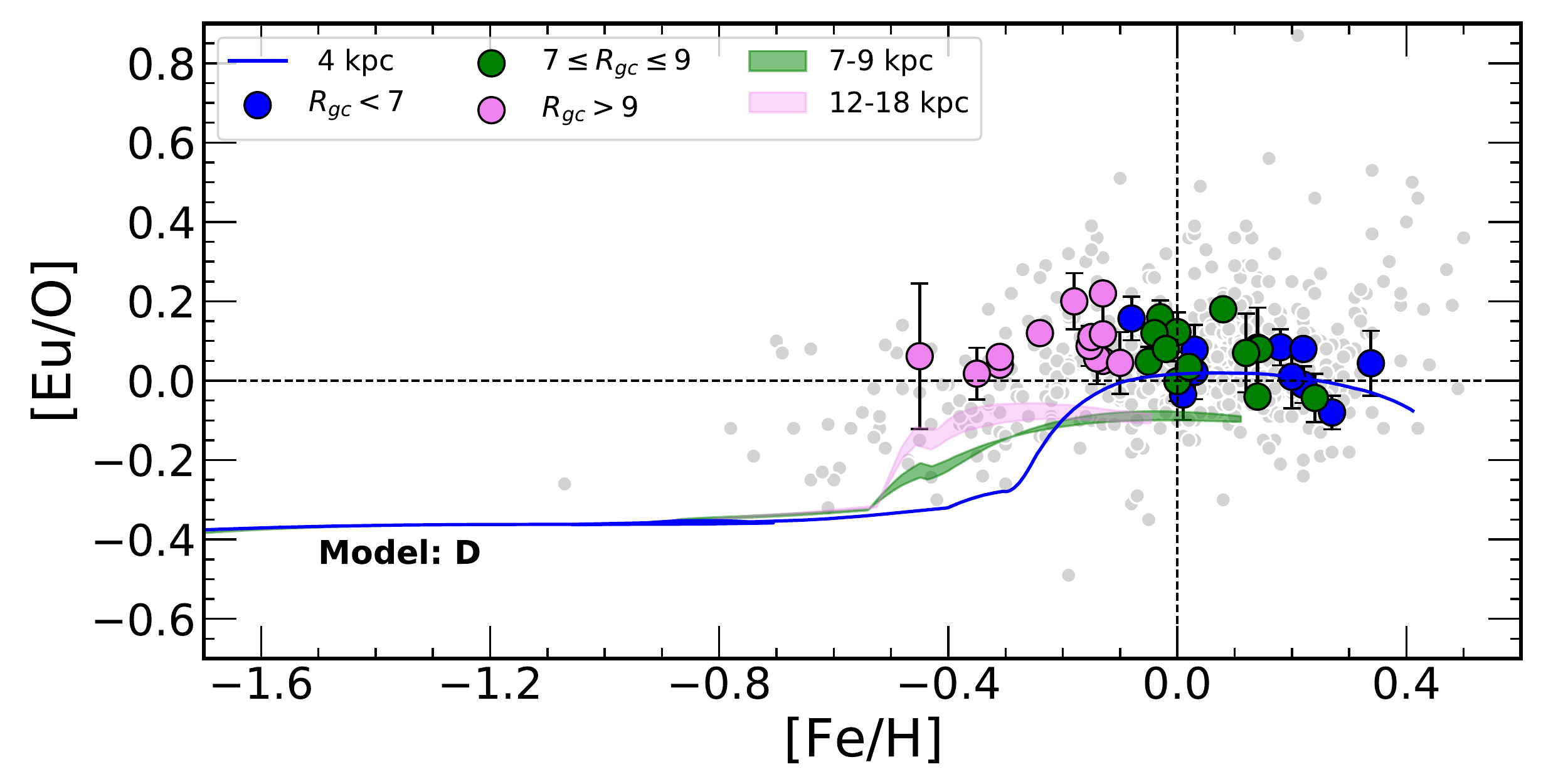}
  \caption{\label{Fig:EuO_vs_FeH}$\abratio{Eu}{O}$ vs. $\abratio{Fe}{H}$ for the field-star sample and the cluster sample. Same symbols and colours as in Fig.~\ref{fig:eufe_feh}. Two models are considered: model C (upper panel), model D (bottom panel).}
\end{figure}

Figure~\ref{Fig:EuO_vs_FeH} shows the evolution of $\abratio{Eu}{O}$ as a function of $\abratio{Fe}{H}$ for the field disc stars, the thin-disc OCs and the two models C and D. Once again, the best-matching model is the model C. The curves corresponding to the three Galactocentric regions under study are indiscernible and are about 0.15 lower than the OC data over the metallicity range $[-0.4, 0.4]$. This under-prediction in the models of the Eu-to-O ratio results from the slightly under-prediction of the Eu-to-Fe ratio seen in Fig.~\ref{fig:eufe_feh} and the over-prediction of O-to-Fe seen in Fig.~\ref{fig:ofe_feh}. On the other hand, the Eu-to-O for the OC sample exhibits a flat trend and only the model C is able to reproduce this feature. The model D, assuming a production of Eu by both \corevun{MRD SNe} and NSMs, seems to under-predict the Eu-to-O ratio by \num{0.2} at $\abratio{Fe}{H} \sim -0.6$ when compared to field stars. After a rapid increase of $\abratio{Eu}{O}$ due to the onset of NSMs, the model D predicts a flattening of Eu-to-O for the three Galactocentric regions. However, only the inner-disc curve matches the corresponding OC data. Thus, the Eu-to-O diagnostic speaks also in favour of a common origin of Eu and O in the thin disc, and therefore favours a rapid production of Eu by \corevun{MRD SNe}.

\begin{figure}
  \centering
  \includegraphics[width=8cm]{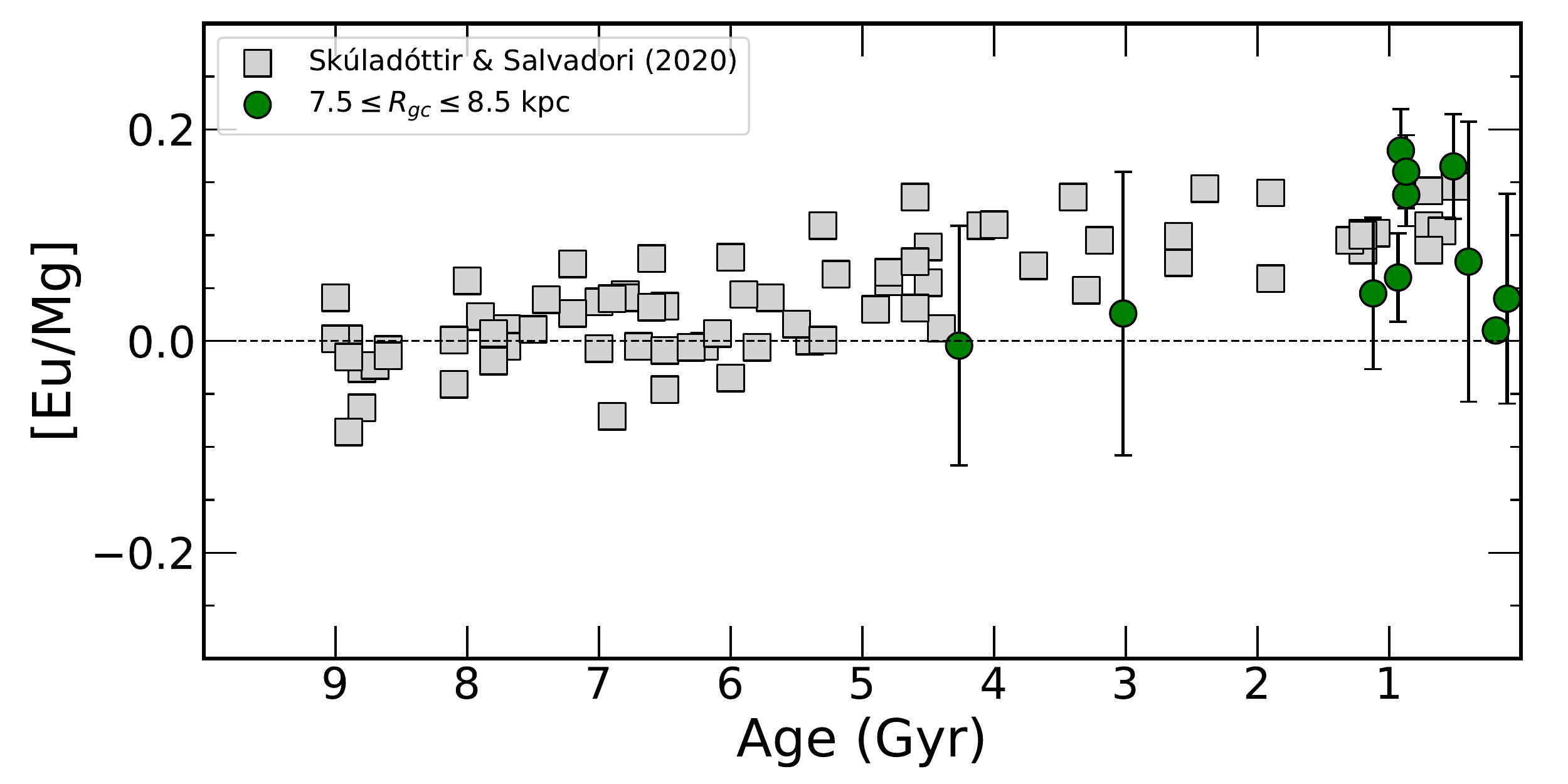}
  \caption{\label{fig:eumg_age_skúladóttir20}$\abratio{Eu}{Mg}$ vs. age (Gy) for our sample of open clusters in a $R_{gc}$ range of \num{7.5}-\SI{8.5}{\kilo\parsec} (green circles) compared with the solar twins (grey squares) from \citet{skuladottir20} with data from \citet{spina18} (ages, Eu abundances) and \citet{bedell18} (Mg abundances).}
\end{figure}

Finally, in Fig.~\ref{fig:eumg_age_skúladóttir20}, we show $\abratio{Eu}{Mg}$ with respect to age for the solar-twins sample discussed in \citet{skuladottir20} (sample based on \citealp{spina18,bedell18}) and for our OC sample, restricted to the solar neighbourhood. More specifically, we selected only OCs in a radial region close to that of the solar twins ($R_{\mathrm{GC}} \sim 7.5-\SI{8.5}{\kilo\parsec}$) and we excluded the clusters likely affected by migration (\object{NGC\,6971}, \object{Berkeley\,44}, and \object{Collinder\,261}; see \citetalias{viscasillas22} for more details on clusters' orbits and migration). The data of solar twins and of open clusters agree in the age range in which they overlap. \citet{skuladottir20} claim to detect a change of slope in the $\abratio{Eu}{Mg}$ vs. age plane occurring \SI{4}{\giga\Year} ago, signing the rise of a the Eu production by NSMs. Given the short age interval spanned by the solar-ring OCs (younger than \SI{6}{\giga\Year}), they cannot be used to investigate this change of slope. However, we would like to stress that the flattening modelled by \citet{skuladottir20} does \corevun{not} appear to be statistically significant. Indeed if we perform an F-test choosing for the null hypothesis $H_{0}$ "the solar-twins distribution is described by a single linear-regression" and the alternative hypothesis $H_{1}$ "the solar-twins distribution is described by two piece-wise linear-regressions", then we cannot reject $H_{0}$ at the \SI{95}{\percent} level (see Table~\ref{Table:linear_regressions_solar_twins}).

\begin{table}
  \caption{\label{Table:linear_regressions_solar_twins} Model significance for the solar-twins sample: assumed regressions and F-value. The critical F-value is given for a right-tailed test, with a false-rejection probability $\alpha$ = 0.05, and the two degrees of freedom are 2 and 75.}
  \centering
  \begin{tabular}{cl}
    \hline
    $H_{0}$ & $\abratio{Eu}{Mg} = 0.0168 \times \mathrm{age} + 0.1328$\\
    \hline
    $H_{1}$ & \makecell{$\abratio{Eu}{Mg} = 0.0076 \times \mathrm{age} + 0.1181$ for $\mathrm{age} \le \SI{4}{\giga\Year}$\\  $\abratio{Eu}{Mg} = 0.0175 \times \mathrm{age} + 0.1367$ for $\mathrm{age} > \SI{4}{\giga\Year}$}\\
    \hline
    F-value & 0.4643\\
    F-critical & 3.119\\
    \hline
  \end{tabular}
\end{table}

\begin{figure}
  \centering
  \includegraphics[width=8cm]{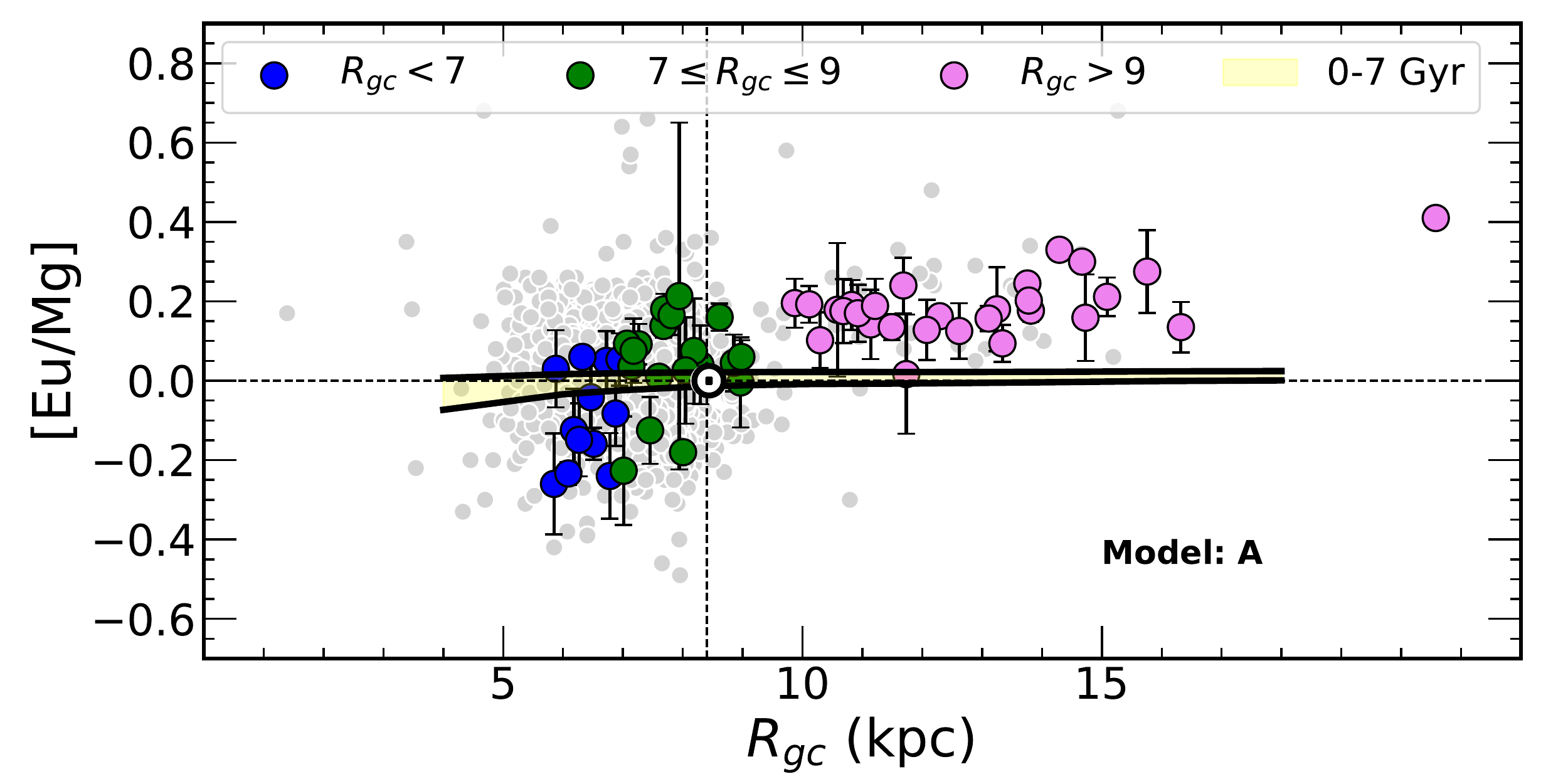}
  \includegraphics[width=8cm]{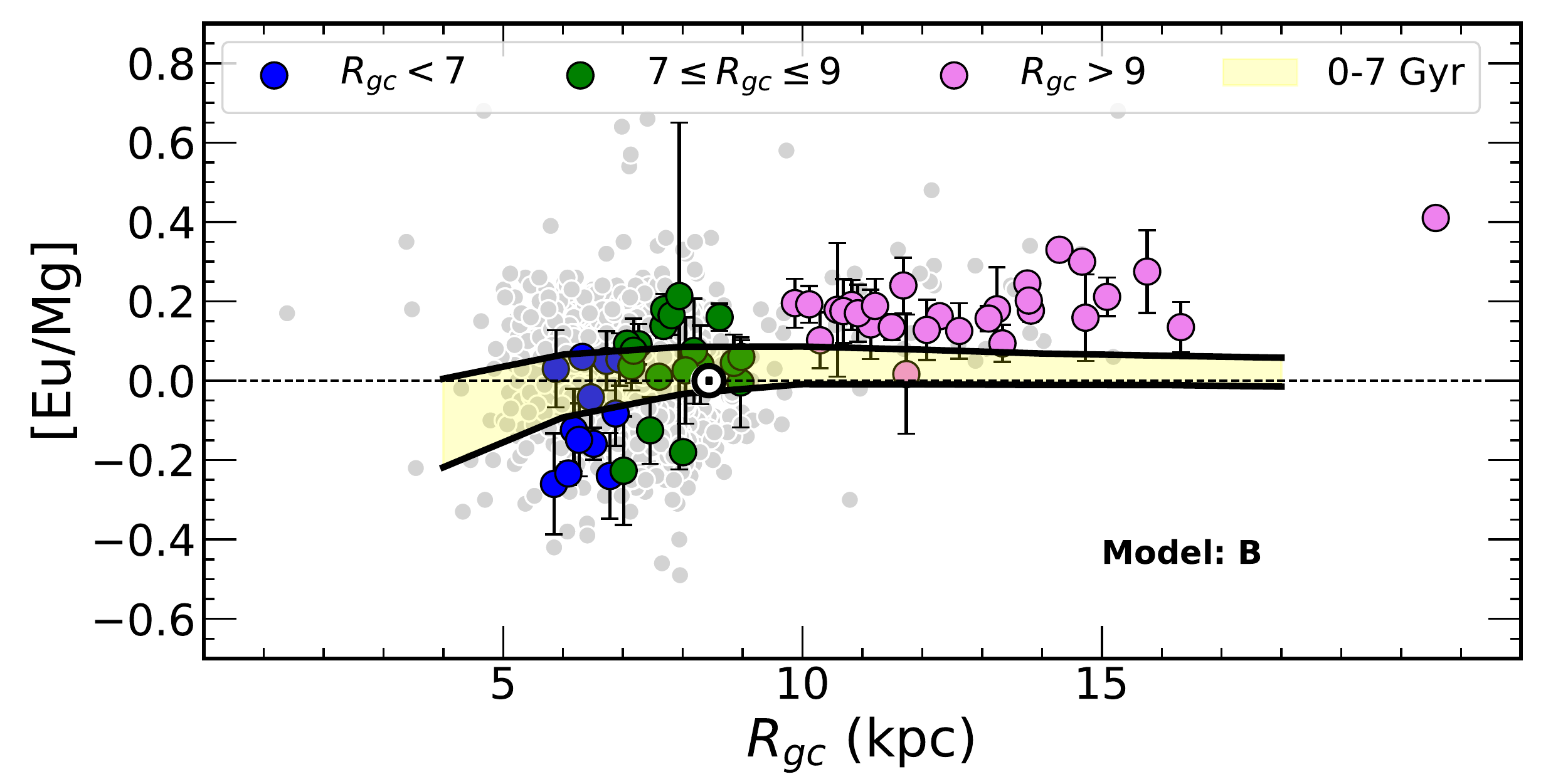}
  \includegraphics[width=8cm]{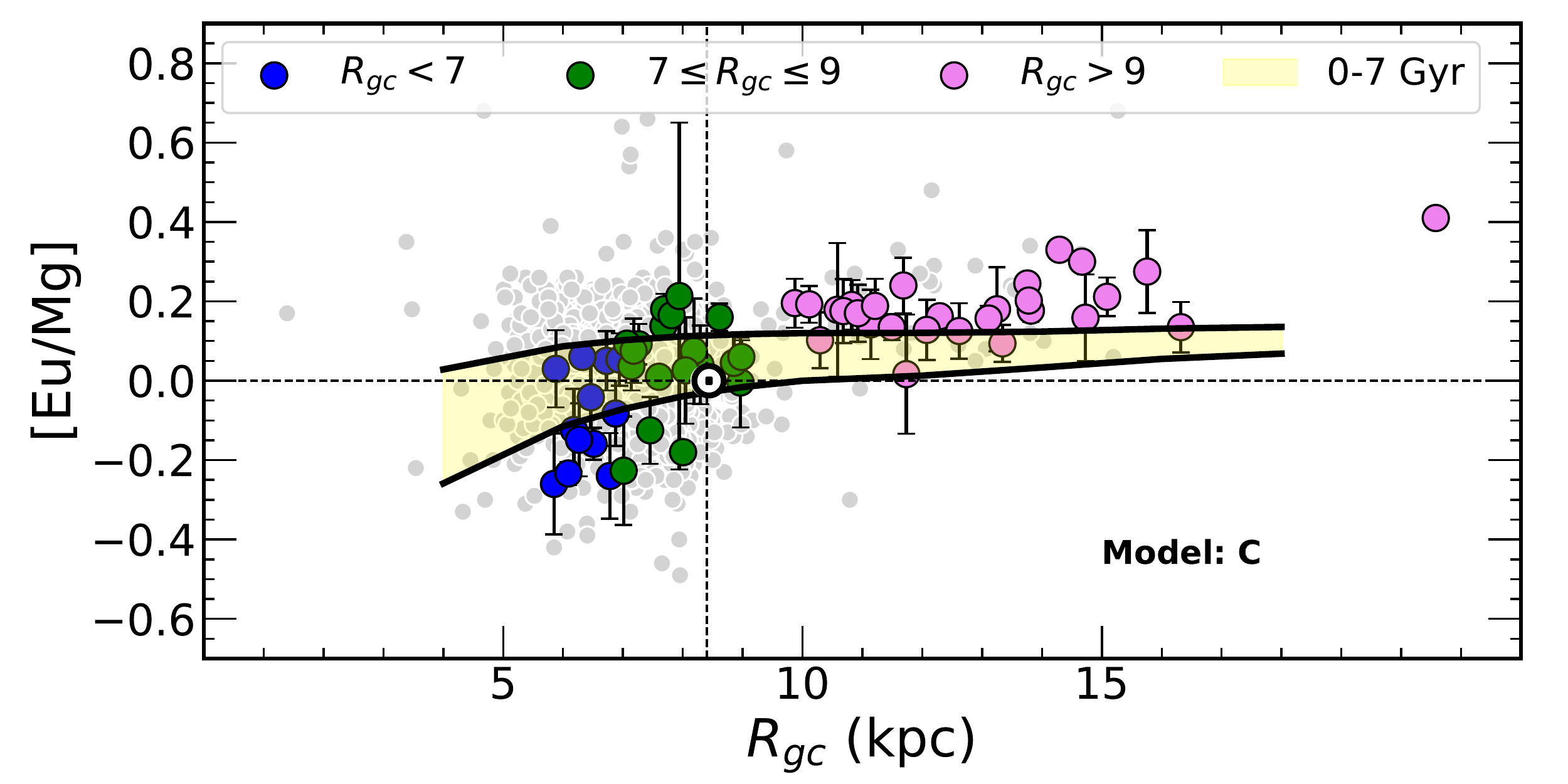}
  \includegraphics[width=8cm]{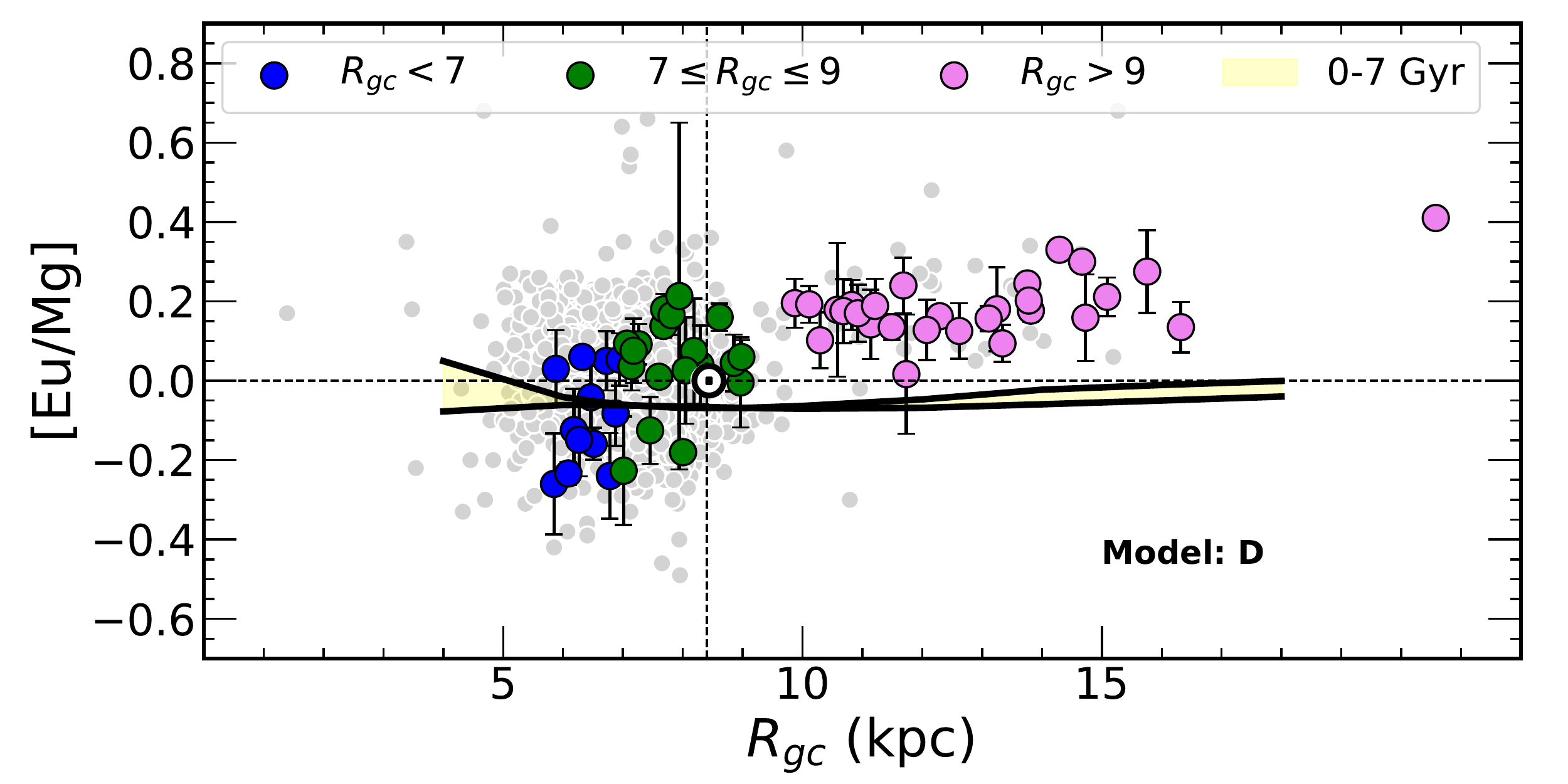}
  \caption{\label{Fig:EuMg_vs_Rgc}$\abratio{Eu}{Mg}$ vs. $R_{GC}$ for the field-star sample and the cluster sample. Same symbols and colours as in Fig.~\ref{fig:eufe_feh}. Four models are considered. From top to bottom: model A, model B, model C and model D.}
\end{figure}

\begin{figure}
  \centering
  \includegraphics[width=8cm]{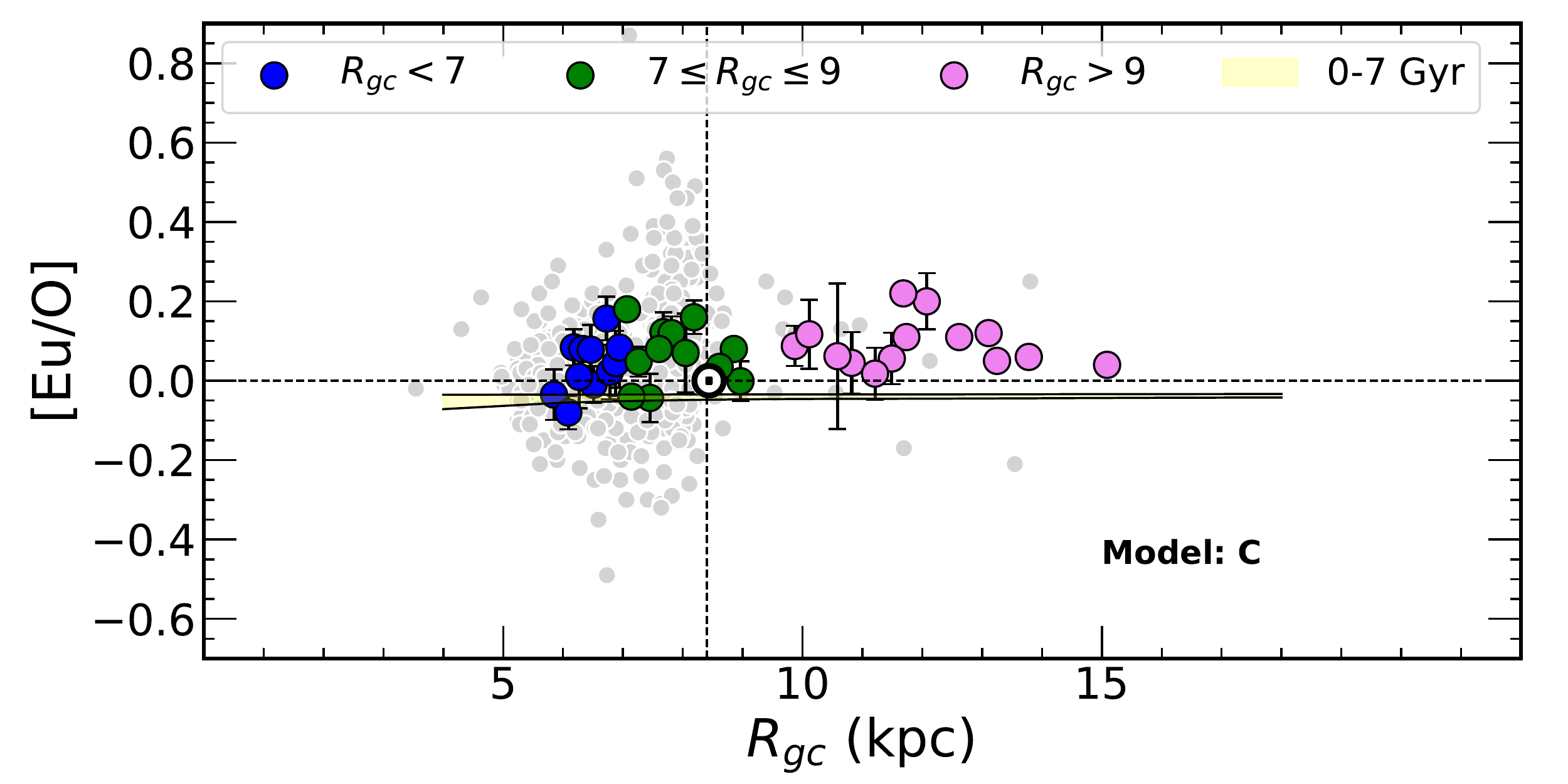}
  \includegraphics[width=8cm]{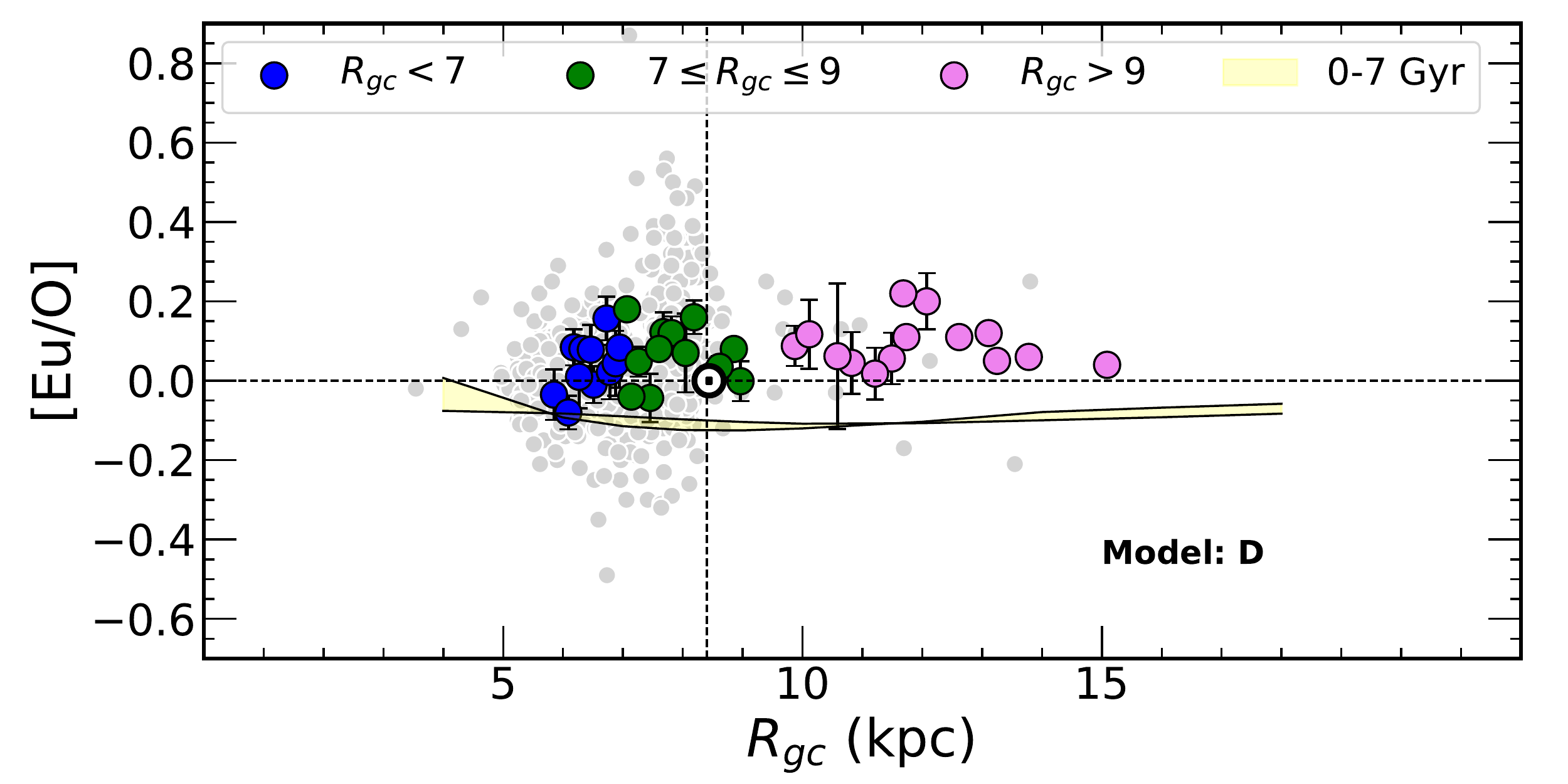}
  \caption{\label{Fig:EuO_vs_Rgc}$\abratio{Eu}{O}$ vs. $R_{GC}$ for the field-star sample and the cluster sample. Same symbols and colours as in Fig.~\ref{fig:eufe_feh}. Two models are considered: model C (upper panel) and model D (bottom panel).}
\end{figure}

Figures~\ref{Fig:EuMg_vs_Rgc} and \ref{Fig:EuO_vs_Rgc} display the radial gradient for $\abratio{Eu}{Mg}$ and $\abratio{Eu}{O}$, respectively. We note an increasing trend of $\abratio{Eu}{Mg}$ with Galactocentric radius from the OC sample. It can be explained by the fact that Mg and Eu are produced via different nucleosynthetic channels: a non-negligible production of Mg by a delayed mechanism (e.g., SNe\,Ia) in the context of inside-out formation of the Galactic disc would explain why $\abratio{Eu}{Mg}$ becomes negative first at smaller radii. On the other hand, $\abratio{Eu}{O}$ is flat over the probed Galactic radii. This is compatible with a scenario where O and Eu are produced by the same progenitors, i.e. CCSNe.

\subsection{Other \emph{r}-process elements}

\corevun{The distributions of heavy elements synthesized by the \emph{s}-process are characterised by the presence of three peaks, corresponding to neutron magic numbers \num{50}, \num{82}, and \num{126}. The \emph{s}-process-dominated elements belonging to the first peak are Sr, Y and Zr. Those belonging to the second peak are Ba, La and Ce. Close to these peaks, there are elements (as Mo, Nd and Pr), whose origin is shared among the \emph{s}-process and the \emph{r}-process. As a matter of fact, the contributions from the two nucleosynthesis processes (and eventually from the \emph{p}-process) are almost equal,} at least in the age and metallicity range of the disc.

Estimates of the contributions of the different processes to their abundances in the Sun vary from one author to another \citep[e.g.][]{arlandini99, Simmerer_2004, Sneden_2008, Bisterzo14, Prantzos_2020}, but they all agree on assigning them a non-negligible percentage of \emph{r}-process, in some cases more than \SI{50}{\percent}. Indeed, the \emph{s}-process dominated elements seem to be placed in the second to fourth IUPAC groups of the periodic table (Sr, Ba, Y, La, Zr, Ce); the mixed elements in the fifth to sixth groups (Pr, Mo, Nd), and the \emph{r}-process dominated elements in the eighth and ninth (Ru, Sm, and Eu). This suggests in most of the cases an increase in the \emph{r}-process component from left to right in the periodic table, group by group, for the aforementioned elements or, equivalently, by increasing its ionization energy. \corevun{In the following, we concentrate on the mixed elements Mo, Nd and Pr.}

For the elements in the present work, by considering everything that does not originate from the \emph{s}-process as produced by the \emph{r}-process, the above-quoted works agree in assigning $\sim \SI{40}{\percent}$ of the \emph{r}-process component to Nd and $\sim \SI{50}{\percent}$ to Pr. For Mo, there is less consensus. \citet{Bisterzo14} attribute more than \SI{60}{\percent} of its origin to the \emph{r}-process component, while \citet{Cowan2021} attribute to Mo an almost complete origin from the \emph{r}-process. On the other hand, \citet{Prantzos_2020} proposed a percentage of \SI{50}{\percent} to \emph{s}-process and \SI{27}{\percent} to the \emph{r}-process, assigning the remaining \SI{23}{\percent} to the \emph{p}-process (in which photo-disintegrations produce proton-rich nuclei starting from pre-existing heavy isotopes; see e.g. \citealt{Mishenina_2019}).

\begin{figure}
  \resizebox{\hsize}{!}{\includegraphics{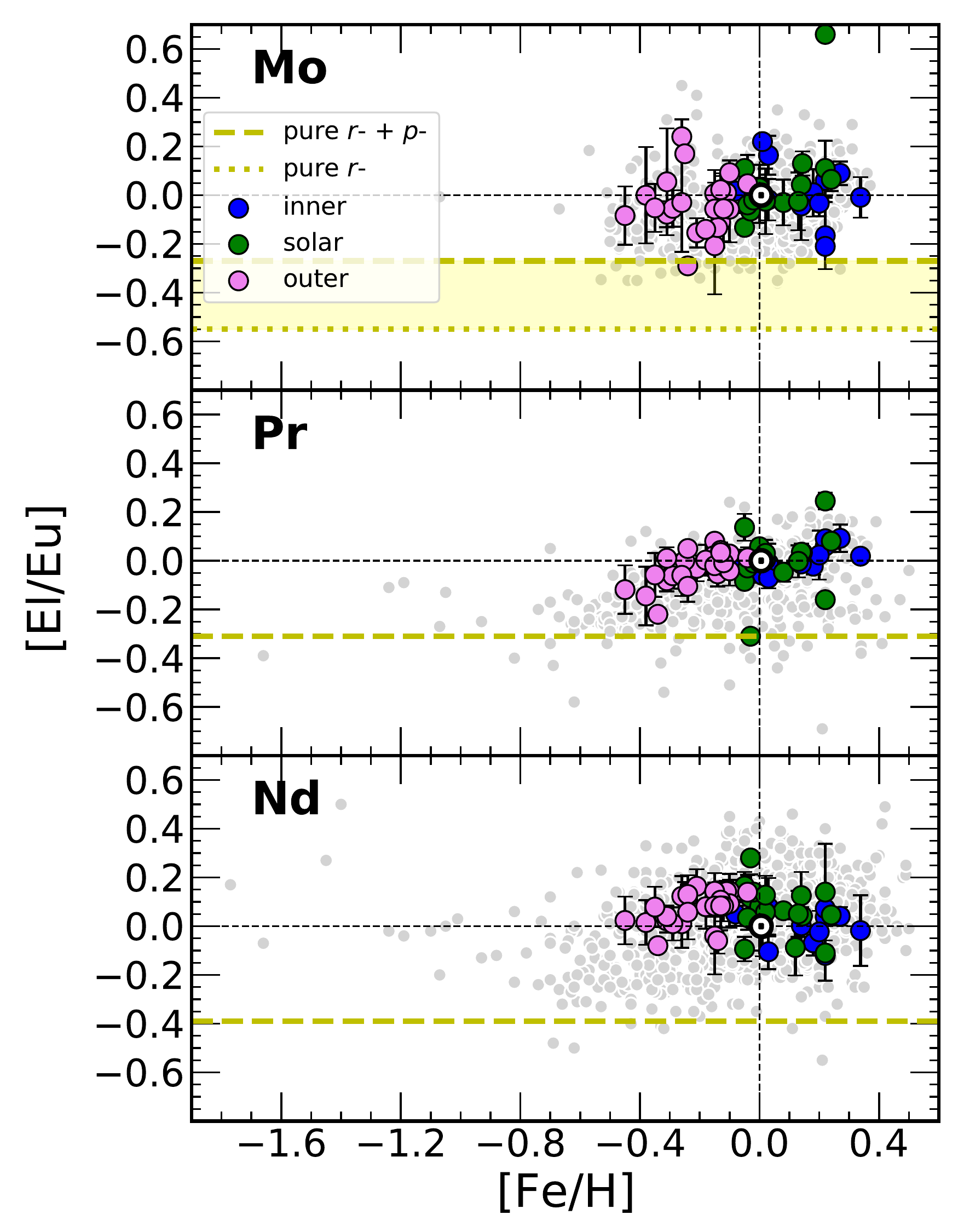}}
  \caption{\label{fig:eleu_mixed}$\abratio{El}{Eu}$ vs. $\abratio{Fe}{H}$ for the mixed elements: Mo, Pr and Nd. Data for both the field-star sample and cluster sample are shown. Same symbols and colours as in Fig.~\ref{fig:eufe_feh}. The yellow dashed lines track the pure contribution of the \emph{r}-process derived using the solar abundances from \citep{Grevesse_2007} and the most recent percentages from \citet{Prantzos_2020} for the \emph{r}-process and the \emph{r}- + \emph{p}-process for Mo.}
\end{figure}

We examine the origin of these elements from an observational point of view, comparing their abundance with that of Eu. In Figure \ref{fig:eleu_mixed}, we present $\abratio{El}{Eu}$ vs. $\abratio{Fe}{H}$ for Mo, Pr, and Nd in both clusters and field stars. In the figure, for each element we also show the \emph{r}-process percentage in the Sun proposed by \citet{Prantzos_2020}. For molybdenum, we also report an intermediate level, determined by \corevun{the} sum of the \emph{r}-component and the \emph{p}-component. \corevun{If the elements were produced only by the \emph{r}-process at all metallicities, we would expect to find their abundances close to the lines which indicate the only \emph{r}-process contribution. To reach the observed abundances at the typical metallicity of the disc, a contribution from the \emph{s}-process is required. The metallicity at which $\abratio{El}{Fe}$ start to increase is different for Mo, Pr and Nd, indicating different time-scales for their production.}

\paragraph{Neodymium:} Among the three elements, Nd has the flattest trend, \corevun{and thus we cannot identify the metallicity corresponding to the transition between the \emph{r}-process-dominated regime and \emph{s}-process-dominated regime, since the contribution of the \emph{s}-process might start at lower metallicities, at least as far as clusters are concerned. Its flat profile with respect to europium and the difference of about \num{0.4} with respect to its \emph{r}-process abundance points to a significant \emph{s}-process contribution of the same order of the \emph{r}-process contribution over the disc metallicity range.}

\paragraph{Praseodymium:} \corevun{The same does not hold for Praseodymium, for which a lower \emph{s}-process contribution is expected. As a matter of fact, a increasing trend of $\abratio{Pr}{Eu}$ with increasing $\abratio{Fe}{H}$ is well defined indicating a recent enrichment by \emph{s}-process (starting from about $\abratio{Fe}{H} \sim -0.4$, as an upper limit). However, the \emph{r}-process component still dominates the Pr production at high metallicity, as recently} reported by \citet{tautvasiene21} for about \num{500} thin and thick disc stars in the solar neighbourhood.

\paragraph{Molybdenum:} \corevun{Finally, Mo abundances show a quite flat trend, characterised by a greater scatter than Pr and Nd ones. We also note that this scatter tend to increase with decreasing metallicity. The contribution from the \emph{s}-process seems to have started at lower metallicities than those we sample with the OCs.} This behaviour was also observed by \citet{Mishenina_2019}. Such a scatter is closely related to the elusive nature of this chemical element \corevun{and to the difficulty in measuring its abundance}. As a matter of fact, different GCE studies reached discordant conclusions on this element, proposing various solutions to reach a better agreement between theory and observations. \citet{Mishenina_2019} concluded that canonical stellar sources of heavy elements do not produce enough Mo, while \citet{koba20} stated that the disagreement can be mitigated by including the $\nu$-wind from nascent neutron stars. Finally, \citet{Prantzos_2020} ascribed to the \emph{p}-process the missing percentage to reproduce the Solar composition. On top of that, recent spectroscopic observations of heavy elements in barium stars (which are thought to have been polluted by the \emph{s}-process at work in the already extinct AGB companion) highlighted that the enhancements of Nb, Mo, and Ru are larger than those expected by current available \emph{s}-process models \citep{roriz21}. For these elements, Ba stars show enrichment definitely larger than those found in field stars, pointing to a process at work in those binary systems (thus excluding a different pristine chemical distribution, more easily attributable to the \emph{r}-process). This would be at odds with the conclusions by \citet{Mishenina_2019}, who excluded the \emph{s}-process as the responsible for such a peculiar chemical feature. Further studies on this topic are urgently needed, possibly focusing on the improvement of the nuclear inputs adopted to run nucleosynthesis models. 

\section{Summary and conclusions}
\label{sec_summary}

With the aim of shedding light on the most prominent sources for the \emph{r}-process in the Galactic disc, we compared the abundance of Eu, that is \corevun{an} element mainly produced by the \emph{r}-process, with those of \corevun{the} two $\alpha$-elements (Mg and O), expected to be originated mostly in core-collapse SNe on short timescales. For this purpose, we rely on a large sample of open clusters from the recently released \emph{Gaia}-ESO iDR6, which gives us the advantage of containing one of the largest and most complete sample of open clusters, distributed in age and Galactocentric distance, in which abundances of neutron capture elements have been measured. We complement our sample of open clusters with that of field stars. As it is known, our ability to obtain ages for field stars is limited, but they can still complement the information we get from clusters, as they have an age distribution that goes to older ages.

We built up a GCE model, in which we make several choices for Eu and Mg nucleosynthesis (models A to D). For Eu, we considered two possible mechanisms of production: a fast production in CCSNe (e.g. magneto-rotationally driven SNe) and a combination of CCSNe and delayed production in neutron-star mergers (NSMs) with a delay of \SI{3}{\giga\Year}. For Mg, we considered three different cases: CCSNe, CCSNe and SNe\,Ia, and finally CCSNe and SNe\,Ia with metallicity-dependent yields. We compared the observations with \corevun{the} results of the model(s) in different planes ($\abratio{El}{Fe}$ vs. age, $\abratio{El}{Fe}$ vs. $\abratio{Fe}{H}$, $\abratio{El1}{El2}$ vs. age and $\abratio{Fe}{H}$). The first conclusion of the model-observation comparison is that for Eu, at the metallicity of the disc, it is not necessary to introduce a delayed component, e.g. from NSMs. The fast production is sufficient to reproduce the observational data. In order to make a meaningful comparison of Eu abundance with those of O and Mg, we have studied their chemical evolution: for oxygen, a rapid source (CCSNe) is sufficient to explain the observations, while for Mg, a growth (flattening in the $\abratio{Mg}{Fe}$ vs. $\abratio{Fe}{H}$ plane) is clearly visible in the data, which we have explained as the contribution of SNe\,Ia at high metallicity. Although not directly related to the main purpose of our work, the differences between O and Mg show that $\alpha$-elements are not interchangeable with each other, and that great care must be taken in their correct use. In particular, Mg has a larger production from SNe\,Ia at high metallicity than usually expected, and it cannot be considered a 'pure' $\alpha$-element, at least in the metallicity range of the Galactic disc.

Once the origin of Mg and O has been established, comparison with Eu gives us a further key to understanding the origin of this element. On the one hand, the observations show a growth of $\abratio{Eu}{Mg}$ at low metallicity, which can be correctly explained by the model only if we consider that Eu and Mg have \corevun{a} different origin. In particular, Eu does not share the same delayed production as Mg at high metallicity (see model C). On the other hand, within the uncertainties, $\abratio{Eu}{O}$ has a flat trend with metallicity, pointing towards a common origin (or, better, towards common timescales) for these two elements. The model with a delayed production of Eu clearly underestimates the $\abratio{Eu}{O}$ ratio at low metallicity. Finally, \corevun{the} observations of star clusters show a positive radial gradient of $\abratio{Eu}{Mg}$ in the disc, which again can be explained by the combination of the inside-out growth of the disc and the delayed extra-production of Mg at high metallicity (not yet reached in the outer disc). The radial gradient of $\abratio{Eu}{O}$ is, on the other hand, almost flat (a small offset between the data and the model is present), indicating again similar timescales for their production. 

We can therefore conclude that the europium we observe in field and cluster populations at the thin disc metallicities is predominantly produced by sources with short lifetimes, such as magneto-rotationally driven SNe or collapsars. The same role can be played by NSMs if their mergers take place with a very short delay \citep{matteucci14} or -- in the contex of an time delay distribution -- if their frequency was higher at low metallicity \citep{simonetti19,cavallo21}. Indeed, with these assumptions, their enrichment can mimic the fast pollution by CCSNe. Introducing the NSMs as additional source can still be an option, but according to our results it appears to be negligible at thin disc metallicities \citep[cf.][]{skuladottir20}.

Finally, we analysed three mixed elements (Mo, Pr and Nd) to which a non-negligible origin in the \emph{r}-process is attributed. For each of them, we \corevun{discuss} the component produced by the \emph{r}-process. The most interesting case is represented by molybdenum, whose cosmic origin is still a debated matter and deserves future dedicated studies.

\begin{acknowledgements}
  \corevun{We thank the anonymous referee for their relevant questions and remarks that helped us in improving the presentation and the discussion of the results.}
  Based on data products from observations made with ESO Telescopes at the La Silla Paranal Observatory under programme ID 188.B-3002, 193.B-0936, 197.B-1074. These data products have been processed by the Cambridge Astronomy Survey Unit (CASU) at the Institute of Astronomy, University of Cambridge, and by the FLAMES/UVES reduction team at INAF/Osservatorio Astrofisico di Arcetri. These data have been obtained from the \emph{Gaia}-ESO Survey Data Archive, prepared and hosted by the Wide Field Astronomy Unit, Institute for Astronomy, University of Edinburgh, which is funded by the UK Science and Technology Facilities Council.
  This work was partly supported by the European Union FP7 programme through ERC grant number 320360 and by the Leverhulme Trust through grant RPG-2012-541. We acknowledge the support from INAF and Ministero dell' Istruzione, dell' Universit\`a' e della Ricerca (MIUR) in the form of the grant "Premiale VLT 2012" and "Premiale 2016 MITiC". The results presented here benefit from discussions held during the \emph{Gaia}-ESO workshops and conferences supported by the ESF (European Science Foundation) through the GREAT Research Network Programme.
  TB was funded by grant No. 621-2009-3911 and grant No. 2018-0485 from The Swedish Research Council.
  FJE acknowledges financial support by the Spanish grant MDM-2017-0737 at Centro de Astrobiolog\'{\i}a (CSIC-INTA), Unidad de Excelencia Mar\'{\i}a de Maeztu.
  TM acknowledges financial support from the Spanish Ministry of Science and Innovation (MICINN) through the Spanish State Research Agency, under the Severo Ochoa Program 2020-2023 (CEX2019-000920-S).
  LS is supported by the Italian Space Agency (ASI) through contract 2018-24-HH.0 to the National Institute for Astrophysics (INAF).
  This work has made use of data from the European Space Agency (ESA) mission \emph{Gaia} (\url{https://www.cosmos.esa.int/gaia}), processed by the \emph{Gaia} Data Processing and Analysis Consortium (DPAC, \url{https://www.cosmos.esa.int/web/gaia/dpac/consortium}). Funding for the DPAC has been provided by national institutions, in particular the institutions participating in the \emph{Gaia} Multilateral Agreement. CVV and LM thank the COST Action CA18104: MW-Gaia. GC and AK acknowledge ChETEC-INFRA (EU project no. 101008324).
  DV acknowledges financial support from the German-Israeli Foundation (GIF No. I-1500-303.7/2019). MB is supported through the Lise Meitner grant from the Max Planck Society. We acknowledge support by the Collaborative Research centre SFB 881 (projects A5, A10), Heidelberg University, of the Deutsche Forschungsgemeinschaft (DFG, German Research Foundation).  This project has received funding from the European Research Council (ERC) under the European Union’s Horizon 2020 research and innovation programme (Grant agreement No. 949173)
\end{acknowledgements}

\bibliographystyle{aa} 
\bibliography{biblio} 

\begin{appendix}
  \section{Additional material}


  \begin{figure*}[h!]
    \centering
    \includegraphics[scale=0.35]{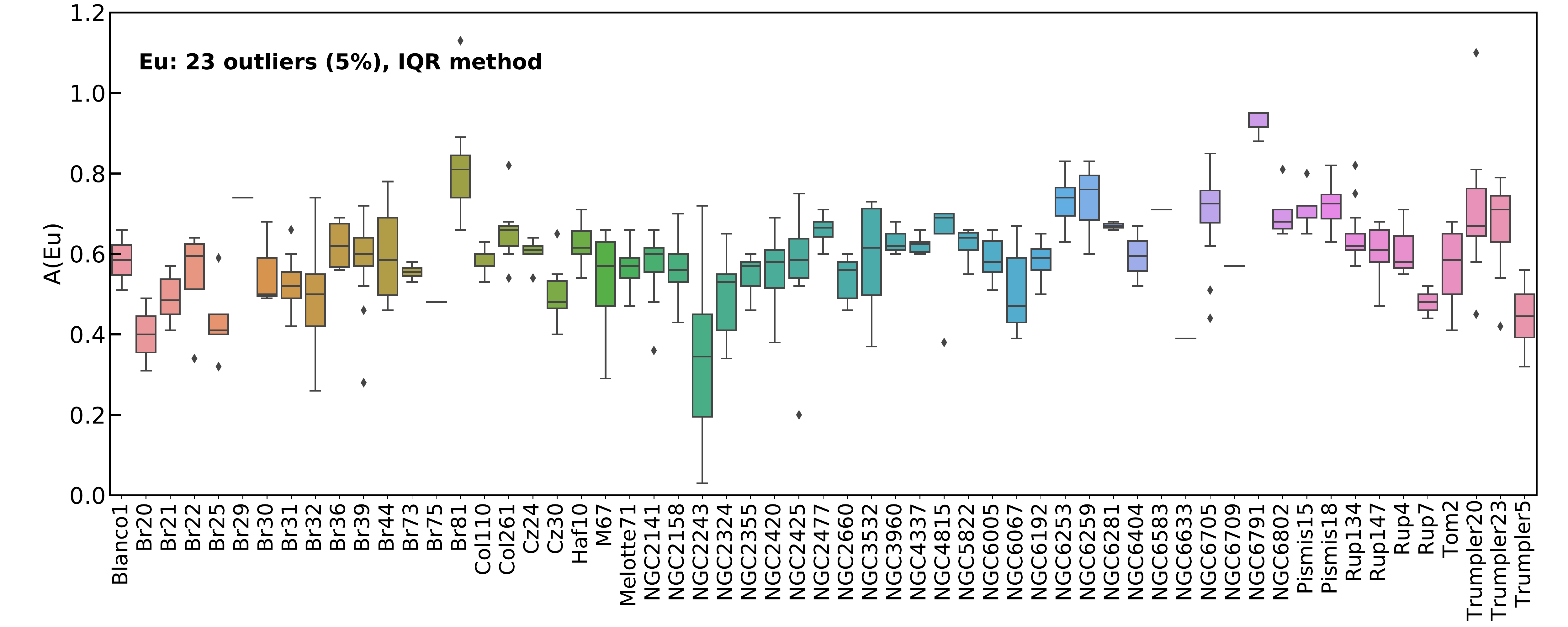}
    \includegraphics[scale=0.35]{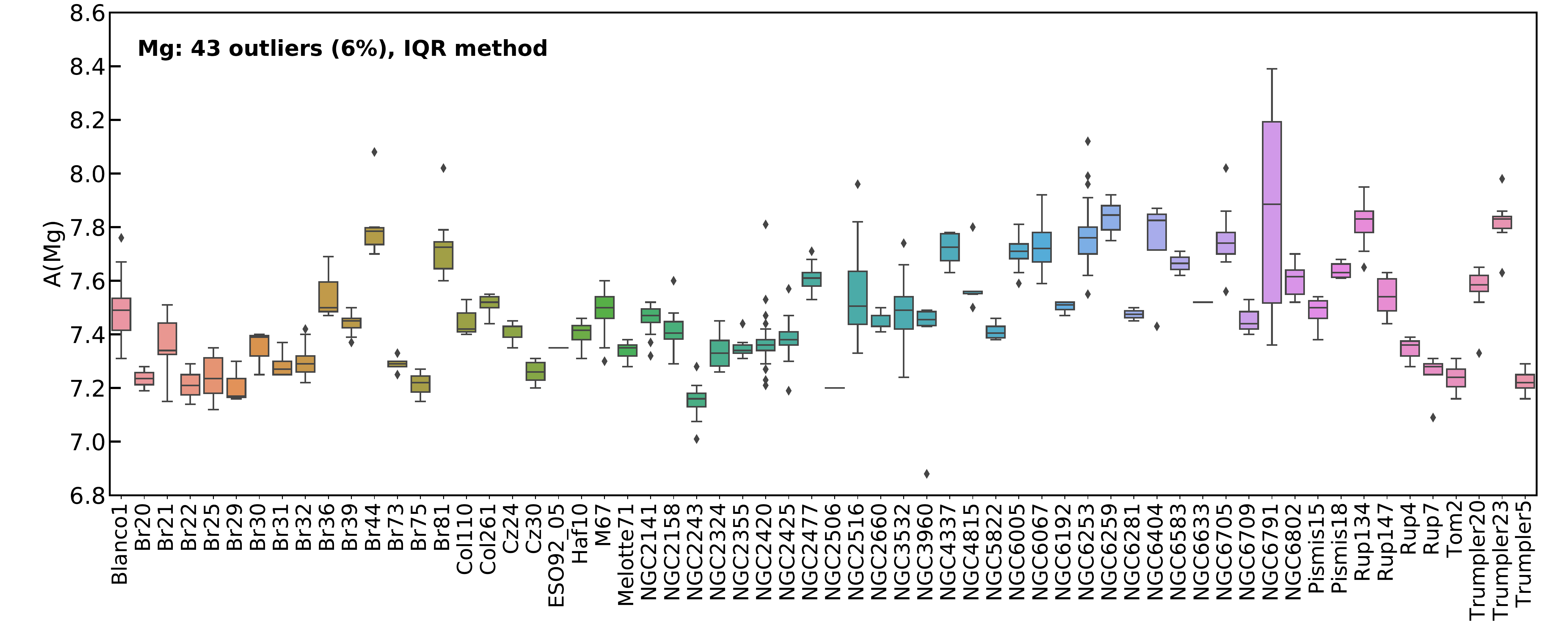}
    \includegraphics[scale=0.35]{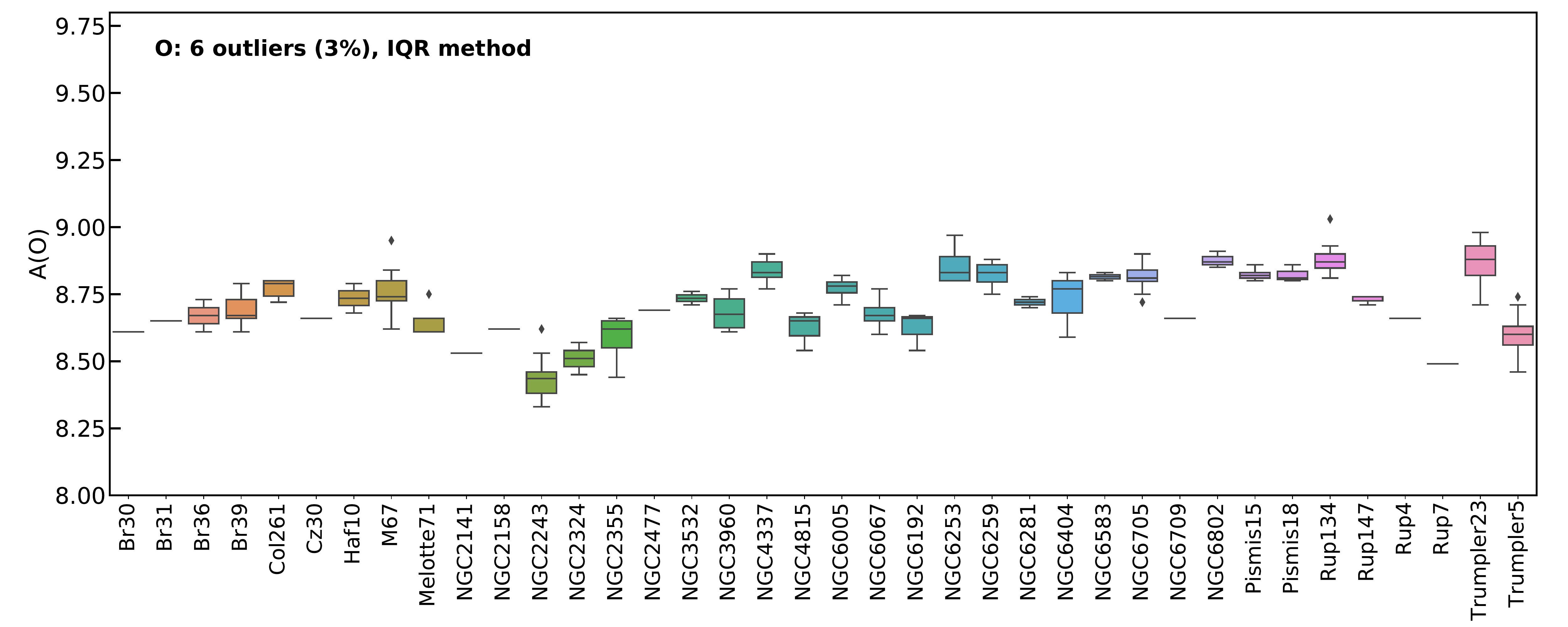}
    \caption{\label{fig:outlier_eu}Boxplots with the interquartile range of the abundance for each of the clusters with the outliers of Eu, Mg and O (observations that fall below Q1 - 1.5 IQR or above Q3 + 1.5 IQR).}
  \end{figure*}


  \begin{table*}
    \caption{\label{clusters_elh}Average [El/H] and $\sigma$[El/H] for our sample of open clusters. The age (Gy) and $R_\mathrm{GC}$ (kpc) are from \citet{CG20} and [Fe/H] from \citet{Randich_2022}.}
    \centering
    \scalebox{0.6}{
      \begin{tabular}{lrrrrrrrrrrrrrrrr}
        \hline\hline
        GES\_FLD & [Fe/H] &  age (Gy) & $R_\mathrm{GC}$ (kpc) & $[\ion{O}{i}/$H] & $\sigma$ $[\ion{O}{i}/$H] & $[\ion{Mg}{i}/$H] & $\sigma$ $[\ion{Mg}{i}/$H] & $[\ion{Mo}{i}/$H] & $\sigma$ $[\ion{Mo}{i}/$H] & $[\ion{Pr}{ii}/$H] & $\sigma$ $[\ion{Pr}{ii}/$H] & $[\ion{Nd}{ii}/$H] & $\sigma$ $[\ion{Nd}{ii}/$H] &  $[\ion{Eu}{ii}/$H] & $\sigma$ $[\ion{Eu}{ii}/$H]  \\
        \hline

        Blanco 1 & -0.03 & 0.1 & 8.3 &  &  & -0.01 & 0.1 &  &  & -0.09 & 0.11 & 0.15 & 0.02 & 0.05 & 0.11\\
        Berkeley 20 & -0.38 & 4.79 & 16.32 &  &  & -0.29 & 0.06 & -0.16 & 0.33 & -0.31 & 0.01 & -0.14 & 0.22 & -0.16 & 0.13\\
        Berkeley 21 & -0.21 & 2.14 & 14.73 &  &  & -0.17 & 0.1 & -0.24 & 0.07 & -0.11 & 0.07 & 0.14 & 0.1 & -0.07 & 0.06\\
        Berkeley 22 & -0.26 & 2.45 & 14.29 &  &  & -0.32 & 0.08 & 0.11 & 0.24 & -0.24 &  & 0.02 & 0.14 & 0.05 & 0.04\\
        Berkeley 25 & -0.25 & 2.45 & 13.81 &  &  & -0.29 & 0.09 & 0.03 & 0.25 & -0.15 & 0.16 & 0.03 & 0.1 & -0.14 & 0.03\\
        Berkeley 29 & -0.36 & 3.09 & 20.58 &  &  & -0.32 & 0.08 & 0.03 &  &  &  & -0.02 & 0.06 & 0.18 & \\
        Berkeley 30 & -0.13 & 0.3 & 13.25 & -0.12 &  & -0.18 & 0.08 & -0.18 &  & 0.02 & 0.09 & 0.12 & 0.08 & -0.0 & 0.11\\
        Berkeley 31 & -0.31 & 2.82 & 15.09 & -0.08 &  & -0.24 & 0.04 & -0.12 & 0.04 & -0.15 & 0.07 & -0.02 & 0.01 & -0.05 & 0.06\\
        Berkeley 32 & -0.29 & 4.9 & 11.14 &  &  & -0.24 & 0.05 & -0.15 & 0.13 & -0.17 & 0.1 & -0.09 & 0.09 & -0.08 & 0.12\\
        Berkeley 36 & -0.15 & 6.76 & 11.73 & -0.06 & 0.08 & 0.02 & 0.12 & -0.19 & 0.13 & 0.08 &  & -0.09 & 0.18 & 0.06 & 0.07\\
        Berkeley 39 & -0.14 & 5.62 & 11.49 & -0.04 & 0.07 & -0.08 & 0.04 & -0.09 & 0.1 & -0.02 & 0.05 & 0.0 & 0.05 & 0.05 & 0.05\\
        Berkeley 44 & 0.22 & 1.45 & 7.01 &  &  & 0.23 & 0.04 & 0.14 & 0.03 & 0.23 & 0.08 & 0.09 & 0.1 & 0.04 & 0.15\\
        Berkeley 73 & -0.26 & 1.41 & 13.76 &  &  & -0.24 & 0.0 & -0.04 & 0.19 & -0.07 & 0.07 & 0.12 & 0.06 & -0.01 & 0.02\\
        Berkeley 75 & -0.34 & 1.7 & 14.67 &  &  & -0.32 & 0.06 &  &  & -0.3 &  & 0.01 & 0.19 & -0.08 & \\
        Berkeley 81 & 0.22 & 1.15 & 5.88 &  &  & 0.16 & 0.07 & 0.05 & 0.08 & 0.23 & 0.01 & 0.25 & 0.07 & 0.22 & 0.07\\
        Collinder 110 & -0.1 & 1.82 & 10.29 &  &  & -0.08 & 0.06 & 0.11 & 0.02 & 0.06 & 0.04 & 0.16 & 0.07 & 0.02 & 0.04\\
        Collinder 261 & -0.05 & 6.31 & 7.26 & 0.04 & 0.04 & -0.02 & 0.04 & -0.04 & 0.02 & 0.03 & 0.13 & -0.01 & 0.02 & 0.09 & 0.03\\
        Czernik 24 & -0.11 & 2.69 & 12.29 &  &  & -0.12 & 0.04 & 0.05 & 0.08 & 0.03 & 0.01 & 0.22 & 0.06 & 0.06 & 0.02\\
        Czernik 30 & -0.31 & 2.88 & 13.78 & -0.07 &  & -0.27 & 0.04 & -0.07 & 0.28 & -0.07 & 0.02 & -0.04 & 0.05 & -0.09 & 0.05\\
        ESO92\_05 & -0.29 & 4.47 & 12.82 &  &  & -0.18 &  & 0.26 &  &  &  & 0.04 &  &  & \\
        Haffner 10 & -0.1 & 3.8 & 10.82 & 0.0 & 0.08 & -0.13 & 0.04 & 0.0 & 0.11 & 0.02 & 0.04 & 0.16 & 0.08 & 0.07 & 0.05\\
        M67 & 0.0 & 4.27 & 8.96 & -0.0 & 0.06 & 0.0 & 0.05 & -0.0 & 0.09 & 0.0 & 0.07 & 0.0 & 0.06 & -0.0 & 0.11\\
        Melotte71 & -0.15 & 0.98 & 9.87 & -0.11 & 0.03 & -0.19 & 0.03 &  &  & 0.02 & 0.04 & 0.07 & 0.03 & 0.0 & 0.06\\
        NGC2141 & -0.04 & 1.86 & 13.34 & -0.2 &  & -0.06 & 0.03 & 0.04 & 0.12 & 0.06 & 0.04 & 0.13 & 0.12 & 0.03 & 0.05\\
        NGC2158 & -0.15 & 1.55 & 12.62 & -0.11 &  & -0.12 & 0.05 & 0.04 & 0.09 & 0.02 & 0.05 & 0.15 & 0.06 & 0.0 & 0.07\\
        NGC2243 & -0.45 & 4.37 & 10.58 & -0.31 & 0.06 & -0.37 & 0.04 & -0.31 & 0.05 & -0.38 & 0.08 & -0.14 & 0.17 & -0.22 & 0.19\\
        NGC2324 & -0.18 & 0.54 & 12.08 & -0.22 & 0.08 & -0.18 & 0.08 & -0.1 & 0.06 & -0.04 & 0.09 & 0.02 & 0.05 & -0.06 & 0.12\\
        NGC2355 & -0.13 & 1.0 & 10.11 & -0.14 & 0.08 & -0.19 & 0.02 & 0.01 & 0.1 & 0.03 & 0.05 & 0.09 & 0.06 & -0.01 & 0.05\\
        NGC2420 & -0.15 & 1.74 & 10.68 &  &  & -0.15 & 0.04 & -0.06 & 0.11 & -0.01 & 0.06 & 0.1 & 0.07 & 0.0 & 0.08\\
        NGC2425 & -0.12 & 2.4 & 10.92 &  &  & -0.15 & 0.05 & -0.03 & 0.14 & 0.02 & 0.05 & 0.13 & 0.06 & 0.04 & 0.07\\
        NGC2477 & 0.14 & 1.12 & 8.85 & -0.04 &  & 0.09 & 0.06 & 0.21 & 0.02 & 0.15 & 0.05 & 0.15 & 0.02 & 0.1 & 0.04\\
        NGC2506 & -0.34 & 1.66 & 10.62 &  &  & -0.33 &  &  &  &  &  &  &  &  & \\
        NGC2516 & -0.04 & 0.24 & 8.32 &  &  & 0.02 & 0.11 &  &  &  &  & 0.45 & 0.08 &  & \\
        NGC2660 & -0.05 & 0.93 & 8.98 &  &  & -0.08 & 0.04 & 0.01 &  & 0.12 & 0.04 & 0.13 & 0.03 & -0.02 & 0.06\\
        NGC3532 & -0.03 & 0.4 & 8.19 & 0.0 & 0.04 & -0.01 & 0.09 & 0.06 & 0.08 & 0.07 & 0.12 & 0.27 & 0.16 & 0.05 & 0.13\\
        NGC3960 & 0.0 & 0.87 & 7.68 & -0.05 & 0.08 & -0.07 & 0.03 & 0.12 & 0.01 & 0.14 & 0.01 & 0.17 & 0.01 & 0.07 & 0.03\\
        NGC4337 & 0.24 & 1.45 & 7.45 & 0.11 & 0.05 & 0.19 & 0.06 & 0.13 & 0.03 & 0.14 & 0.03 & 0.11 & 0.03 & 0.06 & 0.02\\
        NGC4815 & 0.08 & 0.37 & 7.07 & -0.11 & 0.07 & 0.03 & 0.0 & 0.06 & 0.12 & 0.08 & 0.02 & 0.18 & 0.02 & 0.12 & 0.02\\
        NGC5822 & 0.02 & 0.91 & 7.69 &  &  & -0.12 & 0.04 & 0.07 & 0.02 & 0.05 & 0.07 & 0.12 & 0.02 & 0.06 & 0.05\\
        NGC6005 & 0.22 & 1.26 & 6.51 & 0.04 & 0.03 & 0.19 & 0.05 & 0.1 & 0.04 & 0.12 & 0.01 & 0.09 & 0.05 & 0.03 & 0.05\\
        NGC6067 & 0.03 & 0.13 & 6.78 & -0.05 & 0.06 & 0.21 & 0.1 & 0.12 & 0.08 & 0.01 & 0.03 & -0.0 & 0.06 & -0.05 & 0.12\\
        NGC6192 & -0.08 & 0.24 & 6.73 & -0.11 & 0.07 & -0.03 & 0.02 & 0.07 & 0.07 & 0.01 & 0.09 & 0.07 & 0.09 & 0.02 & 0.06\\
        NGC6253 & 0.34 & 3.24 & 6.88 & 0.13 & 0.07 & 0.26 & 0.06 & 0.18 & 0.06 & 0.21 & 0.03 & 0.45 & 0.24 & 0.17 & 0.05\\
        NGC6259 & 0.18 & 0.27 & 6.18 & 0.09 & 0.05 & 0.3 & 0.06 & 0.18 & 0.07 & 0.14 & 0.07 & 0.13 & 0.05 & 0.18 & 0.07\\
        NGC6281 & -0.04 & 0.51 & 7.81 & -0.01 & 0.03 & -0.06 & 0.04 & 0.07 & 0.04 & 0.08 & 0.01 & 0.15 & 0.04 & 0.11 & 0.01\\
        NGC6404 & 0.01 & 0.1 & 5.85 & -0.0 & 0.12 & 0.31 & 0.03 & 0.04 & 0.21 & 0.05 &  & -0.04 & 0.07 & 0.03 & 0.11\\
        NGC6583 & 0.22 & 1.2 & 6.32 & 0.08 & 0.02 & 0.13 & 0.06 & -0.01 & 0.06 &  &  & 0.05 & 0.03 & 0.15 & \\
        NGC6633 & -0.03 & 0.69 & 8.0 &  &  & 0.04 &  &  &  &  &  & 0.27 & 0.18 & -0.14 & \\
        NGC6705 & 0.03 & 0.31 & 6.46 & 0.09 & 0.04 & 0.21 & 0.05 & 0.12 & 0.09 & 0.08 & 0.03 & 0.05 & 0.07 & 0.17 & 0.06\\
        NGC6709 & -0.02 & 0.19 & 7.6 & -0.07 &  & -0.04 & 0.04 & -0.01 &  & 0.0 &  & 0.37 & 0.31 & 0.01 & \\
        NGC6791 & 0.22 & 6.31 & 7.94 &  &  & 0.35 & 0.44 & 0.67 & 0.51 & 0.27 & 0.16 & 0.19 & 0.32 & 0.37 & 0.04\\
        NGC6802 & 0.14 & 0.66 & 7.14 & 0.15 & 0.03 & 0.07 & 0.06 & 0.16 & 0.05 & 0.16 & 0.01 & 0.25 & 0.08 & 0.12 & 0.03\\
        Pismis 15 & 0.02 & 0.87 & 8.62 & 0.09 & 0.02 & -0.05 & 0.06 & 0.11 & 0.13 & 0.17 & 0.07 & 0.28 & 0.1 & 0.13 & 0.03\\
        Pismis 18 & 0.14 & 0.58 & 6.94 & 0.09 & 0.03 & 0.11 & 0.03 & 0.11 & 0.07 & 0.14 & 0.03 & 0.16 & 0.04 & 0.16 & 0.06\\
        Ruprecht 134 & 0.27 & 1.66 & 6.09 & 0.14 & 0.04 & 0.3 & 0.06 & 0.18 & 0.05 & 0.16 & 0.05 & 0.11 & 0.04 & 0.06 & 0.03\\
        Ruprecht 147 & 0.12 & 3.02 & 8.05 & -0.07 & 0.02 & 0.06 & 0.08 &  &  &  &  & -0.01 & 0.04 & 0.07 & 0.08\\
        Ruprecht 4 & -0.13 & 0.85 & 11.68 & -0.07 &  & -0.19 & 0.06 &  &  & 0.09 & 0.05 & 0.14 & 0.02 & 0.05 & 0.09\\
        Ruprecht 7 & -0.24 & 0.23 & 13.11 & -0.24 &  & -0.25 & 0.02 &  &  & -0.07 & 0.01 & 0.09 & 0.06 & -0.08 & 0.04\\
        Tombaugh 2 & -0.24 & 1.62 & 15.76 &  &  & -0.29 & 0.05 & -0.32 &  & -0.01 & 0.09 & 0.07 & 0.04 & 0.0 & 0.12\\
        Trumpler 20 & 0.13 & 1.86 & 7.18 &  &  & 0.06 & 0.04 & 0.11 & 0.08 & 0.13 & 0.05 & 0.18 & 0.05 & 0.13 & 0.07\\
        Trumpler 23 & 0.2 & 0.71 & 6.27 & 0.13 & 0.09 & 0.29 & 0.03 & 0.14 & 0.04 & 0.15 & 0.03 & 0.11 & 0.02 & 0.13 & 0.08\\
        Trumpler 5 & -0.35 & 4.27 & 11.21 & -0.14 & 0.07 & -0.31 & 0.03 & -0.14 & 0.08 & -0.2 & 0.1 & -0.04 & 0.08 & -0.12 & 0.07\\
        \hline
    \end{tabular}}
  \end{table*}

  \begin{table*}
    \caption{\label{clusters_elfe}Average [El/Fe] and $\sigma$[El/Fe] for our sample of open clusters. The age (Gy) and $R_\mathrm{GC}$ (kpc) are from \citet{CG20} and [Fe/H] from \citet{Randich_2022}.}
    \centering
    \scalebox{0.6}{
      \begin{tabular}{lrrrrrrrrrrrrrrrr}
        \hline\hline
        GES\_FLD & [Fe/H] &  Age (Gyr) & R$_{\rm GC}$ (kpc) & $[\ion{O}{i}/$Fe] & $\sigma$ $[\ion{O}{i}/$Fe] & $[\ion{Mg}{i}/$Fe] & $\sigma$ $[\ion{Mg}{i}/$Fe] & $[\ion{Mo}{i}/$Fe] & $\sigma$ $[\ion{Mo}{i}/$Fe] & $[\ion{Pr}{ii}/$Fe] & $\sigma$ $[\ion{Pr}{ii}/$Fe] & $[\ion{Nd}{ii}/$Fe] & $\sigma$ $[\ion{Nd}{ii}/$Fe] &  $[\ion{Eu}{ii}/$Fe] & $\sigma$ $[\ion{Eu}{ii}/$Fe]  \\
        \hline

        Blanco 1 & -0.03 & 0.1 & 8.3 &  &  & 0.02 & 0.06 &  &  & -0.06 & 0.15 & 0.19 & 0.03 & 0.06 & 0.08\\
        Berkeley 20 & -0.38 & 4.79 & 16.32 &  &  & 0.08 & 0.01 & 0.22 & 0.25 & 0.07 & 0.07 & 0.23 & 0.14 & 0.21 & 0.05\\
        Berkeley 21 & -0.21 & 2.14 & 14.73 &  &  & 0.03 & 0.08 & -0.01 & 0.07 & 0.12 & 0.05 & 0.35 & 0.07 & 0.16 & 0.05\\
        Berkeley 22 & -0.26 & 2.45 & 14.29 &  &  & -0.01 & 0.15 & 0.41 & 0.19 & 0.14 &  & 0.28 & 0.12 & 0.29 & 0.05\\
        Berkeley 25 & -0.25 & 2.45 & 13.81 &  &  & 0.01 & 0.04 & 0.34 & 0.11 & 0.18 & 0.03 & 0.32 & 0.09 & 0.17 & 0.02\\
        Berkeley 29 & -0.36 & 3.09 & 20.58 &  &  & 0.07 & 0.11 & 0.47 &  &  &  & 0.4 & 0.09 & 0.58 & \\
        Berkeley 30 & -0.13 & 0.3 & 13.25 & 0.05 &  & -0.04 & 0.09 & -0.01 &  & 0.16 & 0.08 & 0.27 & 0.06 & 0.14 & 0.09\\
        Berkeley 31 & -0.31 & 2.82 & 15.09 & 0.22 &  & 0.04 & 0.06 & 0.21 & 0.04 & 0.17 & 0.05 & 0.3 & 0.03 & 0.27 & 0.06\\
        Berkeley 32 & -0.29 & 4.9 & 11.14 &  &  & 0.05 & 0.06 & 0.15 & 0.1 & 0.12 & 0.06 & 0.2 & 0.06 & 0.2 & 0.08\\
        Berkeley 36 & -0.15 & 6.76 & 11.73 & 0.2 & 0.03 & 0.18 & 0.11 & 0.03 & 0.13 & 0.23 &  & 0.15 & 0.1 & 0.21 & 0.07\\
        Berkeley 39 & -0.14 & 5.62 & 11.49 & 0.12 & 0.05 & 0.06 & 0.06 & 0.04 & 0.1 & 0.13 & 0.05 & 0.14 & 0.05 & 0.19 & 0.04\\
        Berkeley 44 & 0.22 & 1.45 & 7.01 &  &  & 0.02 & 0.11 & -0.06 & 0.06 & -0.01 & 0.17 & -0.08 & 0.14 & -0.12 & 0.1\\
        Berkeley 73 & -0.26 & 1.41 & 13.76 &  &  & 0.01 & 0.01 & 0.22 & 0.15 & 0.19 & 0.08 & 0.38 & 0.07 & 0.25 & 0.05\\
        Berkeley 75 & -0.34 & 1.7 & 14.67 &  &  & 0.03 & 0.08 &  &  & 0.09 &  & 0.36 & 0.13 & 0.31 & \\
        Berkeley 81 & 0.22 & 1.15 & 5.88 &  &  & -0.05 & 0.11 & -0.17 & 0.1 & -0.01 & 0.06 & 0.04 & 0.09 & 0.02 & 0.11\\
        Collinder 110 & -0.1 & 1.82 & 10.29 &  &  & 0.02 & 0.09 & 0.21 & 0.06 & 0.15 & 0.03 & 0.26 & 0.09 & 0.12 & 0.04\\
        Collinder 261 & -0.05 & 6.31 & 7.26 & 0.09 & 0.09 & 0.03 & 0.08 & 0.0 & 0.09 & 0.08 & 0.07 & 0.07 & 0.04 & 0.16 & 0.05\\
        Czernik 24 & -0.11 & 2.69 & 12.29 &  &  & -0.01 & 0.07 & 0.2 & 0.04 & 0.14 & 0.05 & 0.33 & 0.04 & 0.17 & 0.03\\
        Czernik 30 & -0.31 & 2.88 & 13.78 & 0.25 &  & 0.06 & 0.04 & 0.28 & 0.31 & 0.25 & 0.02 & 0.29 & 0.05 & 0.24 & 0.06\\
        ESO92\_05 & -0.29 & 4.47 & 12.82 &  &  & 0.21 &  & 0.53 &  &  &  & 0.31 &  &  & \\
        Haffner 10 & -0.1 & 3.8 & 10.82 & 0.12 & 0.11 & -0.01 & 0.04 & 0.11 & 0.11 & 0.14 & 0.05 & 0.27 & 0.06 & 0.18 & 0.06\\
        M67 & 0.0 & 4.27 & 8.96 & 0.01 & 0.06 & 0.02 & 0.04 & 0.01 & 0.09 & 0.0 & 0.06 & 0.02 & 0.05 & 0.01 & 0.1\\
        Melotte71 & -0.15 & 0.98 & 9.87 & 0.0 & 0.03 & -0.04 & 0.11 &  &  & 0.15 & 0.04 & 0.24 & 0.1 & 0.16 & 0.11\\
        NGC2141 & -0.04 & 1.86 & 13.34 & -0.02 &  & -0.02 & 0.04 & 0.11 & 0.11 & 0.09 & 0.05 & 0.2 & 0.07 & 0.07 & 0.05\\
        NGC2158 & -0.15 & 1.55 & 12.62 & 0.09 &  & 0.04 & 0.04 & 0.19 & 0.07 & 0.18 & 0.04 & 0.31 & 0.03 & 0.17 & 0.05\\
        NGC2243 & -0.45 & 4.37 & 10.58 & 0.17 & 0.06 & 0.1 & 0.06 & 0.15 & 0.07 & 0.1 & 0.08 & 0.31 & 0.13 & 0.26 & 0.17\\
        NGC2324 & -0.18 & 0.54 & 12.08 & -0.04 & 0.08 & -0.01 & 0.07 & 0.08 & 0.06 & 0.14 & 0.09 & 0.2 & 0.05 & 0.12 & 0.12\\
        NGC2355 & -0.13 & 1.0 & 10.11 & -0.03 & 0.05 & -0.09 & 0.04 & 0.14 & 0.08 & 0.13 & 0.05 & 0.2 & 0.04 & 0.09 & 0.05\\
        NGC2420 & -0.15 & 1.74 & 10.68 &  &  & 0.02 & 0.08 & 0.1 & 0.1 & 0.14 & 0.05 & 0.27 & 0.09 & 0.16 & 0.08\\
        NGC2425 & -0.12 & 2.4 & 10.92 &  &  & 0.01 & 0.08 & 0.11 & 0.12 & 0.17 & 0.04 & 0.23 & 0.08 & 0.15 & 0.07\\
        NGC2477 & 0.14 & 1.12 & 8.85 & -0.14 &  & -0.05 & 0.03 & 0.1 & 0.02 & 0.05 & 0.07 & 0.03 & 0.02 & -0.01 & 0.05\\
        NGC2506 & -0.34 & 1.66 & 10.62 &  &  & 0.01 &  &  &  &  &  &  &  &  & \\
        NGC2516 & -0.04 & 0.24 & 8.32 &  &  & 0.06 & 0.07 &  &  &  &  & 0.51 & 0.13 &  & \\
        NGC2660 & -0.05 & 0.93 & 8.98 &  &  & -0.03 & 0.02 & 0.09 &  & 0.16 & 0.02 & 0.2 & 0.02 & 0.03 & 0.05\\
        NGC3532 & -0.03 & 0.4 & 8.19 & -0.02 & 0.01 & -0.0 & 0.09 & 0.03 & 0.08 & 0.07 & 0.09 & 0.27 & 0.16 & 0.06 & 0.11\\
        NGC3960 & 0.0 & 0.87 & 7.68 & -0.05 & 0.06 & -0.07 & 0.03 & 0.12 & 0.01 & 0.13 & 0.01 & 0.16 & 0.01 & 0.07 & 0.02\\
        NGC4337 & 0.24 & 1.45 & 7.45 & -0.15 & 0.04 & -0.07 & 0.08 & -0.12 & 0.03 & -0.11 & 0.02 & -0.14 & 0.02 & -0.19 & 0.03\\
        NGC4815 & 0.08 & 0.37 & 7.07 & -0.07 & 0.03 & 0.03 & 0.07 & 0.06 & 0.07 & 0.06 & 0.03 & 0.18 & 0.02 & 0.11 & 0.04\\
        NGC5822 & 0.02 & 0.91 & 7.69 &  &  & -0.13 & 0.02 & 0.05 & 0.01 & 0.05 & 0.05 & 0.11 & 0.02 & 0.05 & 0.03\\
        NGC6005 & 0.22 & 1.26 & 6.51 & -0.17 & 0.06 & -0.03 & 0.05 & -0.12 & 0.06 & -0.12 & 0.02 & -0.12 & 0.06 & -0.19 & 0.06\\
        NGC6067 & 0.03 & 0.13 & 6.78 & -0.02 & 0.09 & 0.19 & 0.14 & 0.09 & 0.12 & -0.01 & 0.12 & 0.02 & 0.09 & -0.02 & 0.15\\
        NGC6192 & -0.08 & 0.24 & 6.73 & -0.04 & 0.02 & 0.06 & 0.08 & 0.13 & 0.06 & 0.1 & 0.04 & 0.16 & 0.04 & 0.11 & 0.06\\
        NGC6253 & 0.34 & 3.24 & 6.88 & -0.13 & 0.09 & -0.08 & 0.07 & -0.08 & 0.1 & -0.04 & 0.06 & 0.11 & 0.2 & -0.08 & 0.09\\
        NGC6259 & 0.18 & 0.27 & 6.18 & -0.11 & 0.05 & 0.13 & 0.07 & 0.0 & 0.05 & -0.03 & 0.07 & -0.06 & 0.05 & 0.01 & 0.09\\
        NGC6281 & -0.04 & 0.51 & 7.81 & 0.03 & 0.02 & -0.01 & 0.01 & 0.12 & 0.01 & 0.12 & 0.06 & 0.19 & 0.01 & 0.15 & 0.06\\
        NGC6404 & 0.01 & 0.1 & 5.85 & 0.05 & 0.03 & 0.34 & 0.13 & 0.15 & 0.11 & -0.02 &  & 0.03 & 0.07 & 0.02 & 0.02\\
        NGC6583 & 0.22 & 1.2 & 6.32 & -0.14 & 0.01 & -0.09 & 0.05 & -0.23 & 0.05 &  &  & -0.17 & 0.01 & -0.06 & \\
        NGC6633 & -0.03 & 0.69 & 8.0 &  &  & 0.03 &  &  &  &  &  & 0.34 & 0.29 & -0.15 & \\
        NGC6705 & 0.03 & 0.31 & 6.46 & 0.05 & 0.06 & 0.18 & 0.08 & 0.09 & 0.13 & 0.05 & 0.05 & 0.02 & 0.09 & 0.13 & 0.09\\
        NGC6709 & -0.02 & 0.19 & 7.6 & -0.0 &  & -0.0 & 0.06 & 0.06 &  & 0.07 &  & 0.42 & 0.28 & 0.08 & \\
        NGC6791 & 0.22 & 6.31 & 7.94 &  &  & 0.2 & 0.35 & 0.61 & 0.32 & -0.0 & 0.17 & 0.11 & 0.1 & 0.28 & 0.03\\
        NGC6802 & 0.14 & 0.66 & 7.14 & -0.01 & 0.04 & -0.06 & 0.08 & 0.02 & 0.06 & 0.03 & 0.03 & 0.11 & 0.06 & -0.02 & 0.05\\
        Pismis 15 & 0.02 & 0.87 & 8.62 & 0.05 & 0.02 & -0.07 & 0.02 & 0.11 & 0.11 & 0.13 & 0.04 & 0.24 & 0.08 & 0.09 & 0.05\\
        Pismis 18 & 0.14 & 0.58 & 6.94 & -0.07 & 0.02 & -0.04 & 0.03 & -0.04 & 0.05 & -0.02 & 0.07 & 0.02 & 0.02 & 0.02 & 0.09\\
        Ruprecht 134 & 0.27 & 1.66 & 6.09 & -0.12 & 0.07 & 0.02 & 0.06 & -0.09 & 0.07 & -0.11 & 0.05 & -0.17 & 0.04 & -0.2 & 0.04\\
        Ruprecht 147 & 0.12 & 3.02 & 8.05 & -0.2 & 0.02 & -0.06 & 0.06 &  &  &  &  & -0.13 & 0.03 & -0.04 & 0.09\\
        Ruprecht 4 & -0.13 & 0.85 & 11.68 & 0.05 &  & -0.05 & 0.06 &  &  & 0.22 & 0.04 & 0.27 & 0.03 & 0.19 & 0.07\\
        Ruprecht 7 & -0.24 & 0.23 & 13.11 & -0.05 &  & -0.02 & 0.03 &  &  & 0.18 & 0.08 & 0.33 & 0.09 & 0.14 & 0.07\\
        Tombaugh 2 & -0.24 & 1.62 & 15.76 &  &  & -0.06 & 0.08 & -0.06 &  & 0.19 & 0.03 & 0.32 & 0.07 & 0.24 & 0.07\\
        Trumpler 20 & 0.13 & 1.86 & 7.18 &  &  & -0.07 & 0.04 & -0.01 & 0.08 & 0.0 & 0.06 & 0.06 & 0.06 & 0.01 & 0.1\\
        Trumpler 23 & 0.2 & 0.71 & 6.27 & -0.08 & 0.09 & 0.08 & 0.04 & -0.07 & 0.06 & -0.07 & 0.05 & -0.1 & 0.03 & -0.08 & 0.09\\
        Trumpler 5 & -0.35 & 4.27 & 11.21 & 0.21 & 0.06 & 0.04 & 0.03 & 0.2 & 0.07 & 0.15 & 0.1 & 0.31 & 0.07 & 0.23 & 0.07\\
        \hline
    \end{tabular}}
  \end{table*}

  \begin{table}
    \caption{\label{tab:outliers_eu}23 member stars with values of $A(\mathrm{Eu})$ enhanced or decreased and classified as outliers in their respective open clusters according to the IQR method.}
    \centering  
    \begin{tabular}{lll}
      \hline\hline
      CNAME & GES\_FLD & A(Eu)
      \\
      \hline\hline
      05582595+0746114 & Berkeley 22 & 0.34\\
      06411680-1630203 & Berkeley 25 & 0.59\\
      06412105-1629038 & Berkeley 25 & 0.32\\
      06573668+0810127 & Berkeley 31 & 0.66\\
      07464760-0439563 & Berkeley 39 & 0.28\\
      07464911-0441557 & Berkeley 39 & 0.46\\
      19013651-0027021 & Berkeley 81 & 1.13\\
      12381233-6820314 & Collinder 261 & 0.54\\
      12381261-6821494 & Collinder 261 & 0.82\\
      05552710+2052163 & Czernik 24 & 0.54\\
      07310960-0957183 & Czernik 30 & 0.65\\
      06025078+1030280 & NGC2141 & 0.36\\
      07382342-1453123 & NGC2425 & 0.2\\
      12572442-6455173 & NGC4815 & 0.38\\
      18504737-0617184 & NGC6705 & 0.51\\
      18511116-0614340 & NGC6705 & 0.44\\
      19303309+2015442 & NGC6802 & 0.81\\
      09345191-4800467 & Pismis15 & 0.8\\
      17523054-2930564 & Ruprecht 134 & 0.75\\
      17524742-2931471 & Ruprecht 134 & 0.82\\
      12390476-6041475 & Trumpler20 & 1.1\\
      12391113-6036528 & Trumpler20 & 0.45\\
      16004035-5333047 & Trumpler23 & 0.42\\
      \hline

    \end{tabular}
  \end{table}

\end{appendix}

\end{document}